\documentclass[fleqn,usenatbib]{mnras}

\usepackage{newtxtext,newtxmath}

\usepackage[T1]{fontenc}
\usepackage{color}
\usepackage{xcolor}
\DeclareRobustCommand{\VAN}[3]{#2}
\let\VANthebibliography\thebibliography
\def\thebibliography{\DeclareRobustCommand{\VAN}[3]{##3}\VANthebibliography}

\usepackage{graphicx}	
\usepackage{amsmath}	

\title[A new step forward in realistic cluster modelling]{A new step forward in realistic cluster lens mass modelling: Analysis of Hubble Frontier Field Cluster Abell S1063 from joint lensing, X-ray and galaxy kinematics data}

\author[Beauchesne et al.]{
Beauchesne Benjamin$^{1,2}$\thanks{E-mail: benjamin.beauchesne@epfl.ch},
Benjamin Cl\'ement$^{1}$,
Pascale Hibon$^{2}$,
Marceau Limousin$^{3}$,
Dominique Eckert$^{4}$,
\newauthor Jean-Paul Kneib$^{1,3}$,
Johan Richard$^{4}$,
Priyamvada Natarajan$^{6,7}$,
Mathilde Jauzac$^{8,9,10,11}$,
Mireia Montes$^{12,13}$,
\newauthor Guillaume Mahler$^{8,9}$,
Adélaïde Claeyssens$^{14}$,
Alexandre Jeanneau$^{5}$,
Anton M.~Koekemoer$^{15}$,
David Lagattuta$^{8,9}$,
\newauthor Amanda Pagul$^{15,16}$,
Javier Sánchez$^{15,17}$
\\
$^{1}$Institute of Physics, Laboratory of Astrophysics, Ecole Polytechnique Fédérale de Lausanne (EPFL), Observatoire de Sauverny, 1290 Versoix, Switzerland\\
$^{2}$ESO, Alonso de Córdova 3107, Vitacura, Santiago, Chile\\
$^{3}$Aix Marseille Univ, CNRS, CNES, LAM, Marseille, France \\
$^{4}$Department of Astronomy, University of Geneva, Ch. d’Ecogia 16, 1290 Versoix, Switzerland\\
$^{5}$Univ Lyon, Univ Lyon1, Ens de Lyon, CNRS, Centre de Recherche Astrophysique de Lyon (CRAL) UMR5574, F-69230 Saint-Genis-Laval, France\\
$^{6}$Department of Astronomy, Yale University, New Haven, CT 06511, USA\\
$^{7}$Department of Physics, Yale University, New Haven, CT 06520, USA\\
$^{8}$Centre for Extragalactic Astronomy, Department of Physics, Durham University, South Road, Durham DH1 3LE, UK\\
$^{9}$Institute for Computational Cosmology, Department of Physics, Durham University, South Road, Durham DH1 3LE, UK\\
$^{10}$Astrophysics Research Centre, University of KwaZulu-Natal, Westville Campus, Durban 4041, South Africa \\
$^{11}$School of Mathematics, Statistics \& Computer Science, University of KwaZulu-Natal, Westville Campus, Durban 4041, South Africa\\
$^{12}$Instituto de Astrof\'{\i}sica de Canarias, c/ V\'{\i}a L\'actea s/n, E-38205 - La Laguna, Tenerife, Spain\\
$^{13}$Departamento de Astrof\'isica, Universidad de La Laguna, E-38205 - La Laguna, Tenerife, Spain\\
$^{14}$The Oskar Klein Centre, Department of Astronomy, Stockholm University, AlbaNova, SE-10691 Stockholm, Sweden\\
$^{15}$Space Telescope Science Institute, 3700 San Martin Dr.,Baltimore, MD 21218, USA\\
$^{16}$Department of Physics and Astronomy, University of California Riverside, Pierce Hall, Riverside, CA, USA\\
$^{17}$Fermi National Accelerator Laboratory, P. O. Box 500, Batavia, IL 60510, USA\\
}

\date{Accepted XXX. Received YYY; in original form ZZZ}

\pubyear{2023}

\begin{document}
\label{firstpage}
\pagerange{\pageref{firstpage}--\pageref{lastpage}}
\maketitle

\begin{abstract}
We present a new method to simultaneously and self-consistently model the mass distribution of galaxy clusters that combines constraints from strong lensing features, X-ray emission and galaxy kinematics measurements. We are able to successfully decompose clusters into their collisionless and collisional mass components thanks to the X-ray surface brightness, as well as use the dynamics of cluster members to obtain more accurate masses exploiting the fundamental plane of elliptical galaxies. Knowledge from all observables is included through a consistent Bayesian approach in the likelihood or in physically motivated priors. We apply this method to the galaxy cluster Abell S1063 and produce a mass model that we publicly release with this paper. The resulting mass distribution presents different ellipticities for the intra-cluster gas and the other large-scale mass components; and deviation from elliptical symmetry in the main halo. We assess the ability of our method to recover the masses of the different elements of the cluster using a mock cluster based on a simplified version of our Abell S1063 model. Thanks to the wealth of mutli-wavelength information provided by the mass model and the detected X-ray emission, we also found evidence for an on-going merger event with gas sloshing from a smaller infalling structure into the main cluster. In agreement with previous findings, the total mass, gas profile and gas mass fraction are all consistent with small deviations from the hydrostatic equilibrium. This new mass model for Abell S1063 is publicly available, as the \textsc{Lenstool} extension used to construct it.
\end{abstract}




\section{Introduction}

Galaxy clusters are amongst the largest structures bound by gravity and, as such, represent a formidable probe at the cross-roads between cosmology and astrophysics. From their formation and growth to the evolution of the galaxies within them as well as the physics of plasma with the intra-cluster gas heated by the strong gravitational interaction ruling them, galaxy clusters provide a laboratory for astrophysics on all scales. It is possible to use them to constrain models of structure formation and evolution \citep[see][for review]{Allen2011,Kravtsov2012} or probe possible deviations from General relativity \citep{Clowe2006,Lam2012,Cataneo2018} as well as estimators of cosmological parameters \citep{Jullo2010,Acebron2017}. In the context of the $\Lambda$ Cold Dark Matter ($\Lambda$CDM) paradigm, they represent one of the best tools to study Dark Matter (DM) properties \citep{Clowe2004,Bradac2008,Natarajan2017} and divergences to the collisionless model \citep{Harvey2015,Meneghetti2020,Limousin2022}. Such analyses rely on accurate measurements of the cluster mass distribution through different physical phenomena, among which gravitational lensing has an unquestionable place. 

In the framework of General Relativity, gravitational lensing refers to the bending of light rays passing near massive objects such as galaxy clusters called lenses. The most prominent advantage of using this phenomenon  is its ability to account for the total mass, making the DM indirectly visible  \citep[see review by ][]{KneibPN2011}. It also requires fewer assumptions on the studied cluster, besides the thin lens approximation, compared to other methods based on hydrostatic equilibrium \citep{Ettori2010} or galaxy kinematics \citep{Mamon2013}. The lensing effect can be divided into two regimes of different intensities. The weak regime refers to a statistical effect on the apparent shapes of background galaxies which is mainly used to study the outskirts of galaxy clusters \citep{Jauzac2012,Jullo2014}. Closer to the cluster core, the lensing effect intensifies, and features in the strong regime appear, such as lensed galaxies distorted into giant arcs or an apparent presence of the same object multiple times in the case of a multiply-imaged system. Different modelling techniques have been developed to use the strong lensing regime to recover the mass distribution, particularly with multiple images positions and/or shapes \citep[Lenstool, Glafic, LTM, WSLAP;][]{Jullo2007,Oguri2010,Zitrin2009,Diego2007}. All of them have different biases given their specific assumptions. Still, these independent modelling methods obtain total mass profiles of clusters with only a few per cent error \citep{Meneghetti2017,Jauzac2014} and yet offer one of the most robust probes of the $\Lambda$CDM paradigm \citep{Robertson2019,Meneghetti2020}.  

Though lensing analyses are now almost unavoidable to study galaxy clusters in detail, they are not yet providing exhaustive pictures of clusters as additional multi-wavelength information that is available needs to be and can be incorporated. Indeed, even if the DM largely dominates the mass content, that profile can only be extracted accurately with further knowledge of the baryonic matter, i.e. the intra-cluster gas and the stars belonging to the cluster galaxies. The intra-cluster gas, typically expected to be in hydrostatic equilibrium in the cluster potential is heated to high temperature and emits X-ray photons through a Bremsstrahlung emission that allows us to obtain its distribution. Thanks to the optical and spectroscopic observations needed to acquire lensing information, it is possible to recover the mass of cluster members. Therefore, together with the X-ray observations, we can almost fully disentangle the masses of the various cluster components - DM, hot gas and star content. Such decompositions are able to probe in detail the behaviour of the different components, allowing us to see any displacement between them and constraining potential self-interactions of DM particles, like in the case of merging clusters \citep{Markevitch2004,Harvey2015} and test the assumption of hydrostatic equilibrium \citep{Cerini2022}. Combining different observables to estimate the mass also reduces systematic effects arising inherently from each method, and allows us to increase the complexity of the mass reconstructions as more information can be included \citep{Bonamigo2018,Granata2022}.

Recent work by \citet{Bonamigo2017,Bonamigo2018} incorporated an independent gas component from X-ray observables to the lensing modelling software \textsc{Lenstool} \citep{Jullo2007}. This component is added with fixed parameters, and the rest of the mass components are optimised with the systems of multiple images and this extra halo. In particular, this adds asymmetry to the reconstruction that would be difficult to obtain with strong lensing alone due to the sparsity of the data, and the low contribution of the gas to the overall mass budget \citep{Bonamigo2018}. However, this two-step approach does not allow us to include the full information provided by the X-ray emission as only the best-fit model for the gas is explicitly included. This precludes the assessment of possible degeneracies between the different components and obtaining overall errors for the entire model through a single posterior distribution sampling. This method does not therefore provide a balance between the different pieces of information; the best-fit model with gas-only is a priori not the best-fit model according to the combination of the X-ray and lensing observables. This prior work has paved the way by developing the necessary tools, and set the stage for merging the two steps with a single, simultaneous joint analysis that would address the previous weaknesses and allow comprehensive studies of galaxy clusters truly combining X-ray and lensing.

Continuing these previous improvements to \textsc{Lenstool}, the modelling of cluster members has also been enhanced by explicitly including spectroscopic information to calibrate the scaling relations based on the Faber \& Jackson law \citep[][hereafter B19]{Bergamini2019} and combining it with more complex scaling relations such as the so-called fundamental plane of elliptical galaxies \citep[][hereafter G22]{Granata2022}. The analysis of G22 enables the addition of a scatter in the possible mass of each galaxy based on their optical morphology and star kinematics. Notably, both of these works include the gas component from \citet{Bonamigo2018}, and thus provide a more accurate decomposition of the cluster components. In comparison to G22, B19 were able to include the full information from the Faber \& Jackson relation calibration in the form of Bayesian priors because this scaling relation was the only one currently implemented in \textsc{Lenstool}. Indeed, as for \citet{Bonamigo2017}, the fundamental plane measurement was only represented by the best-fit values, and the scaling between the light and mass profiles was adjusted outside the parameter sampling. It was fixed for a specific run of the sampler, and the final value was chosen by selecting the one producing the best-fitting model among all runs. It is then difficult to assess the proper influence of these assumptions on other parts of the model as this amounts to slicing the likelihood on some parameters instead of marginalising on it as a fully consistent Bayesian method would do. 

Taking inspiration from these previous works, we present in this paper a new method aimed at tackling the previously mentioned drawbacks on the modelling of the gas and cluster galaxy members by integrating both components in a single self-consistent homogeneous process. Our method can be applied to all other strong lensing mass reconstruction methods, but we have implemented it in the publicly available software package \textsc{Lenstool} to make it broadly available to the rest of the community. Thus, we start the paper with details of the observational data of the cluster  Abell S1063 in Sect.\ref{sect:AS1063-observations}. We explain in Sect.~\ref{sect:Pre-modelling-analysis} the analysis of the X-ray and galaxy spectroscopic data that are required as part of this new method. We, then, detail the building blocks of the mass model in Sect.~\ref{sect:Mass-modelling}, followed by the definition of our joint X-ray and lensing likelihood in Sect.~\ref{sect:likelihood_def}. In Sect.~\ref{sect:mass-of-as1063} and Sect.~\ref{sect:mock-mass-result}, we analyse the mass distribution of Abell S1063 and the reconstruction of a mock cluster for comparison. We finally discuss the deviation from the hydrostatic equilibrium; a possible merging event and interrogate our modelling hypotheses in Sect.~\ref{sect:discussion}, and concluding with prospects for the application of this new method in Sect.~\ref{sect:conclusion}.

We adopt a flat $\Lambda$CDM cosmology with $\Omega_\Lambda=0.7$, $\Omega_m=0.3$ and $H_0=70$ km~s$^{-1}$Mpc$^{-1}$, and we use magnitudes quoted in the AB system throughout this paper. Regarding the statistical treatment of all the analyses, we compute the uncertainties with the median-centred credible intervals ($\rm CI$) based on the posterior distribution of the considered random variable. We choose the size of these intervals in analogy with the $\sigma$ levels of a normal distribution; thus, uncertainties are expressed in intervals that contain $100\times \rm erf\left(\frac{n}{\sqrt{2}}\right)$ per cent of the posterior with $n$ an integer. To avoid heavy notations and to make clear that we are considering $\rm CI$, we denote them $"n\sigma"\, \rm CI$. We only use $\sigma$ to refer to the actual standard deviation. 

\section{Abell S1063: Observational data}
\label{sect:AS1063-observations}
\begin{figure*}
    \centering
    \includegraphics[width=\linewidth]{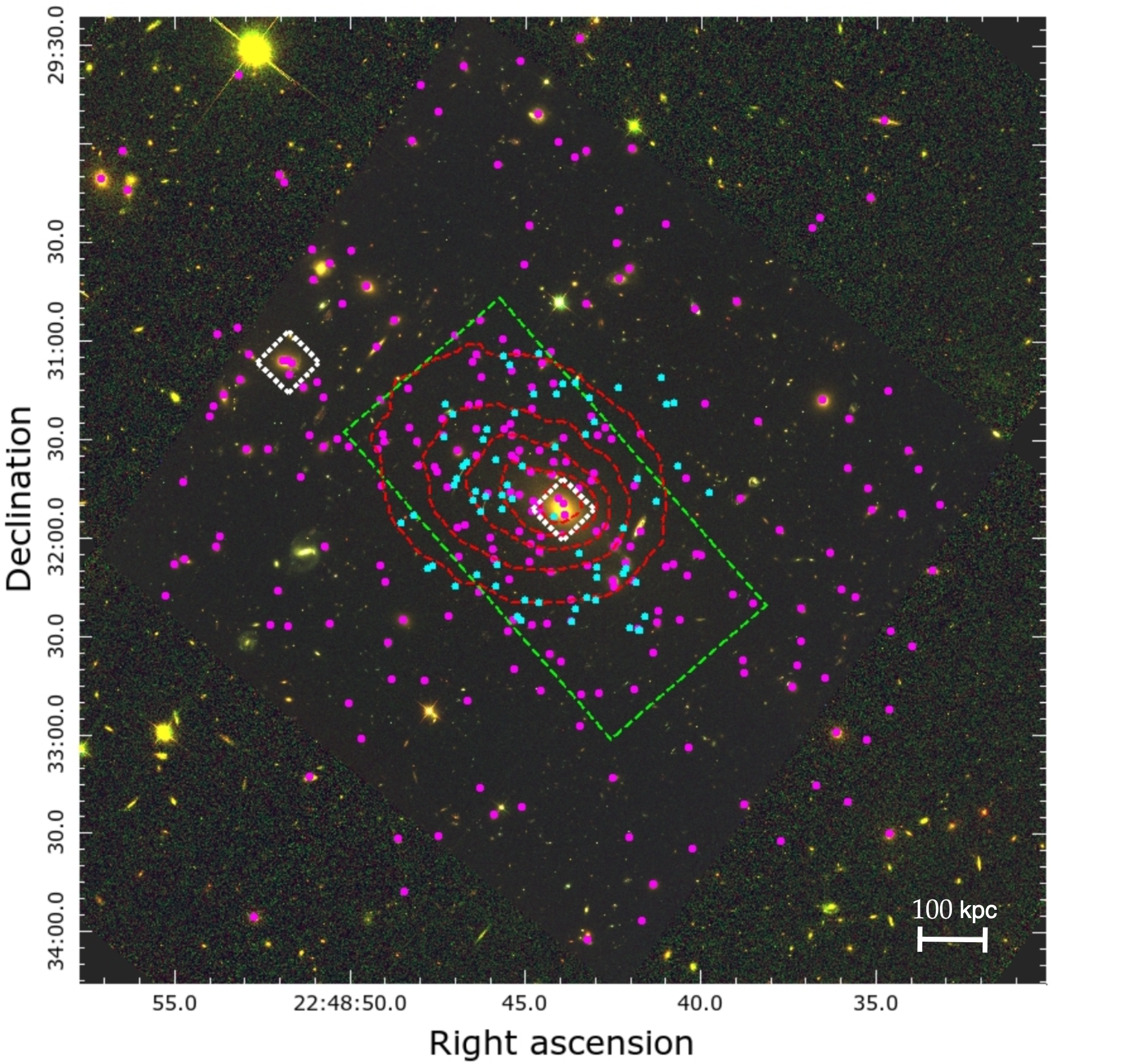}
    \caption{Composite colour image of AS1063 made with \textit{HST} filters ${\rm F435W}$ (blue), ${\rm F606W}$ (green) and ${\rm F814W}$ (red).  Magenta and cyan dots highlight the positions of the modelled cluster members and multiple images, respectively. The two dashed white boxes show the position of the two main mass haloes as detailed in Sect.~\ref{sect:DM-dist}. Finally, dashed red contours and dashed green box present the X-ray surface brightness seen by the \textit{Chandra observatory} and the footprint of the available VLT/MUSE observations, respectively.}
    \label{fig:summary_S1063}
\end{figure*}

Abell S1063 (AS1063; also known as RXC J2248.7-4431 and SPT-CL J2248-4431) is a massive galaxy cluster at $z=0.3475$ first identified by \citet{Abell1989}. It is a bright X-ray source with one of the highest X-ray temperatures measured, $L_X\approx2.5\times10^{45}$ erg $s^{-1}$ in the $[0.1,2.4]$ keV band and $T\approx12$ keV, respectively \citep{Rahaman2021}. It also presents a strong Sunyaev–Zel’dovich (SZ) effect detection in the South Pole Telescope survey \citep{Williamson2011}, with an estimated SZ mass at the virial radius of $M_{200}=2.9\times 10^{15}$ $M_{\odot}$. Fig.~\ref{fig:summary_S1063} shows a colour image of the cluster core made with the \textit{Hubble Space Telescope} (\textit{HST}) broad band filters ${\rm F435W}$ (blue), ${\rm F606W}$ (green) and ${\rm F814W}$ (red).

AS1063 is well-suited for our analysis as it has been observed in multiple wavelengths -X-ray and optical - from space, as well as with an integral field spectrograph from the ground. Thus, we have all the necessary observational data to constrain the gas distribution and the galaxy scaling laws in addition to the lensing observables. Its morphology is also simpler in comparison to other cluster lenses with the same wealth of data. For starters, the cluster mass distribution comprises a single peak, i.e. it is unimodal. In addition, although it appears to be relaxed at first sight, previous analyses have favoured a slightly perturbed cluster. As proposed by \citet{Gomez2012} and supported by \citet{Rahaman2021}, \citet{deOliveira2021}, \citet{Xie2020} and \citet{Mercurio2021}, AS1063 X-ray emission; intra-cluster light \citep[ICL][for a review]{Montes2022}; the presence of a giant radio halo, and galaxy kinematics seem to show that the cluster had undergone a recent off-axis merger.

\subsection{X-ray observations}
\label{sect:X-ray-observations}

As we are focused on a strong lensing analysis, we will model the core of AS1063. Thus, we choose observations from the \textit{Chandra X-ray observatory} as it provides the best spatial resolution amongst X-ray telescopes with a field of view (FoV) covering the entire area of interest. In particular, we use three combined archival pointings of this cluster with a total exposure time of $123$ ks in very faint \footnote{\url{https://cxc.cfa.harvard.edu/ciao/why/aciscleanvf.html}} (\textsc{VFAINT}) mode only, with data from proposals IDs: 4966 (PI: Romer, 2004), 18818 and 18611 (PI: Kraft, 2016)

As the cluster covers a large part of the FoV of each observation, we also retrieved the \textit{Blank-sky} backgrounds \citep{Hickox2006} associated with them to estimate the emission from the background sky. All of these data have been obtained with the Advanced CCD Imaging Spectrometer (ACIS;\citet{Garmire2003}) on board \textit{Chandra}. 

We used a python wrapper of \textsc{CIAO}\footnote{\url{https://cxc.cfa.harvard.edu/ciao/}} $4.13$ \citep{ciao} and \textsc{CALDB} $4.9.6$ provided in the \textsc{Lenstool} public repository\footnote{\url{https://git-cral.univ-lyon1.fr/lenstool/lenstool}} to reduce the data. We proceed first by removing point sources through a combination of visual inspection and the \textsc{wavedetect} routine.  Then, we correct for background flares by successively cleaning them in the $[9.5,12]$ keV bands and in the whole energy range with the \textsc{deflare} tool. We finally remove the area covered by the point sources from the background observations to avoid having negative number counts when subtracting.

These data are then used to create four images binned to four times the initial spatial resolution in the $\left[4.0,7.0\right]$ keV band ($27531$ observed and $12512$ background photon counts in the considered area) but also in the soft ($\left[0.5,1.2\right]$ keV; $29256$ photon counts), medium ($\left[1.2,2.0\right]$ keV; $50977$ photon counts) and hard ($\left[2.0,7.0\right]$ keV; $69660$ photon counts) science energy bands as defined in the \textit{Chandra} Source Catalog \citep{Evans2010}. The former is used in the analysis of the X-ray observations to produce a tessellation of the FoV to map the thermodynamic properties of the intra-cluster gas as detailed in Sect.~\ref{sect:X-ray-data-analysis}. As for the images in the soft, medium and hard bands, they will be fed to \textsc{Lenstool} to perform the fit of the X-ray emitting gas, which is outlined in Sect.~\ref{sect:Mass-modelling}. The background level amongst these three bands (or the broad band) is of $32838$ counts after re-scaling it to the observations.

\subsection{Photometric data}
\label{sect:photo-cat}
AS1063 has been widely observed with \textit{HST} through the Cluster Lensing and Supernova Survey (CLASH; \citet{Postman2012}); the Hubble Frontier Fields (HFF; \citet{lotz2017}) and the Beyond Ultra-deep Frontier Fields And Legacy Observations (BUFFALO; \citet{steinhardt2020}). We use fully reprocessed mosaic images produced by the BUFFALO collaboration combining all available observations and including the HFF dataset. For more details about the different available filters, FoV and depth of the \textit{HST} observations, we refer the reader to \citet{steinhardt2020}.

We use the publicly available measurements on galaxies in the FoV made on the HFF data for this cluster from three different catalogues. In particular, we use the multi-wavelength photometric catalogue produced by \citet{Pagul2023}, combined with two catalogues of galaxy structural properties from \citet{Tortorelli2018} and \citet{nedkova2021}. We primarily use the morphology measurements from \citet{Tortorelli2018} that are made by fitting a single Sérsic light profile with \textsc{Galapagos} and \textsc{Galfit} (i.e. the single band version; \citet{Peng2010}) in a two-step process. As per this process, profiles are first optimised on postage-stamp size images of each object. In the second step, all galaxies are fitted at the same time to the whole image, which improves the estimation of the local background, in particular in the central region where the postage-stamp images are dominated by the light from the bright cluster members. The second structural catalogue presented in \citet{nedkova2021} has been produced as part of the DeepSpace project and relies on a similar method to the one described above, but with only postage-stamp images included in the optimisation with the multi-wavelength fitting tools \textsc{GalfitM} and \textsc{Galapagos-2}, developed as part of the MegaMorph project \citep{haussler2013,vika2013}. Hence, as it has a lower robustness to the intra-cluster light and the bright cluster members, we only use these data if they are not available from the first morphology catalogue to maximise our available knowledge at galaxy-scales.

\subsection{Spectroscopic data}
\label{sect:spectra-cat}

AS1063 was observed by the Multi Unit Spectroscopic Explorer (MUSE) mounted on the Very Large Telescope (VLT) \citep{Bacon2010} with two pointings covering the South-West (SW) and North-East (NE) regions of the cluster (see the green dashed box in Fig.~\ref{fig:summary_S1063} for the combined FoV of the MUSE observations) with the following proposals IDs:\\
60.A-9345(A) (PI: Caputi \& Grillo) and 095.A-0653(A) (PI: Caputi)

These data have already been reduced and analysed in \citet{Karman2015,Karman2017} and \citet{Caminha2016}, wherein a redshift catalogue of the cluster core has been presented. However, we re-reduce and re-analyse the data set with the improved pipeline detailed in \citet{Richard2021}, which has been specifically developed with a focus on cluster fields, to take into account more accurately the ICL and the edge of each integral field unit. Going forward, we use this new redshift catalogue and new MUSE data cube for the rest of this work. We note $10$ per cent of the objects have their first redshift measurement from this re-analysis.

In addition, AS1063 is a target of the CLASH-VLT program (ESO ID: 186.A-0798; PI: P. Rosati) that obtained $200$ hours of observations on clusters from the CLASH sample with the VIsible Multi-Object Spectrograph (VIMOS) previously installed on the VLT before its decommissioning. A catalogue containing almost $4000$ redshifts has been produced by \citet{Mercurio2021} and released publicly. Thus, we use it to complete the MUSE catalogue mentioned previously on the HFF footprint.

\section{Pre-modelling analysis}
\label{sect:Pre-modelling-analysis}
The approach used for this work is based on the one developed for the \textsc{Lenstool} software \citep{Jullo2007} enhanced with the following refinements:
\begin{itemize}
    \item Modelling of the X-ray emitting intra-cluster gas.
    \item Inclusion of the kinematics of Galaxy-scale perturbers.
    \item Addition of a perturbative surface of B-spline on the lensing potential \citep[see][and \ref{sect:B-spline-dist} for details]{Beauchesne2021}.
\end{itemize}
Hence, in addition to the usual modelling analysis presented in Sect.~\ref{sect:Mass-modelling}, we need measurements of the temperature and metallicity of the intra-cluster gas to fully define our plasma emission model. This process is detailed in Sect.~\ref{sect:X-ray-data-analysis}. We also recover the kinematics of cluster members described in Sect.~\ref{sect:MUSE-data-analysis}.

\subsection{X-ray data analysis}
\label{sect:X-ray-data-analysis}

\begin{figure*}
    \centering
    \includegraphics[width=\linewidth]{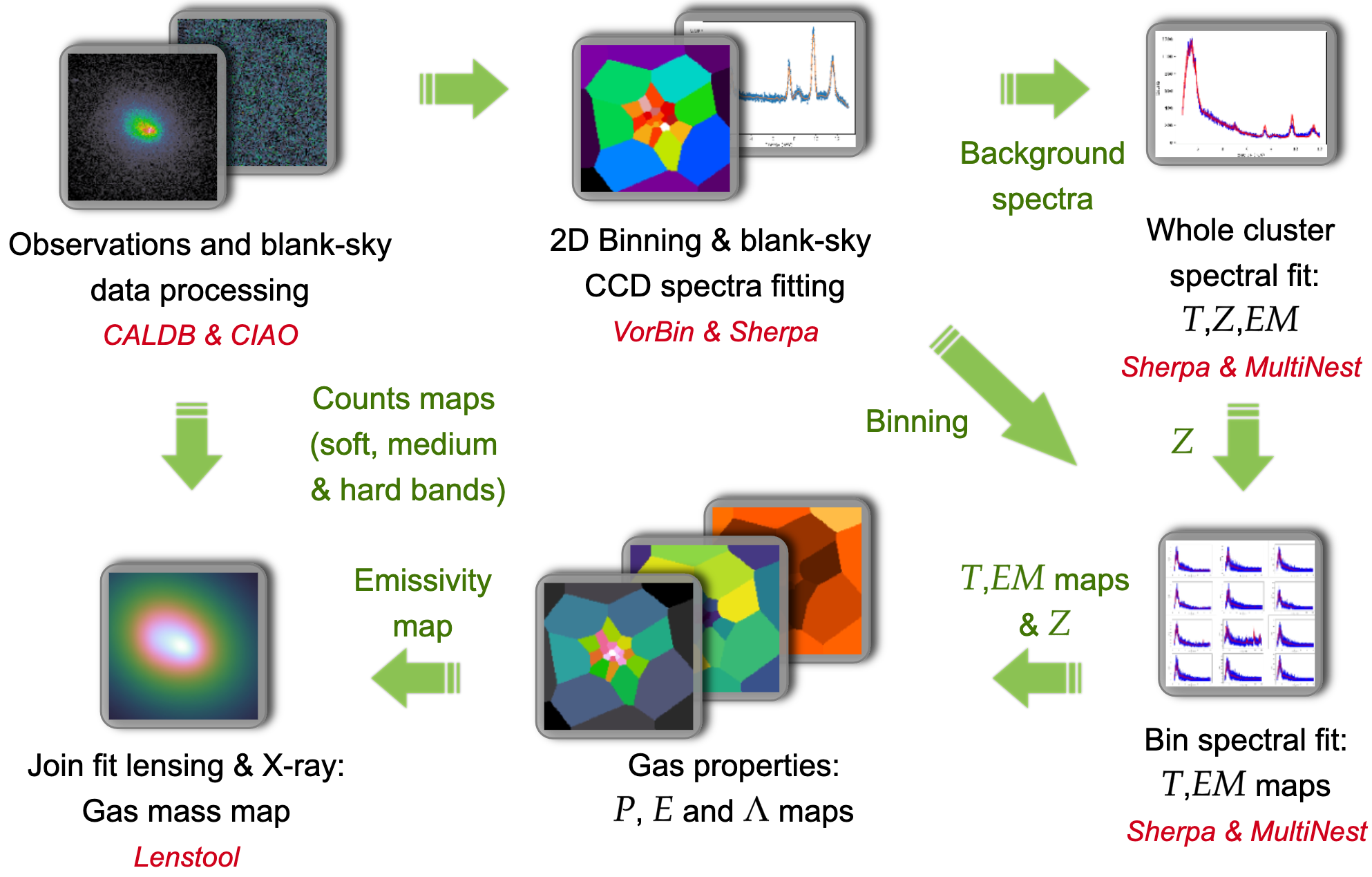}
    \caption{Flowchart presenting the process applied on X-ray data to recover the intra-cluster gas properties and the mass distribution. Each step is detailed in Sect.~\ref{sect:X-ray-data-analysis}.}
    \label{fig:x-ray-flowchart}
\end{figure*}
We are interested in modelling the intra-cluster gas mass distribution using the X-ray surface brightness ($S_X$) maps that are presented in Sect.~\ref{sect:X-ray-observations}, defined as:

\begin{equation}
    S_X(x,y)=\frac{\Lambda (T,Z)}{4\pi(1+z)^4}\int_{\mathbb{R}} n_e(x,y,z)n_p(x,y,z) dz
    \label{Eq:S_X}
\end{equation}

Where $(x,y,z)$ are the observer coordinates with the $z$-axis aligned with the Line-Of-Sight (LOS). Parameters $z$, $T$, $Z$, $n_e$ and $n_p$, are the cluster redshift, temperature, metallicity, electron density, and proton density, respectively. $\Lambda(T,Z)$ is the associated cooling function \citep{Boehringer&Hensler1989,Sutherland&Dopita1993}. To derive $S_X$ from the mass modelling (i.e. $n_e$ and $n_p$), we need to extract the thermodynamic properties of the gas. The method associated with this extraction is outlined in the following two sections, and the multiple steps are shown schematically in the flowchart in Fig.~\ref{fig:x-ray-flowchart}.

\subsubsection{X-ray spectra fitting}

To obtain the thermodynamic properties of the gas from its X-ray spectrum, we follow a similar approach to the one detailed in \citet{Rossetti2007}. We make a Voronoi tessellation of the X-rays images of the cluster field with the \textsc{VorBin} python packages \citep{capellari2003} from the photon counts associated with the cluster in the $\left[4,7\right]$ keV band. We need to rely on a tesselation to obtain enough photons on the X-ray spectra. We remove the signal associated from the background of the total signal, and we use the \citet{Gehrels1986} approximation to obtain the uncertainty as the sum of the two associated variances. A signal-to-noise (S/N) threshold parametrises this tessellation, for which we use an S/N of $10$ as it seems to be the best compromise between the uncertainty and the mapping resolution in the different S/N that we tried. We obtained $9962^{+2489}_{-2747}$ observed counts per bin on the $\left[0.5,12.5\right]$ keV band amongst all observations with this S/N as well as $16913^{+72685}_{-15387}$ for the blank-sky background before rescaling. We also reproduce the spectral fitting for different S/N to obtain more details on gas properties and insight into the induced bias on the gas distribution in Appendix.~\ref{app:SN_threshold}.

We assumed the metallicity to be constant across the whole cluster field, and as the temperature is high in AS1063 ($<8$~keV), this assumption should not bias the result on the mass fitting. In that range of temperature, the emission lines have only a limited contribution to the overall spectra. Other parameters of the gas emission model are mapped in the field based on the tessellation described above. All these quantities are fitted with the Astrophysical Plasma Emission Code (APEC) model\footnote{\url{http://atomdb.org/}} combined with a photoelectric-absorption (PHABS) model\footnote{\url{https://heasarc.gsfc.nasa.gov/docs/xanadu/xspec/manual/XSmodelPhabs.html}} to account for the foreground galactic gas. For the APEC model, we assume the abundance ratio provided by \citet{Asplund2009} and keep the other parameters free. The hydrogen column density ($n_H$) needed by the PHABS model is fixed to the value measured by \citet{HI4PI}. Before proceeding to the two successive fitting procedures shown in the flowchart in Fig.\ref{fig:x-ray-flowchart}, we empirically model the instrumental and sky background provided by the blank-sky observations with the B-spline functions. This background modelling approach is detailed in Appendix~\ref{app:X-ray-background-model}.

Our fitting process is based on the fitting environment \textit{Sherpa} \citep{sherpa} combined with the nested sampling method \textit{pyMultiNest} \citep{multinest,pymultinest}. We adopt the following Poisson log-likelihood, $\mathcal{L}_{\rm X-ray}$, which is similar to Xspec \textsc{Cstat}\footnote{\url{https://cxc.harvard.edu/sherpa4.13/ahelp/cstat.html}} without the Sterling approximation:
\begin{equation}
   \mathcal{L}_{\rm X-ray}=\sum_{i} D_i\log\left(M_i\right)-M_i-\log\left(D_i!\right)
   \label{Eq:poisson-likelihood}
\end{equation}
where $D_i$ and $M_i$ represent the observed and model number of counts in each spectral bin, respectively. The term, $\log\left(D_i!\right)$, is computed before and kept in memory. We benefit from the following relation to reduce the computational cost:
\begin{equation}
\log\left(D_{i}!\right)=\log\left(D_{i-1}!\right)+\log\left(D_{i}\right)
\end{equation}
We first fit the whole cluster field X-ray spectrum to extract the metallicity of the gas. We then fix it to the best-fit value and perform the same spectral fit on each individual bin. All spectra are taken in the $\left[0.5,12.0\right]$ keV range, the high energy part being dedicated to the normalisation of the background model. Hence, the sky background is assumed to follow the instrumental one, as this normalisation is mainly constrained on energy levels where the effective area of the telescope goes to zero. We also check the suitability of our assumed $n_H$ values by allowing it free to optimise, which leads to $n_H=(2.285^{+0.496}_{-0.681})\times10^{20}$~cm$^{-2}$ for the $"1\sigma"\, \rm CI$. The measured value of $n_H=1.293\times10^{20}$~cm$^{-2}$ from \citet{HI4PI} is included in the $"2\sigma"\, \rm CI$. Hence, it is a fair assumption to use the measured values for the rest of our analysis.

\subsubsection{Gas properties}

The fitting approach above provides us with the temperature, $T$, the metallicity, $Z$, and the APEC norm, $N$, that we use to obtain the pseudo-pressure, $P$, and the pseudo entropy, $E$, with the following formula (e.g. similar to \citet{Rossetti2007} with different units):
\begin{align}
    P&=T\sqrt{\frac{10^{-14}\pi {\rm nhc}}{4.6656\times10^{8}(1+z)^2}\frac{N}{Area_{\rm bin}}}\\
    E&=T\left(\frac{10^{-14}\pi {\rm nhc}}{4.6656\times10^{8}(1+z)^2}\frac{N}{Area_{\rm bin}}\right)^{-1/3}
\end{align}
Where $Area_{\rm bin}$ is the area covered by each bin in arcsec, and ${\rm nhc}$ is the conversion factor from $n_e$ to $n_p$. Regarding $\Lambda(T,Z)$, it is obtained as a single emission measure for the associated plasma model with a \textit{Sherpa} routine. We compute it for the soft, medium and hard energy bands that we multiply with the exposure maps associated before summing them. The obtained maps are then fed in to \textsc{Lenstool} with the total counts among all bands and the \textit{blank-sky} backgrounds as detailed in Sect.~\ref{sect:X-ray-constraint}.

\subsection{Cluster member kinematics}
\label{sect:MUSE-data-analysis}
In this section, we detail how we measure the stellar line-of-sight velocity dispersion (hereafter LOSVD) of cluster members from the MUSE datacube. Our approach is similar to the one outlined in B19, with some modifications needed for our modelling choices as presented in Sect.~\ref{sect:Mass-modelling}.

\subsubsection{LOSVD measurements}
We extract galaxy spectra from the MUSE datacube with an elliptical aperture, $R_{ap}=R_e$, where $R_e$ is the half-light radius of the galaxy. To create this elliptical mask, we measure the semi-major axis, \textsc{A\_WORLD}, semi-minor axis, \textsc{B\_WORLD} and position angles, \textsc{THETA\_J2000}, using \textsc{SExtractor}\footnote{\url{https://sextractor.readthedocs.io/en/latest/Position.html}} \citep{Bertin1996} in the HFF images which have the same astrometry as our datacube. We use $R_e$ from the structural catalogues presented in Sect.\ref{sect:spectra-cat}, using \textsc{FLUX\_RADIUS} (i.e. $R_e$ proxy estimation) when the previous ones are not available. Thus, we obtained spectra for all the $107$ spectroscopically confirmed cluster members in the MUSE FoV. As the BUFFALO program mainly expands the HFF observations spatially, these measurements should not be affected by the use of the later program images. The extraction of the spectra is performed with the python package \textsc{MPDAF} \citep{Bacon2016} developed for the analysis of MUSE data. We use the optimal extraction algorithm for CCD spectroscopy implemented \citep{horne1986}. Then, we limit our wavelengths of analysis to $4850$~nm-$7160$~nm in the observer frame, which corresponds to the range used by B19.

Measurements of the LOSVD are performed with the python package \textsc{pPXF} \citep{Cappellari2017} modified to use the nested sampling engine \textit{pyMultiNest} \citep{multinest,pymultinest} as the non-linear optimiser method. LOSVD parameters are obtained from a cross-correlation between the galaxy and a set of star spectra templates. In the original package, parameters are optimised through the minimisation of a $\chi^2$ statistic that has been modified here to maximise the associated Gaussian log-likelihood. In particular, we sample the LOS velocity with respect to the cluster redshift, $V$, the velocity dispersion, $\sigma_e$, and the two first Hermite moments, ($h_3,\,h_4$), from Equation~$13$ of \citet{Cappellari2017}. We also use a Legendre polynomial of degree $3$ to modify in a multiplicative way the continuum as showed in Equation~$11$ from the same reference. Thus, we fit seven parameters for each spectrum, where we use gaussian priors for $V$, $h_3$ and $h_4$ with the following law $\mathcal{N}(0,250)$, $\mathcal{N}(0,0.05)$ and $\mathcal{N}(0,0.05)$, respectively. The Legendre polynomial coefficient and $\sigma_e$, have uniform priors with the $[-0.3,0.3]$ interval (i.e. default bounds) for the first, and in the range $[0,1000]$ for the second. For the star templates, we use the Indo-US Library of Coudé Feed Stellar Spectra \citep{valdes2004}, which have a FWHM of $1.35$~$\mathring{\rm A}$, and a pixel-scale of $0.44$~$\mathring{\rm A}$ pixel$^{-1}$. To automatically correct for the emission line contamination, we use the \textsc{clean} parameter of \textsc{pPXF} in a two-step approach. We first run \textsc{pPXF} with a geometrical optimiser that is in our case, the Trust Region Reflective algorithm implemented in \textsc{Scipy} \citep{2020SciPy-NMeth} with \textsc{pPXF} cleaning mode activated. This mode rejects outliers recursively until it converges to a stable number of masked data points. Then, we use the previous mask to perform the nested sampling. This is a compelling computing optimization, as running the nested sampling iteratively to improve the masking is time expensive.

We restrict our sample to galaxies with $S/N>3$ for their spectra, and a redshift estimation with more than $80$ per cent reliability (i.e. $z_{conf}\geq2$ for MUSE redshift; see \citet{Bacon2015} for details). We analyse the bias of these measurements in Appendix~\ref{app:LOSVD-bias} and we also apply the correction defined in Equation~\ref{Eq:sigma_poly}.

\subsubsection{The fundamental plane of elliptical galaxies and the Faber \& Jackson relation}
\label{sec:fundamental_plane_calibration}

\begin{figure*}
    \begin{minipage}{0.48\linewidth}
    \centering
    \includegraphics[width=\linewidth]{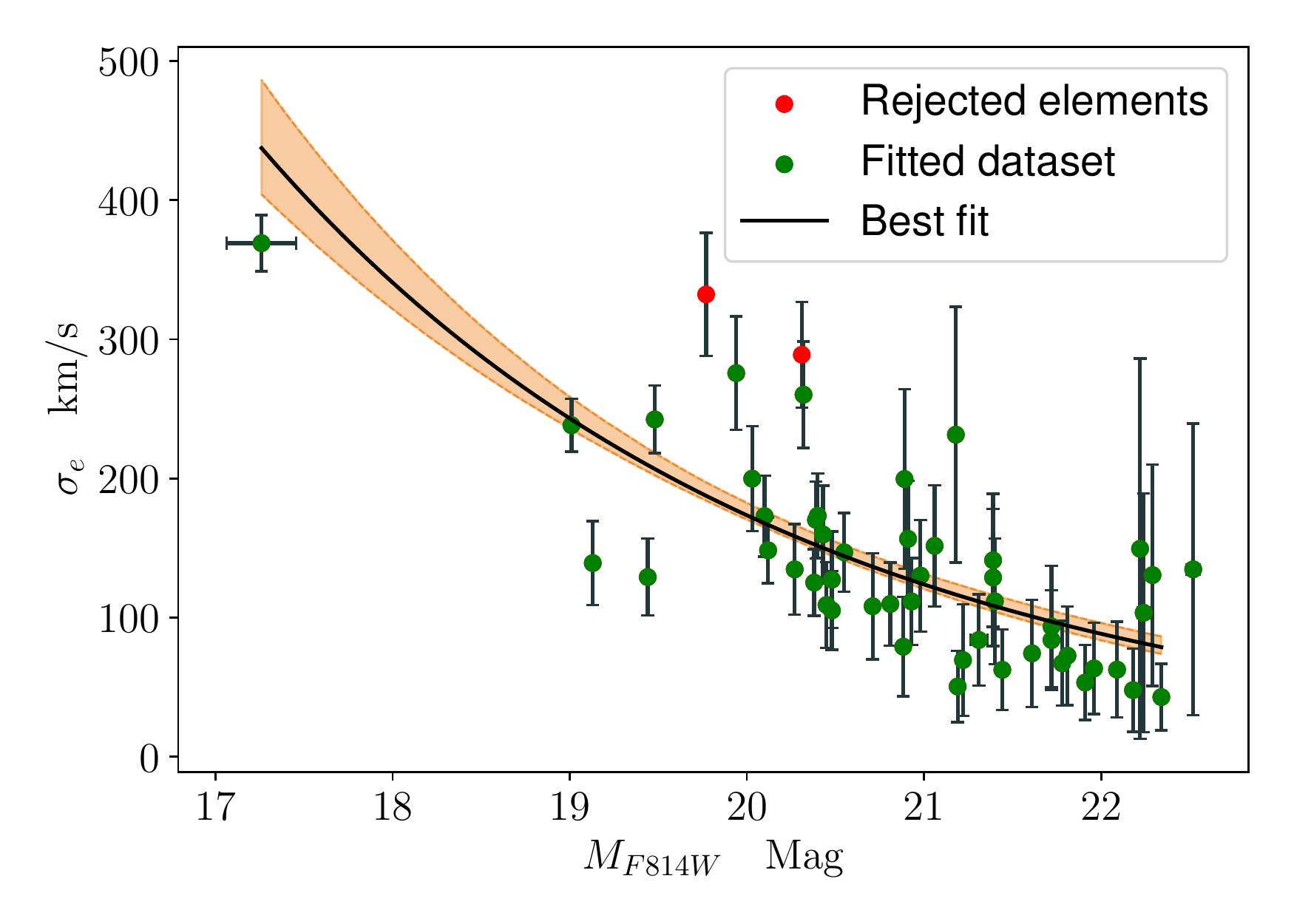}
    \end{minipage}
    \begin{minipage}{0.48\linewidth}
    \centering
    \includegraphics[width=\linewidth]{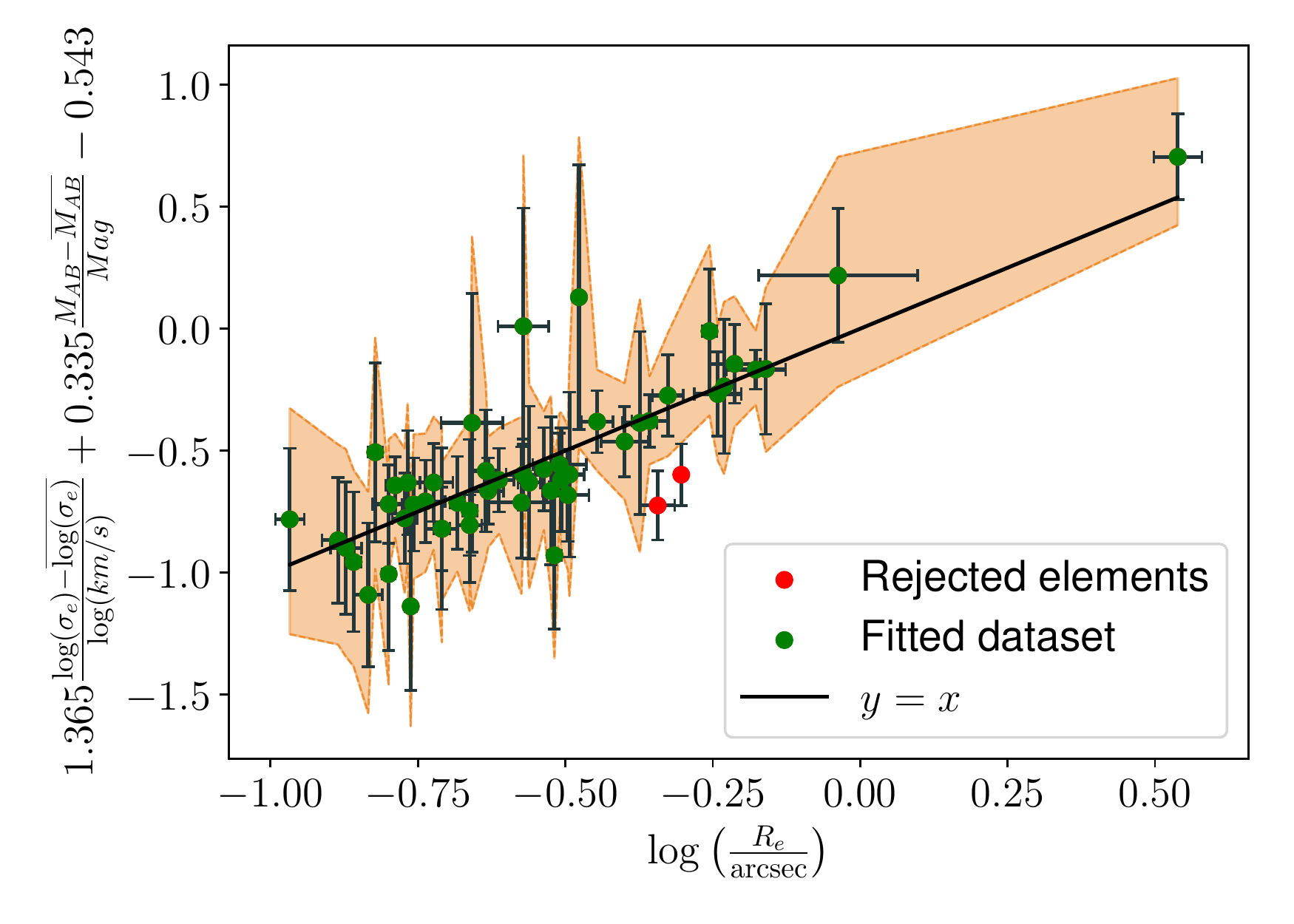}
    \end{minipage}
    \caption{\textit{Left panel:} Graph of the LOSVD measurements as a function of the magnitude in the \textit{HST} ${\rm F814W}$ filter of the associated cluster member. LOSVD errors are the standard deviation of the distribution obtained by the nested sampling run, and the magnitude ones are provided by the \citet{Pagul2023} catalogue. The plain black line and the light orange area represent the best-fit relations as well as the $"1\sigma"\, \rm CI$ of the whole sampled relations. \textit{Right panel:} Plot of the fundamental plane against the logarithm of the half-light radius. Errors represent the standard deviation and have been taken from the morphological catalogues detailed in Sect.~\ref{sect:photo-cat} or computed from a sampling run. The black plain line shows the best fit of the fundamental plane, and the light orange area represents the position of the green dots among the $"1\sigma"\, \rm CI$ of the sample of fundamental plane parameters. For both panels, green dots highlight objects considered in the final fit when the blue ones have been rejected from the calibration process.}
    \label{fig:FJ-FP}
\end{figure*}

Thanks to high-resolution images provided by \textit{HST} within the HFF program, the structural properties of cluster members have been measured. With the addition of the velocity dispersion, $\sigma_e$, measurement, we can calibrate the fundamental plane of elliptical galaxies as defined in \citet{Hyde2009} as follows:

\begin{equation}
    \log_{10}(R_e)=a +b\mu_e+c\log_{10}(\sigma_e)
\end{equation}

Where $\mu_e$ is the averaged surface brightness inside $R_e$, in mag~arcsec$^{-2}$, and $a$, $b$ and $c$ are the parametrization of the plane. Regarding the units, we adopt arcseconds and km/s for $R_e$ and $\sigma_e$, respectively. We use $R_e$ and $\mu_e$ provided by the structural catalogues that we previously obtained as outlined in Sect.~\ref{sect:photo-cat}. 

From the expression of the fundamental plane, we can derive the Faber \& Jackson relation \citep{Faber1976}, which has the following form:
\begin{equation}
    L\propto\sigma_e^{\alpha}
\end{equation}
Where $L$ is the galaxy luminosity, and $\alpha$ is associated with the coefficient $c$. Indeed, we have:
\begin{align}
    L\propto&\, R_e^2 10^{-\mu_e/2.5}\\
    \Rightarrow L\propto&\, 10^{-\mu_e/2.5} 10^{2(a +b\mu_e)}\sigma_e^{2c}\\
    \Rightarrow L\propto&\,\sigma_e^{2c}\text{, assuming }\mu_e={\rm Cst.}
\end{align}
Following B19, we calibrate the Faber \& Jackson relation through the associated scaling relation usually defined in \textsc{Lenstool} \citep{PNKneib1997,Richard2010}, which is:
\begin{align}
    \sigma_e&=\sigma_e^{*}\left(\frac{L}{L^*}\right)^{1/2c}\\
    \Rightarrow\log(\sigma_e)&=\log(\sigma_e^{*})+\frac{1}{2c}\log\left(\frac{L}{L^*}\right)
\end{align}
Where $L^*$ and $\sigma_e^{*}$ are the luminosity and the velocity dispersion of a galaxy which is typical of the galaxy population at that redshift. $L^*$ is usually chosen where the elliptical galaxy luminosity function cuts off \citep{schechter76}. Thus, the calibration of this relation is identical to fitting a line in the log space. Joint optimisation of both relations can be done with four parameters, three for the fundamental plane and one for the Faber \& Jackson relation.

To perform the calibration of the plane and the Faber \& Jackson relation, we use a combined approach inspired by the method outlined in \citet{Cappellari2013}. We use the same $\chi^2$ statistics that are defined to fit a plane or a line, taking into account the scatter on all datasets, but we sum the two associated likelihoods to perform the joint optimisation with \textsc{pyMultiNest}. We assume the statistical independence of both relations in the combined likelihood, which is a priori not true, but allows us to derive a consistent slope between both relations. As shown in G22 and B19, the separate fit can lead to different slopes in the mass versus $\sigma$ plane, and thus inconsistent modelling of cluster members depending on which relation is used. Regarding the outliers, we use an iterative approach applied on the fundamental plane only, where all data points that are at more than $2.5\sigma$ than the associated best-fit relations are rejected. This rejection is performed until there are no more data points to exclude. However, this method results in an increased intrinsic scatter on this second relation. Regarding this scatter, it is estimated at each step of the rejection by running the nested sampler a first time with scatter fixed to zero and choosing it in a way that the associated reduced $\chi^2_{\rm best\,fit}=1$. The actual sampling is then done with this value.

\begin{table}
\centering
\begin{tabular}{ccc}
    \hline
    Parameters & Mean & $\sigma$\\
    \hline
    $a$&$-0.543$& $0.029$\\
    $b$&$0.335$& $0.028$ \\
    $c$&$1.365$& $0.122$ \\
    $\sigma_e^*$&$195.4$& $6.8$\\
    $b_{\rm FB}$&$[0.142634]$& $-$\\
    $b_{\rm FP}$&$[0.159140]$& $-$\\
    $\overline{\log(\sigma_{e})}$&$[2.08949]$& $-$\\
    $\overline{M_{AB}}$&$[19.62834]$& $-$\\
    \hline
\end{tabular}
\caption{Parameters of the Faber \& Jackson scaling relations and the fundamental plane of elliptical galaxies used or sampled in this work.}
\label{Tab:rest_fundamental_plane}
\end{table}

Figure~\ref{fig:FJ-FP} presents the fundamental plane (left panel) as well as the Faber \& Jackson law (right panel) sampling results, and the parameter posterior distribution statistics are shown in Table~\ref{Tab:rest_fundamental_plane}. The left plot shows that only two sources have been rejected from the fundamental plane (i.e. red dots), signalling that our selection of cluster members lies well into that plane, unlike the Faber \& Jackson relations where some galaxies show a larger scatter from the relation. In comparison to G22 and B19, we have a higher estimation of $c$ at $3.1 \sigma$, and a similar one at less than $1 \sigma$, respectively. However, we used different apertures than B19 and G22 to extract cluster member spectra which can partially explain the difference with G22 as well as in the joint optimisation. Both plane estimations are in agreement with blank field results at less than $2 \sigma$ in the R-band \citep{Hyde2009}. Regarding our $b$ estimation, our results agree with the two other studies at less than $2 \sigma$, depending on the case. Therefore, we conclude that our calibration process is able to recover estimates consistent with the previously derived values available in the literature.

\subsubsection{Parameters estimation through the fundamental plane}
\label{sect:FP_prediction_accuracy}

The quality of our fundamental plane calibration can be assessed through its prediction, $R_{e,\rm FP}$, against the measured one, $R_{e}$, and the same can be done for $\sigma_{e,\rm FP}$ and $\sigma_e$. Except for a few outliers, most predictions are in good agreement with the measured values. The $R_e$ and $\sigma_e$ predictions show a mean relative error on the considered sample of $4.0$ and $-3.6$ per cent, respectively. However, the uncertainties based on based on the calibrations show larger uncertainties. In fact, it is a factor of four of the observed errors for $\sigma_e$ and twice for $R_e$. As we are using the fundamental plane to recover the mass of cluster members, we are more interested in the accuracy and precision of the relations for a combination of $\sigma_e$ and $R_e$ rather than them independently as the combination appears in the mass estimate. In particular, the total mass of a model following a mass profile of a dual Pseudo-Isothermal Elliptical (dPIE) \citep{Limousin2005} follows $M_{\rm tot,dPIE}\propto R_e\sigma_e^2$, if we assume that the central velocity, $\sigma_0$, and the cut radius, $r_{\rm cut}$, are proportional to $\sigma_e$ and $R_e$, respectively.

We need two quantities from the plane to predict the third one, though only $R_e$ and $\sigma_e$ are relevant for the mass modelling. Thus, we use the combination of $\sigma_e$ and $\mu_e$ to predict $R_e$ (i.e. $R_{e,\rm FP}$) and combine $R_e$ and $\mu_e$ to obtain $\sigma_{e,\rm FP}$. From these two uses of the fundamental plane, we create two mass proxies. We estimate $R_e$ and $\sigma_e$ from the fundamental plane, which leads to which leads us to the two mass proxies, $R_{e,\rm FP}\sigma_{e}^2$ and $R_{e}\sigma_{e,\rm FP}^2$. Both proxies agree well with the equivalent based on measured values only. Notably, the averaged relative error for $R_{e}\sigma_{e,\rm FP}^2$, and $R_{e,\rm FP}\sigma_{e}^2$, are of $-4.4$ and $8.0$ per cent. Regarding the uncertainties, we assume that for the measurements-based proxies, that $R_e$ and $\sigma_e$ are statistically independent, and we sample them according to their measurement uncertainties. The fundamental plane based proxies uncertainties are assumed to be only from $\sigma_{e,\rm FP}$ or $R_{e,\rm FP}$, as $R_e$ or $\sigma_e$ will be fixed. Hence, for both proxies, the propagated uncertainties are $\sim4-5$ times smaller than the observed ones. Though, these are large enough to be in agreement at less than the $0.6  \sigma$ level with their measured counterparts. We note here that a proper comparison with the observational uncertainty will require the correlation between $R_e$ and $\sigma_e$ but it is beyond the scope of this paper.

\section{Mass modelling}
\label{sect:Mass-modelling}

As previously mentioned, our modelling incorporates the usual cluster-scale model represented mainly by the DM, the intra-cluster gas, cluster members through the fundamental plane of elliptical galaxies, as well as the Faber \& Jackson law and an additional perturbative component. We detail the modelling of each in order as: Sect.~\ref{sect:DM-dist} for the cluster-scale DM and Sect.~\ref{sect:gas-dist} for the gas. We finish with the galaxy-scale elements and the perturbative piece in Sect.~\ref{sect:clus-memb-dist} and Sect.~\ref{sect:B-spline-dist}, respectively.

Except for the modelling of the additional perturbation, all the building blocks of our models are dPIE potentials \citep{Limousin2005}. Each of the different components is composed of a sum of this analytical profile with specific assumptions on their parameters. Such potentials are defined by $7$ parameters, the central coordinates, $(x_c,y_c)$, the position angle, $\theta$, the ellipticity, $\epsilon$, a central velocity dispersion, $\sigma_0$, a core radius, $r_{\rm core}$, and a cut radius, $r_{\rm cut}$. The 3D mass density, $\rho$, follows the relation:

\begin{equation}
    \rho(r)=\frac{r_{\rm core}+r_{\rm cut}}{2\pi G r_{\rm core}^2r_{\rm cut}}\times \frac{\sigma_0^2}{\left(1+\left(\frac{r}{r_{\rm core}}\right)^2\right)\left(1+\left(\frac{r}{r_{\rm cut}}\right)^2\right)}
\end{equation}

With $r$, the elliptical radius in the coordinates of the potential and $G$, the universal constant of gravitation. Now, we outline how we parametrize these dPIE potentials for {\bf each of the cluster components}, followed by the definition of the small B-spline perturbation that we add on top of the overall parametric modelling of the various components.

\subsection{Large scale dark matter distribution}
\label{sect:DM-dist}

Similarly to previous studies on AS1063 \citep{Caminha2016,Limousin2022}, we use two haloes to represent the smooth DM component of the cluster. The main potential is associated with the brightest cluster galaxy (BCG) to account for most of the cluster DM, while the second one is placed near a higher concentration of galaxies in the North-East (NE) of the cluster. This placement is shown as the dashed white diamonds in Fig.~\ref{fig:summary_S1063}. The main cluster halo has all of its parameters free to vary except its position, and $r_{\rm cut}$, which are fixed to the BCG centre, and to $3$ $\rm Mpc$ (beyond the virial radius of the cluster), respectively. The parameter $r_{\rm cut}$, is in fact, ill-constrained due to the lack of strong lensing constraints in the cluster outskirts. Regarding the potential in the NE, we use a Bayesian regularisation on its position. We assume it to be centred around the light distribution with Gaussian priors on its centre coordinate. It is similar to the assumptions made in \citet{Limousin2022} and avoids dPIE halo positions to be inconsistent with the luminous distribution. We assume a standard deviation of $2$~$\rm arcsec$ around the centre of the group of three galaxies. As there are no spectroscopically confirmed multiple images near the NE halo, its $r_{\rm core}$ is ill-constrained, and fixed to $0.5$~$\rm arcsec$. The remaining parameters of this halo are assigned uniform priors.

\subsection{Intra-cluster gas}
\label{sect:gas-dist}

To model the X-ray emitting gas in the cluster, we use a similar approach to \citet{Bonamigo2017,Bonamigo2018} for the definition of the gas mass component, which consists of a free-form modelling of the distribution with elliptical dPIE potentials. We deviate significantly from these previous works in the other aspects of the modelling, such as during parameter optimisation or in model discrimination described in Sect.~\ref{sect:likelihoods} and Sect.~\ref{sect:model-discri}, respectively. The shapes of these potentials are constrained by the X-ray surface brightness in the form of the number of photon counts, and by strong lensing through their contributions to the total mass. We will detail how we discriminate the number of potentials to be used in Sect.~\ref{sect:model-discri}.

All parameters of the dPIE potentials are free to vary except for the cut radius, $r_{\rm cut}$. When using more than one potential, this parameter is degenerated, and our optimisation process does not converge. Thus, we fix it to the large value of $1250$ $\rm kpc$ for all of the potentials. This choice corresponds to the radius at which the X-ray signal starts to be equivalent to the noise. As we only use a few dPIE profiles, we only model  the cluster-scale distribution of the gas. Hence, small scale variations like gas sloshing, shock front or micro-physics processes are not taken into account in our current approach. If such processes have enough amplitude, they will leave traces in the count residual between the observations and our models.

\subsection{Cluster member masses}
\label{sect:clus-memb-dist}

\begin{figure*}
    \begin{minipage}{0.48\linewidth}
    \centering
    \includegraphics[width=\linewidth]{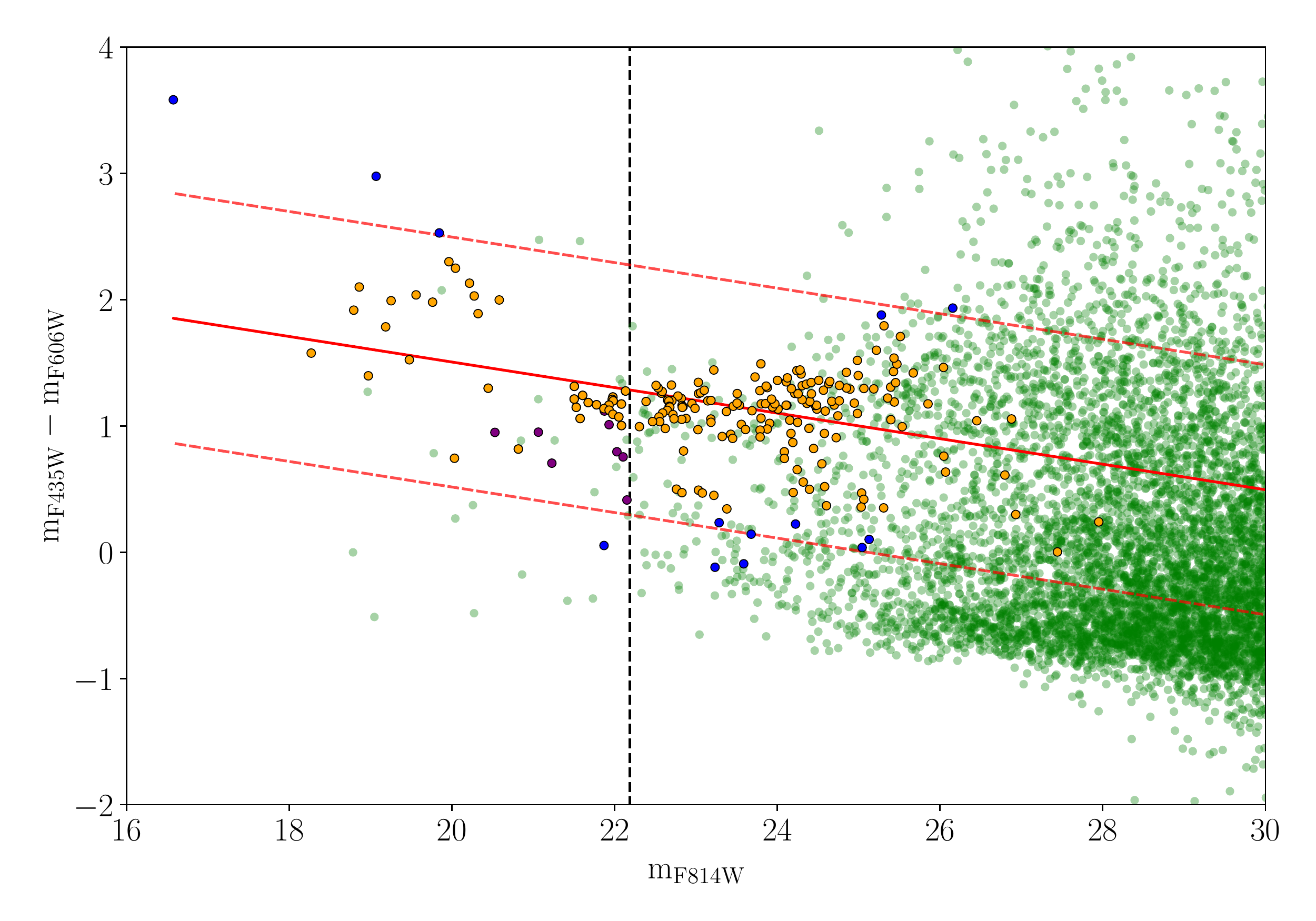}
    \end{minipage}
    \begin{minipage}{0.48\linewidth}
    \centering
    \includegraphics[width=\linewidth]{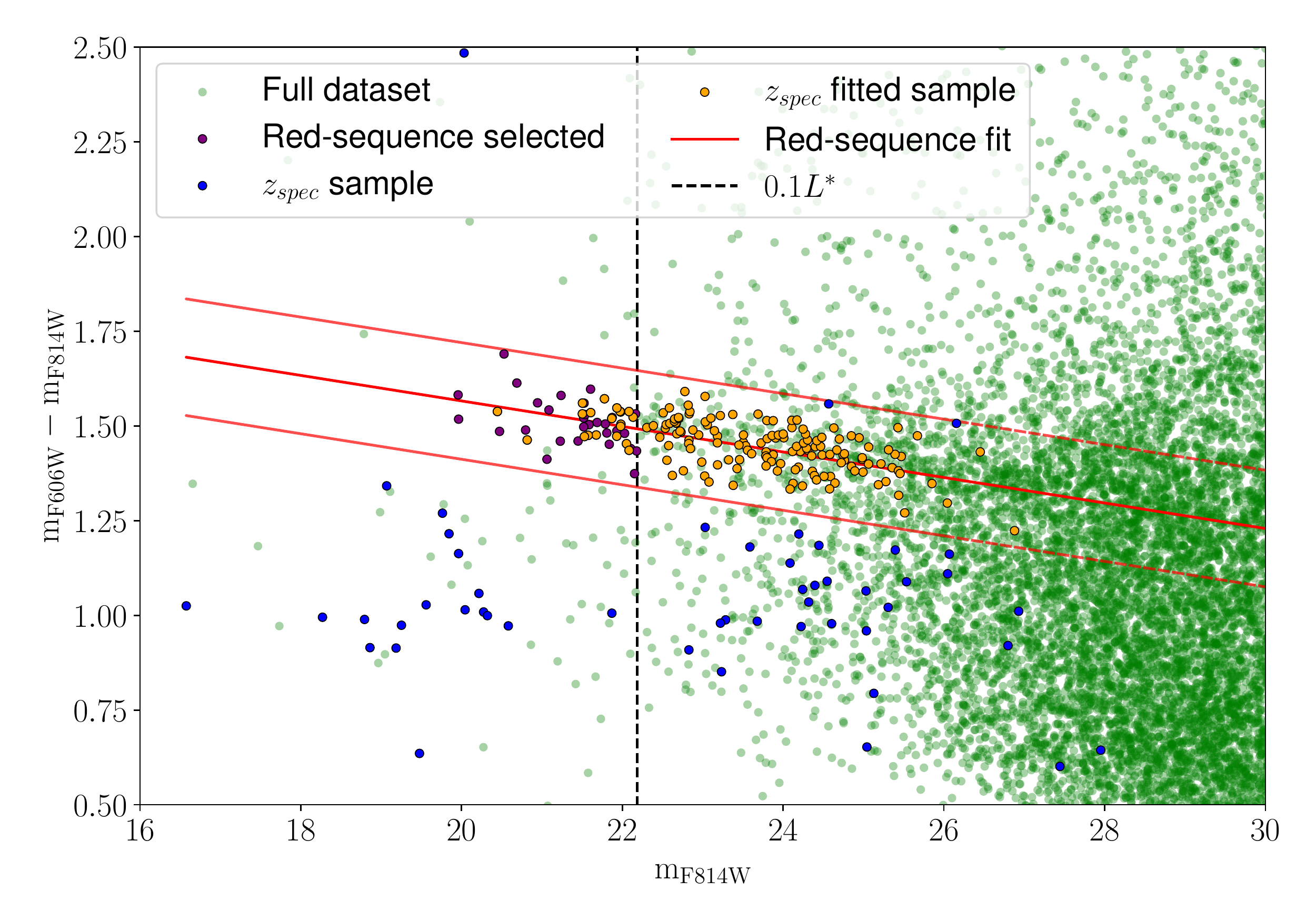}
    \end{minipage}
    \caption{\textit{Left panel:} colour-magnitude diagram of $\rm m_{F435W}-m_{F606W}$ versus $m_{F814W}$. The blue and orange dots highlight the spectroscopically confirmed cluster members, and the orange ones represent the objects that were not rejected by the iterative procedure to fit the red sequence. The green and purple dots show the full photometric catalogue. The former represents all objects, while the second represents the galaxies that have been selected by both colour-magnitude diagrams. The red plain and dashed lines represent the best-fit red sequence and the $3\sigma$ bounds used to select galaxies, respectively. The dashed black lines show the magnitude threshold of $0.1L^{*}$. \textit{Right panel:} Same as the left panel but for the colour-magnitude diagram of $\rm  m_{F606W}-m_{F814W}$ versus $m_{F814W}$. There are more purple dots in this panel than in the left one, as some objects do not have a $\rm m_{F435W}$ measurement.}
    \label{fig:redsequencefit}
\end{figure*}

We select the cluster members using a spectroscopic confirmation combined with a red sequence selection, if spectroscopy is not available. We include all galaxies that have a redshift, $z$, in the range of $\left[0.327,0.360\right]$, and are in the HFF footprint, independently of their type, similar to \citet{Lagattuta2017}. We use this galaxy sample to calibrate the red sequence and select the cluster member without spectroscopic confirmation. We perform a red sequence fitting on the $({\rm m_{F606W}-m_{F814W},m_{F814W}})$ and the $({\rm m_{F435W}-m_{F606W},m_{F814W}})$ planes. We use a similar iterative approach as in our fit for the fundamental plane and Faber \& Jackson scaling relations. We fit a linear model in the considered plane, rejecting all elements that are at more than $2.5\sigma$ of the fitted lines. We reiterate this procedure until no galaxies are rejected. We found that this value of $2.5\sigma$ leads to fast convergence of the rejection process. To constitute the red sequence selected sample, we consider all galaxies that are brighter than $0.1L^{*}$ and are within $3\sigma$ of the relation of fitted on both planes. In case some galaxies have no measurement for $\rm m_{F435W}$, we select galaxies with the other relation only. With the addition of these photometrically selected galaxies, we reach a total of $250$ elements. Fig.~\ref{fig:redsequencefit} presents the two colour-magnitude diagrams previously mentioned with the spectroscopic and photometric samples. The red plain and dashed lines show the two red sequence relations, representing the best-fit relation and the $3\sigma$ range. As we can see by the number of purple dots (i.e. red sequence selected cluster member) in comparison to the spectroscopic sample (e.g. blue and orange dots), most of the cluster members used in the modelling are spectroscopically confirmed.

We assume, to some extent, a light-traces-mass hypothesis for profiles associated with selected cluster members. Indeed, we follow the same geometry by fixing $(x_c,y_c)$, $\theta$ and $\epsilon$, to the values measured in the light distribution with \textsc{SExtractor}, and use a dPIE profile to represent both. For the remaining parameters, we suppose that light and mass profiles have the same shape with only a scaling as a difference. This allows us to link $\sigma_0$ from the measurement of $\sigma_e$ with the projection factor $c_p$ (its calculation is detailed in Appendix~\ref{app:c_p_explanation}). The scaling is defined as follows on the three relevant parameters:
\begin{align}
    &\sigma_{0,\rm light}=\sigma_{0}\\
    &\nu r_{\rm cut,light}=r_{\rm cut}\\
    &\nu r_{\rm core,light}=r_{\rm core}
\end{align}
Where $\nu$ is the scaling factor between mass and light profiles. $\sigma_{0,\rm light}$, $r_{\rm cut,light}$ and $r_{\rm core,light}$ are dPIE parameters associated with the light profile. These hypotheses imply a constant mass-to-light ratio of $\nu$ as well as a proportionality relation between the $R_e$ and half-mass radius, $R_{M/2}$, in the form of $R_{M/2}=\nu R_e$. Regarding the proper scaling of $r_{\rm cut}$, $r_{\rm core}$ and $\sigma_0$, we apply different relations depending on the available observational data. We split the cluster member selection into three groups depending on the photometric measurements that are publicly available. If no light profile has been fitted, meaning that we only have the integrated luminosity, $L$, we use a Faber \& Jackson scaling, and if we have it, we take benefit of the fundamental plane with one relation, of our two different scalings.

\subsubsection{Case 1: if only $L$ is available} 
In this case, we rely on the Faber \& Jackson law with small differences in the power law in comparison to previous works such as \citet{Richard2010} and \citet{Bergamini2019}. We use the following formulae:
\begin{align}
    &r_{\rm core}=r_{\rm core}^{*}\left(\frac{L}{L^*}\right)^{1-1/c}\\
    &r_{\rm cut}=r_{\rm cut}^{*}\left(\frac{L}{L^*}\right)^{1-1/c}\\
    &\sigma_0=\frac{\sigma_e^{*}}{c_p\left(r_{\rm cut}^{*}/r_{\rm core}^{*},\nu\right)}\left(\frac{L}{L^*}\right)^{1/2c}
\end{align}
Where $r_{\rm core}^{*}$, $r_{\rm cut}^*$ and $\sigma_e^*$ are the parameters associated with the galaxy representing the population at that redshift defined in Sect.~\ref{sec:fundamental_plane_calibration}. In our case, we follow previous studies \citep{Richard2010,Mahler2018,Lagattuta2019} and we fix $r_{\rm core}^{*}$ to $0.15$ ${\rm kpc}$ as its effect is negligible on strong lensing constraints used here.

\subsubsection{Case 2: if $\sigma_e$ and the light-profile are available} 
If available, we use the $\sigma_e$ measurement and we rely on the fundamental plane prediction for $R_{e\rm FP}$ combined with our assumption on the relationship between light and mass profiles. Thus, we obtain the following relations:
\begin{align}
    &r_{\rm core}=\frac{r_{\rm core}^{*}}{r_{\rm cut}^{*}}r_{\rm cut}\\
    &r_{\rm cut}=\frac{\nu 10^{a+b\mu_e+c\log\sigma_e}}{\frac{3}{4}\sqrt{1+\frac{10}{3}\frac{r_{\rm core}^{*}}{r_{\rm cut}^{*}}+\left(\frac{r_{\rm core}^{*}}{r_{\rm cut}^{*}}\right)^2}}\\
    &\sigma_0=\frac{\sigma_e}{c_p\left(r_{\rm cut}^{*}/r_{\rm core}^{*},\nu\right)}
\end{align}
The denominator of the $r_{\rm cut}$ relation comes from the transformation of $R_{M/2}$ to $r_{\rm cut}$ with the constant ratio assumed between $r_{\rm core}$ and $r_{\rm cut}$. Hence, we do not make direct use of $R_e$ measurement, and we instead use the fundamental plane estimation, which allows us to propagate partially the observational error on these measurements in our MCMC sampling. The proper solution would be to consider $\sigma_e$ and $R_e$ as random variables, but this would overwhelmingly increase the computational cost of the analysis.

\subsubsection{Case 3: if the light-profile is available} 
We use a similar scaling as the previous case, and the only difference is that we are using $\sigma_{e,\rm FP}$ instead as we do not have a measurement in this case. Thus, now the relations are the following:
\begin{align}
    &r_{\rm core}=\frac{r_{\rm core}^{*}}{r_{\rm cut}^{*}}r_{\rm cut}\\
    &r_{\rm cut}=\frac{\nu R_e}{\frac{3}{4}\sqrt{1+\frac{10}{3}\frac{r_{\rm core}^{*}}{r_{\rm cut}^{*}}+\left(\frac{r_{\rm core}^{*}}{r_{\rm cut}^{*}}\right)^2}}\\
    &\sigma_0=\frac{10^{(-a-b\mu_e+\log R_e)/c}}{c_p\left(r_{\rm cut}^{*}/r_{\rm core}^{*},\nu\right)}
\end{align}

Between the two relations in case~$2$ and case~$3$, we prioritise the case~$2$, which is motivated by the accuracy of the associated mass proxies in recovering the true parameter value from the fundamental plane, as discussed in Sect.~\ref{sect:FP_prediction_accuracy}. We use the same treatment for both of them as we use the same prior based on our calibration for the fundamental plane parameters (i.e. Gaussian priors based on the mean and standard deviation of the calibration). Notably, both proxies present similar trends but with a different amplitude in the discrepancy between the propagated uncertainty and the observed one. We choose not to treat them differently as the insights from our mass proxy analyses are not robust enough at the present time because we lack the covariance between the $R_e$ and $\sigma_e$.

From all the parameters mentioned, only $r_{\rm cut}^*$ and $\nu$ are optimised with uniform priors as they are not involved in the calibration process. This ensures consistency with the galaxies modelled with the fundamental planes and Faber \& Jackson relation, as in both cases, the parameter that we optimise with minimum prior knowledge is $r_{\rm cut}$.

Regarding the BCG, we apply a special treatment as it is not modelled according to the previous relations. Its associated dPIE has its shape and position parameters fixed by its luminous distribution as for other cluster members, but its $r_{\rm core}$, $r_{\rm cut}$ and $\sigma_{0}$ are optimised. To define physically motivated priors, we ran a nested sampling algorithm to fit its LOSVD with data points measured at different radii for the BCG. We used the  $\sigma_{0}$ posterior distribution to obtain a Gaussian prior for the sampling using the lensing and X-ray constraint. $r_{\rm core}$ and $r_{\rm cut}$ did not converge in the previous run; thus, we assign uniform priors to them.

\subsection{Additional Perturbation}
\label{sect:B-spline-dist}
To complete our modelling, we consider a perturbation under the form of a B-spline surface added to the lensing potential \citep{Beauchesne2021}. It allows us to incorporate effects like complex asymmetries that can not be captured with the number of large-scale haloes considered to represent the smooth DM distribution. The surface expression, $\Delta\psi$, is the following:
\begin{equation}
    \Delta\psi(x,y)=\frac{D_{ls}}{D_{s}}\sum^m_{j,l=1} C_{j,l} B_{j,p,t_x}(x) B_{l,p,t_y}(y)
    \label{eq:pert_surf}
\end{equation}
where $B_{j,p,t_x}$ are the $j^{th}$ $1$D B-spline basis functions of polynomial degree, $p$, associated with the knot vector, $t_x$. $C_{j,l}$ are the coefficients of each B-spline basis, and $m$ is the number of B-spline functions per axis. $D_{\rm ls}$ and $D_{\rm s}$ are the angular distance between the lens and the source, and the observer and the source, respectively. 

This component is added to the model at a different step than the previous ones as we follow a two-step optimisation \citep{Beauchesne2021}, where we first run a nested sampling with the priors previously defined on the dPIEs parameters without the perturbation. We use the obtained posterior distributions to define Gaussian priors on each parameter for the second sampling run. We use the best-fit values as the mean, and three times the empirical standard deviation. For the physically motivated priors related to cluster members, we use the mean of the posterior and not the best-fit values as it does not take into account the bias from the priors and only one time the empirical standard deviation, as the widths of the posterior distributions are similar to the ones of their priors.

We tested the addition of an external shear \citep{Mahler2018,Lagattuta2019} at the same step to take into account the effect of haloes on the cluster outskirts. However, it did not provide improvements on the reconstruction of the lensing constraints, so we choose not to include it in the final method. In the case of cluster lenses such as Abell 2744 \citep{Mahler2018} where structures surrounding the cluster core are clearly seen, it is important to include this kind of physically supported perturbation.

\section{Joint lensing and X-ray likelihood}
\label{sect:likelihood_def}
We now outline how we combine the two probes of the mass distribution of the galaxy cluster by detailing what quantities we use as constrain, and how the likelihoods are defined and combined. We start with the X-ray and lensing constraints in Sect.~\ref{sect:X-ray-constraint} and Sect.~\ref{sect:multiple-images}, respectively, and then go on to describe the likelihoods in Sect.~\ref{sect:likelihoods}.

\subsection{X-ray emission: Constraints and modelling}
\label{sect:X-ray-constraint}

As indicated in Sect.\ref{sect:X-ray-data-analysis}, we fit the X-ray counts in the $0.5$ to $7$ $\rm keV$ band, which corresponds to the combined range of the soft, medium and hard science energy bands of \textit{Chandra}. To define a mask where the X-ray emission is taken into account, we use the radius at which the X-ray signal is equivalent to the associated noise. We obtain this radius by making circularly averaged profiles of the counts from both the source and the blank-sky background, and measuring where the signal is going below two times the noise (2$\sigma$). Thus, we use a square mask of $1250$ $\rm kpc$ centred on the position of the centroid of the BCG light distribution.

To constrain the gas distribution, we associate the dPIE potentials with the X-ray surface brightness. We start by obtaining, $\int \rho_{\rm gas}^2$, along the line-of-sight with the analytical formulae computed by \citet{Bonamigo2017}. To switch from the previous quantity to the definition of the surface brightness outlined in Eq.\ref{Eq:S_X}, we need to transform $\rho_{\rm gas}^2$, with $\rho_{\rm gas}$ the gas mass density, into $n_e n_p$ for a fully ionised gas, which is done in the following way:

\begin{align}
    &\rho_{\rm gas}=\mu(n_e+n_p)\\
    &\Rightarrow n_e n_p=\frac{\rho^2}{\mu^2 \left(1+{\rm nhc}\right)\left(1+\frac{1}{\rm nhc}\right)}
\end{align}

With $\mu$, the mean molecular weight per particle in a fully ionised gas, and ${\rm nhc}$, the conversion factor from $n_p$ to $n_e$. For both parameters, we assume the value tabulated in \citet{Asplund2009}. The total count number is obtained by multiplying $\int n_e n_p$ by the exposure time, the effective area of the telescope, and the cooling function $\Lambda(Z,T,E)$. $\Lambda(Z,T,E)$ used in this analysis has been computed with a binning of the FoV made with a SN threshold of $10$ as detailed in Sect.\ref{sect:X-ray-data-analysis}, but we analyse its possible bias in Appendix~\ref{app:SN_threshold} with SN thresholds of $6$ and $14$. As we fit the count number on the broad band of \textit{Chandra} science band, we estimate the effective area of the CCD pixels at different energies. In particular, we use the following approximation:

\begin{align}
    \int t_{\rm exp} A_{\rm eff}(E)\Lambda (Z,T,E) dE&=\sum_{\rm Band} t_{\rm exp} A_{\rm eff}(E_{\rm Band})\\
    &\times\int_{E_{\rm min,Band}}^{E_{\rm max,Band}}\Lambda (Z,T,E)dE
\end{align}

Where $\rm Band$ refers to the soft, medium and hard energy bands of \textit{Chandra}, $t_{\rm exp}$, the exposure time, $A_{\rm eff}$, effective area of the CCD pixels. $E_{\rm Band}$ is the energy value associated with each of these bands for the computation of the exposure map, and $E_{\rm min/max,Band}$ are the associated bounds. We finally add the blank-sky background to obtain the count model.

\subsection{Multiply-imaged systems}
\label{sect:multiple-images}

Regarding strong lensing constraints, we build our set of multiply-imaged systems from the one put in place by the HFF modelling teams. In particular, we use the \textit{Gold} sample \citep[see Section~4.2 from ][ for the explanation of all HFF labels]{Lagattuta2019}, which is a set of $50$ images over $19$ systems all spectroscopically confirmed. We complete it with other additional spectroscopically confirmed systems presented in \citet{Caminha2016}, which brings the total to $61$ multiple images from $22$ systems. Notably, we have three more images and one more system than previous works, which is due to different selection of multiple image since the objects are present in both sets.

Thanks to our new reduction of the MUSE datacube, we are able to measure the spectroscopic redshift of some objects for the first time, which allows us to also identify a new pair of multiple images. One object (system $303.1$) is already present in previous spectroscopic catalogues, while the other comes from these new measurements. In addition, we identify a close pair as a multiply-imaged system that was previously considered as a single image \citep[$SW-54$ in][and system $304$ in this work]{Karman2015,Karman2017} and a second pair from one confirmed image and the MUSE narrow band datacube (system $305$). These new constraints bring our final dataset to a total of $67$ images for $25$ systems that are indicated with red circles in Fig.~\ref{fig:summary_S1063}. The spectra and optical images of these new systems are presented in Appendix~\ref{app:New_im_sys} as well as their coordinates and redshifts. The method used to find these new systems is detailed in \citet{Richard2021}.

\subsection{Likelihoods}
\label{sect:likelihoods}
In our optimisation process, we simultaneously constrain the mass distribution with the strong lensing and the X-ray surface brightness. We obtain the combined likelihood, $\mathcal{L}_{\rm tot}$, by assuming that both observables are independent, which leads to the following expression:
\begin{equation}
    \mathcal{L}_{\rm tot}=\mathcal{L}_{\rm SL}\times\mathcal{L}_{\rm X-ray}
\end{equation}
where $\mathcal{L}_{\rm SL}$ and $\mathcal{L}_{\rm X-ray}$ are the likelihoods for the strong lensing only and the X-ray only, respectively. For the latter one, we use a Poisson-Gamma mixture likelihood. For the former, we use a Gaussian likelihood slightly modified in comparison to previous works based on \textsc{Lenstool} \citep{Mahler2018,Lagattuta2017}.

\subsubsection{Constructing the X-ray likelihood function} 

As our modelling of the X-ray gas mass solely uses a couple of dPIE potentials, we can only reproduce the large-scale distribution, excluding all microphysical gas phenomena. These small-scale events in the plasma are linked to surface brightness fluctuations \citep{Eckert2017}, that we take into account by adding an uncertainty on the mean, $\lambda_i$, of the Poisson distribution of the observed photon counts in the $i^{th}$ bin. We assume that $\lambda_i$ follows a Gamma distribution, $\Gamma(\mu_i,\sigma_{\rm X}^2)$, parameterised by its expected value $\mu_i$ and its variance $\sigma_{\rm X}^2$ instead of the usual shape and scale parameters. $\mu_i$ is our predicted count model excluding the microphysics, whereas, $\sigma_{\rm X}^2$ represents the induced uncertainty assumed to be the same for all bins. As we only intend to model the count model uncertainty statistically, we marginalise on the realization of the Gamma distribution, which leads to the following likelihood in the $i^{th}$ bin to observe $k_i$ count:
\begin{align}
    \mathcal{L}_{\rm X-ray,i}=\int^\infty_0 \mathbb{P}_i(X_{\lambda}=k)f_{\Gamma}(\lambda,\mu_i,\sigma_{\rm X}^2)d\lambda
\end{align}
where $\mathbb{P}_i$ is the probability mass function of the Poisson distribution and $f_{\Gamma}$ is the probability density function of $\Gamma(\mu_i,\sigma_{\rm X}^2)$. This compound probability distribution is known to be a continuous extension of a negative binomial distribution $\mathbb{NB}(r=\mu_i^2/\sigma_{\rm X}^2,p=\mu_i/(\mu_i+\sigma_{\rm X}^2))$ with the following expression:
\begin{align}
    \mathcal{L}_{\rm X-ray,i}=\frac{\Gamma(k_i+r)}{k_i!\Gamma(r)}p^r\left(1-p\right)^k_i
\end{align}
The expected value of $\mathbb{NB}$ is $\mu_i$, and the variance is $\mu_i+\sigma_{\rm X}^2$, which implies an over-dispersion of the Poisson model in the case of $\sigma_{\rm X}>0$. We fall back to a Poisson modelling when the uncertainty is zero, with the expected value being equal to the variance. We can now obtain our final likelihood by injecting $r$ and $p$ expressions in the previous definitions and assuming that each bin is statistically independent of the others:
\begin{align}
    \mathcal{L}_{\rm X-ray}=\prod_i \frac{\Gamma\left(k_i+\frac{\mu_i^2}{\sigma_{\rm X}^2}\right)}{k_i!\Gamma\left(\frac{\mu_i^2}{\sigma_{\rm X}^2}\right)}\left(\frac{\mu_i}{\mu_i+\sigma_{\rm X}^2}\right)^\frac{\mu_i^2}{\sigma_{\rm X}^2}\left(\frac{\sigma_{\rm X}^2}{\mu_i+\sigma_{\rm X}^2}\right)^{k_i}
\end{align}

Preliminary tests show that results obtained with this likelihood are consistent with the ones given with a Poisson likelihood without uncertainties on its expected value. Indeed, the best-fit count model obtained through an X-ray-only fit has a relative difference of less than $1$ per cent on average with a maximum of less than $10$ per cent.

\subsubsection{Constructing the strong lensing likelihood function} 

The expression of $\mathcal{L}_{\rm SL}$ is the following \citep{Mahler2018,Lagattuta2017}:
\begin{equation}
    \mathcal{L}_{\rm SL}=\sum^{N_{\rm sys}}_j \frac{1}{\prod^{N_{\rm im,j}}_i \sigma_{ij}\sqrt{2\pi}} \exp\left(-\chi^2_j/2\right)
\end{equation}
with $N_{\rm sys}$ the total number of multiply-imaged systems, $N_{\rm im,j}$, the number of images in the $j^{th}$ system, and $\sigma_{ij}$, the observational error on the $i^{th}$ images of the $j^{th}$ system. $\chi^2_j$ is the $\chi^2$ statistics associated with the $j^{th}$ system and read as follows:
\begin{equation}
    \chi^2_j=\sum^{N_{\rm im,j}}_i \frac{\left|\left|\vec{\theta}^{\rm obs}_{i,j}-\vec{\theta}^{\rm pred}_{i,j}\right|\right|^2}{\sigma_{ij}^2}
\end{equation}
where $\vec{\theta}^{\rm obs}_{i,j}$ is the observed multiple images positions, and $\vec{\theta}^{\rm pred}_{i,j}$ is the predicted ones. The difference between our likelihood and previous works based on the \textsc{Lenstool} software is hidden in the $\vec{\theta}^{\rm pred}_{i,j}$ term, and specifically in the position of the multiply-imaged system source. Previously, the intrinsic source position used to solve the lens equation was chosen to be the barycentre or amplification weighted barycentre of the sources associated with each image individually. This solution is convenient as it does not add extra calculations to the computationally expensive part of this process which is solving the lens equation, but it introduces a bias to the modelling workflow. Indeed, choosing a different intrinsic source position that will better suit the multiply-imaged systems directly modifies the likelihood value by reducing it drastically in some specific cases, for example, when an image is near its associated critical lines. This difference will also modify the Bayesian criteria used to discriminate models. Thus, we add an extra computational step, which consists of optimising the intrinsic source position to reduce the associated positional error on each image of the system. We do this operation at a fixed model for each step of the nested sampling run, and we solve the non-linear least square problem associated with the Levenberg-Marquardt algorithm implemented in the GNU Scientific Library. The starting point of the Newton method is the previously defined barycentre.

\section{Mass reconstrucion of Abell S1063}
\label{sect:mass-of-as1063}

Multiple models are possible with our new method with different complexity. Hence, we describe how discrimination between models is performed in Sect.~\ref{sect:model-discri}. The reproduction of the observational constraints and the obtained mass distribution for each cluster component is presented in Sect.~\ref{sect:constraints_repro} and Sect.~\ref{sect:res-mass}, respectively.

\subsection{Model discrimination}
\label{sect:model-discri}
Given that our modelling method relies on X-ray and lensing data information of the cluster, we have to adapt the model discrimination to include this new type of non-homogeneous constraints. We choose to follow a similar two-step process as described in \citet{Beauchesne2021}. The first step is modified to include the X-ray part, while the second one includes some changes related to the physically motivated priors on the cluster member parameters as explained in Sect.~\ref{sect:B-spline-dist}. The original aim of this approach was to solve convergence issues when adding the B-spline surfaces as the number of free parameters increased by a significant number. Indeed, the parametric part of the model struggles to converge to high-likelihood areas. The first step aims at discriminating between models on the parametric side only, while the second one allows us to define the best number of B-spline basis functions.

\subsubsection{Parametric model discrimination}

To discriminate between parametric models, we treat separately the gas distribution constrained by the X-rays and the rest of the model that relies solely on lensing data. We use a disjoint process as, in our case, the numerical values of the two likelihoods have three to four orders of magnitude difference in favour of the X-ray one. As the gas distribution only represents a small part of the overall mass budget (we know a priori that the gas mass is roughly of the order of $\sim 10\%$ of the total mass), we assume that by discriminating the lensing-only part without the gas mass will not introduce a significant bias. 

For the lensing-only part, we rely on previous studies of AS1063 which mostly agree on a model with two large-scale dPIEs as detailed in Sect~\ref{sect:DM-dist} \citep{Limousin2022,Bergamini2019,Bonamigo2018}. In the case of less studied galaxy clusters where previous models may not be available, we would have based our choice on Bayesian criteria such as the Bayesian evidence (i.e. marginal likelihood) to select which model best describes the data.

For the gas distribution, we use a different approach as the statistical knowledge of the constraints is well known contrary to the lensing ones. Actually, the observational errors on the position of the multiple images integrate both errors due to physical phenomena such as line-of-sight perturbers and/or systematic limitations due to the parametric approach, for example. Hence, we developed a measure of the goodness of fit adapted to our Poisson-Gamma mixture likelihood with a similar approach as in \citet{Kaastra2017}. The only differences are the statistics used as well as a numerical approach of the measure instead of an analytical one. We consider that a gas distribution model is explaining well the data when the likelihood of the observations is at least included in the $"5\sigma"\,\rm CI$ of the expected value of the X-ray likelihood for the considered model. 

We obtain the distribution of the X-ray likelihood through a Monte Carlo approach, where we sample the Poisson-Gamma mixture model based on the parameter of the best-fit. We proceed by adding one dPIE at a time, and we stop at the model with less complexity that satisfies the previous condition (i.e. included in the $"5\sigma"\,\rm CI$ of the expected value of the likelihood). In our case, we end up with a model composed of three dPIEs as in \citet{Bonamigo2018}, but our selection method assures that we reach the right complexity for the data we have. This is only possible through the use of the Poisson-Gamma mixture, as the same model under-fits the data and is outside the $"5\sigma"\,\rm CI$ with a Poisson statistic. The bound of the $"5\sigma"\,\rm CI$ is at $1.5\sigma$ from the obtained best-fit likelihood, but as we are not considering a Gaussian distribution, the $"n\sigma"\,\rm CI$ does not scale linearly with the standard deviation.

When both parametric components are defined, we merge them and proceed to a sampling run that is used as a starting point for the next step. This run, and the following, are performed with the dynamic nested sampling algorithm \textsc{MLFriends} \citep{Buchner2014,Buchner2019} implemented in the python package \textsc{UltraNest}\footnote{\url{https://johannesbuchner.github.io/UltraNest/}} \citep{Buchner2021}.

\subsubsection{Perturbative model discrimination}

All variants of perturbative modelling are created in the same way as described in Sect.~\ref{sect:B-spline-dist}. The discrimination is done with the Bayesian evidence, $\log\mathcal{E}$ (i.e. marginal likelihood), provided by the nested sampling engine. We select the model that gives the best $\log\mathcal{E}$ in the case of a parabolic-like shape, or the first modelling reaching the $\log\mathcal{E}$ plateau as seen in \citet{Beauchesne2021}. For all the runs including a B-spline perturbation, we use an identical observational error estimate for all multiple images as $0.2$ arcsec. 

We try models with a mesh of $3\times3$, $4\times4$, $5\times5$ and $6\times6$ of B-spline basis. The associated $\log\mathcal{E}$ shows that B-spline modellings reach a plateau starting at $4\times4$ with $\log\mathcal{E}=-33309.71$. $5\times5$ and $6\times6$ have values close to the previous one with a $\log\mathcal{E}$ of $-33310.04$ and $-33304.65$, respectively. Models with $3\times3$ mesh of B-spline have a significantly lower marginal log-likelihood with a value of $-33331.95$. As in \citet{Beauchesne2021}, we consider that a model is better than the other if it is more supported by $\log\mathcal{E}$ at a difference equivalent to $5\sigma$ (i.e. a difference of $12.5$ between $\log\mathcal{E}$ estimations). Thus, the B-spline variant with a mesh of  $4\times4$ function is the best according to this selection procedure. The associated best-fit model has a Root Mean Squared (RMS) error of $0.35$ arcsec. Therefore, we construct a new sampling run with this error as the observational one to account for the model systematic uncertainty.

\subsection{Reproduction of the observational constraints}
\label{sect:constraints_repro}

Regarding the reproduction of the lensing constraints, our modelling reaches an RMS of $0.42$~$\rm arcsec$ for the best-fit model. In comparison to models by B19 and G22, we obtain a better fit of the constraints as these previous works had RMSs of $0.55$ and $0.60$~$\rm arcsec$, respectively. The main explanation for this difference is the inclusion of the B-spline surfaces which increase the flexibility of the mass reconstruction. Indeed, by excluding the B-spline surfaces, the RMS of our model is $0.70$~$\rm arcsec$. Note that  we include more multiple images than B19 and G22. In particular, the triple system found by \citet{Vanzella2016a} (i.e. system $11$ in our ID system), has been hard to model accurately \citep{Caminha2016}. Our new likelihood scheme has been able to improve the modelling with the inclusion of system $11$, as the barycentre approximation for the source position can struggle with non-linearity due to the presence of critical lines near the images. Our result is also close to the RMS of $0.36$~$\rm arcsec$ reported by \citet{Limousin2022} that also use a surface of B-spline but with a mesh of $5\times5$ basis functions, and the same set of constraints as B19 and G22. Moving away from models produced with Lenstool, \citet{Raney2020} report an RMS of $0.34$~$\rm arcsec$ with the inclusion of LOS galaxies. Adding these galaxies does not seem to particularly improve the overall fit as detailed in their study, in particular, system $11$ mentioned above that was removed due to potential bias from a LOS galaxy. Previous studies \citep{Gilman2020,Aloisio2014} tend to favour that LOS structure does not perturb mass models at discernable levels given the current data-sets. 

Regarding the RMS of our model without B-spline (i.e. $0.70$~$\rm arcsec$), it is higher than models that include  system $11$ made with a similar method as the Cluster As TelescopeS team\footnote{\url{https://archive.stsci.edu/pub/hlsp/frontier/abells1063/models/cats/}} (PIs: Natarajan \& Kneib) as part of the Frontier Field lensing modelling effort. Indeed, the RMS obtained for this modelling is of $0.49$~$\rm arcsec$, but it includes an external shear component with a high shear amplitude (i.e. $\gamma=0.1039$ for the best-fit model) that seems unlikely to be the sole product of some massive clumps in the cluster outskirts. Removing that component leads to a significant increase in the error of the predicted position of the multiple images, reaching values similar to $~0.80$~$\rm arcsec$. This increase is particularly important for system $11$ with a jump from $0.83$~$\rm arcsec$ to $1.97$~$\rm arcsec$. However, one of our new multiply-imaged systems (e.g. system $304$ in Appendix~\ref{app:New_im_sys}) broke this trend and prevented the external shear from having such undue influence on system $11$, and thus on the global RMS. It is partially a reason for the non-inclusion of that extra component, as it did not enhance the model with or without B-spline.

\begin{figure*}
    
    \includegraphics[width=\linewidth]{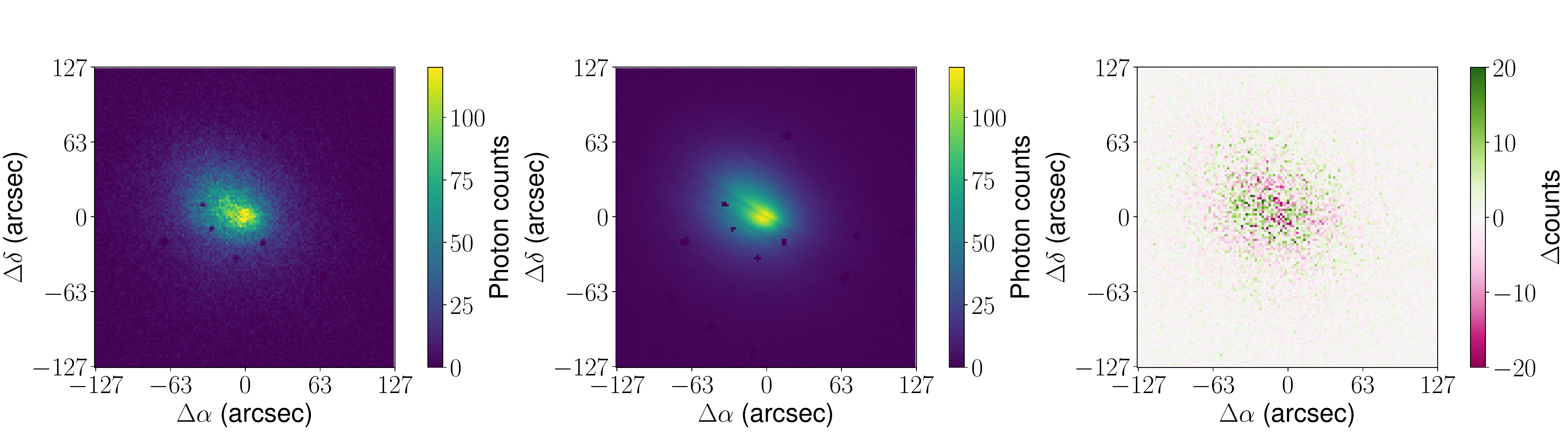}
    \caption{\textit{From left to right:} Maps of the observed X-ray counts, the counts model from our best-fit model and the residuals between the two previous maps (i.e. observation minus model). In both the observation and the model, the X-ray point sources have been masked.}
    \label{fig:X-ray-fit}
\end{figure*}

As we are constraining our model with  X-ray data in addition to the lensing information, we have to estimate the ability of our modelling to recover the X-ray surface brightness map. The quality of the reconstruction of the X-ray observations can be seen in Fig.~\ref{fig:X-ray-fit}, which shows, from the left to the right, the observed photon counts with point sources masked, the count model and the residual. Thanks to the use of multiple dPIE, our gas model has been able to capture the main asymmetry observed in the count map and provide a good fit of the data. However, a banana-shaped pattern of count underestimation can be observed in pink in the residual map below the cluster centre, which indicates the limitation due to the dPIE scale. Such a pattern combined with the overestimation above in green can indicate the presence of gas sloshing from the NE clump to the main one \citep{Paterno-Mahler2013}. Results from \citet{Bonamigo2018} also present the same artefact in their residuals (see Figure $1$ in their analysis).

\label{sect:X-ray-error}
\begin{figure}
    \centering
    \includegraphics[width=\linewidth]{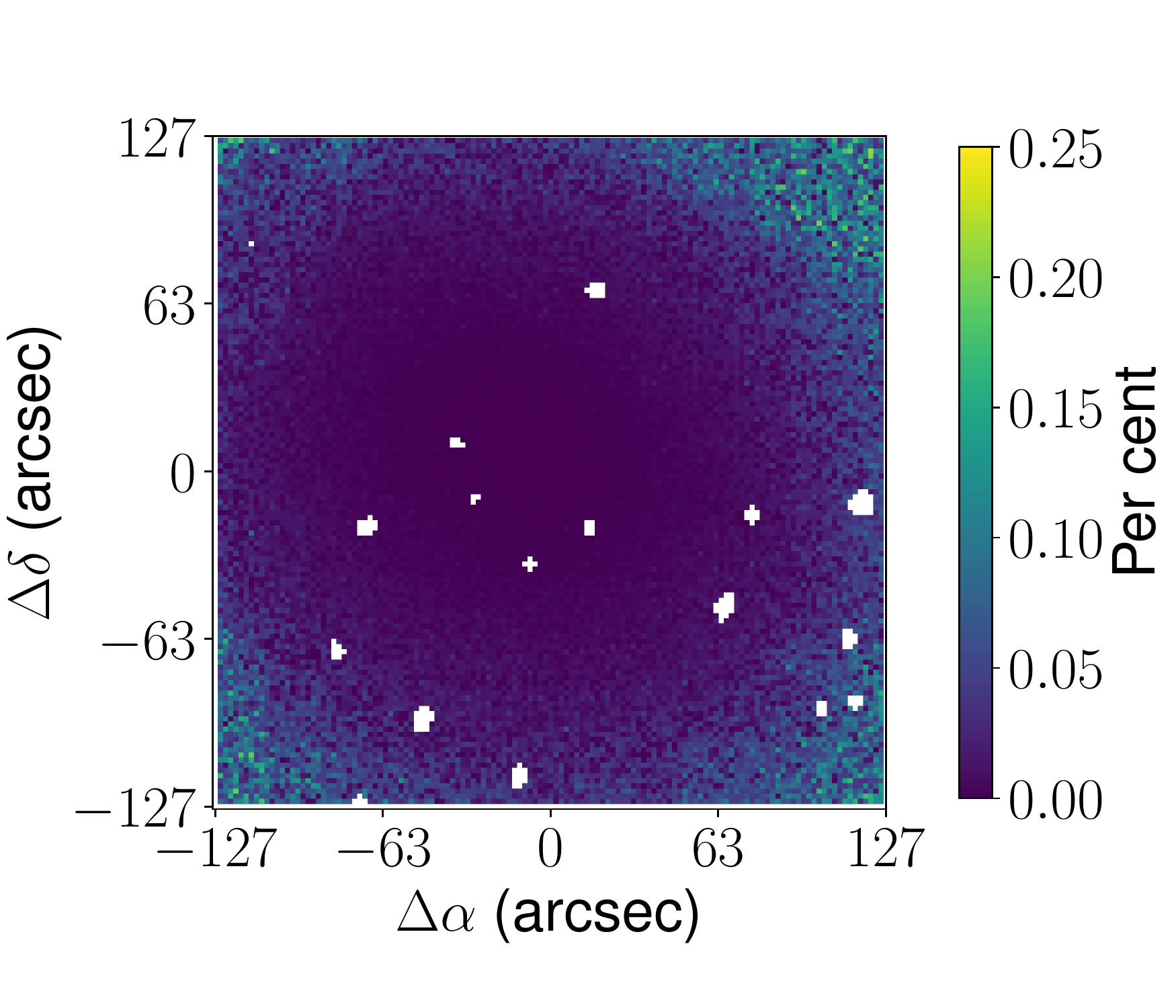}    
    \caption{Map of the absolute relative differences between the X-ray loglikelihood computed with a $\sigma_X=0.48$ counts and the best-fit value. The white areas are masked point-like sources.}
    \label{fig:X-ray-error}
\end{figure}

A special feature in our approach is the inclusion of an uncertainty on our count model, to account for small-scale fluctuations in surface brightness. In contrast to previous similar studies \citep{Bonamigo2017,Bonamigo2018}, we choose not to use a Poisson statistic to fit the X-ray photon counts. We instead use a Gamma-Poisson mixture model, where the uncertainty, $\sigma_{\rm X}$, is explicitly included with the Gamma distribution. During the sampling of the joint likelihood, we obtain an estimation of $\sigma_{\rm X}$ through the standard deviation of the Gamma statistics, which is estimated to be $0.263^{+0.027}_{-0.028}$~counts. If, theoretically, we obtain information on small-scale phenomena such as the micro-physics of the gas, in principle we could assess the robustness of the error estimation. Our goodness of fit procedure is able to give us insights into how well our model can describe the observations. Indeed, this Monte Carlo method samples the possible count realization for a given model. Depending on where the likelihood for the best-fit model is, we may know how much of the observations are plausible according to it. Our results show that this likelihood is between the limits of the $"3\sigma"\,\rm CI$ and $"5\sigma"\,\rm CI$ on the underfitting side of the previously mentioned distribution. Hence, if the real observations are plausible according to the best-fit model, they are still unlikely.

However, $\sigma_{\rm X}$ values have a significative influence on the distribution of the plausible X-ray likelihood values, which consequently changes the position of the best-fit model likelihood in the different $"n\sigma"\,\rm CI$. Increasing $\sigma_{\rm X}$ to values between $0.45$ and $0.50$ photon counts moves the considered likelihood to be close to the expected value, inside the $"1\sigma"\,\rm CI$. It shows that our method seems to underestimate $\sigma_{\rm X}$, in the sense that we are not obtaining the model that is the best at explaining the observations. In particular, Fig.~\ref{fig:X-ray-error} presents the absolute relative differences of the log-likelihood per pixel between the best-fit $\sigma_{\rm X}$ estimation, and a value of $0.48$ counts. As we can see, the discrepancies between both uncertainties are mainly in the low counts area where micro-physics events from say turbulence are not expected to occur, or at least, with a lower amplitude. This highlights the simplicity of our scheme as the central area, and the outskirt would be better treated differently in agreement with the amplitude of the different processes in the gas. Such improvements are needed to extract information on small-scale phenomena with satisfactory reliability in the estimation of their amplitude.

\subsection{The Mass distribution}
\label{sect:res-mass}
\begin{figure*}
    \centering
    \includegraphics[width=\linewidth]{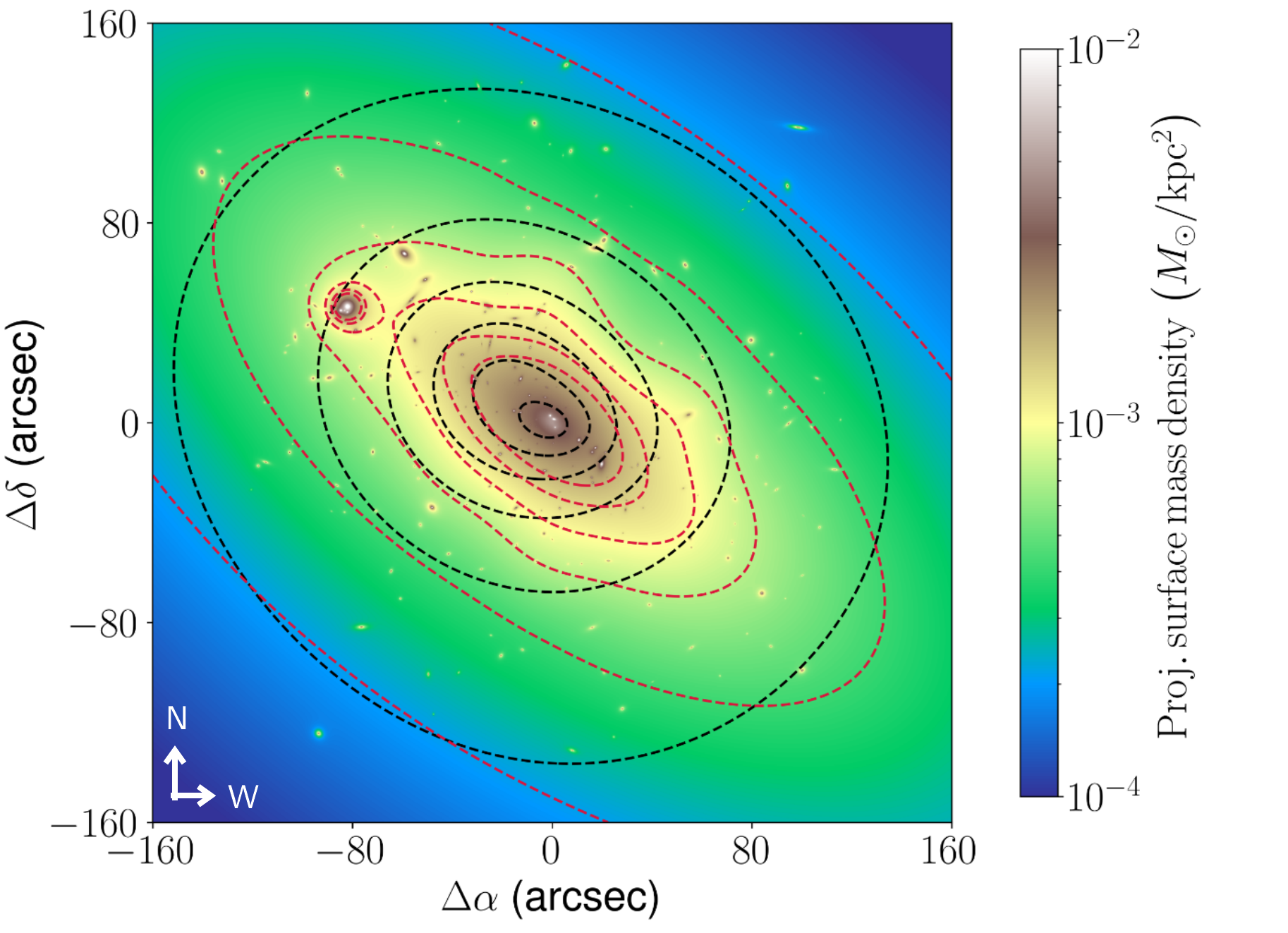}    
    \caption{Map of the projected surface mass density from our best-fit model where the dashed black and crimson contours represent the surface mass density of the X-ray emitting gas only and the smooth mass component (i.e. all mass components including the BCG but without the other cluster members), respectively.}
    \label{fig:mass_summary}
\end{figure*}

Figure~\ref{fig:mass_summary} presents the total mass distribution of AS1063 obtained from our best-fit model. The black and crimson dashed lines indicate the contours of the gas distribution and the smooth mass components, respectively. The statistics derived from the posterior distribution of dPIEs parameters are shown in Appendix~\ref{sect:dPIE-parameter}. The dominant component of the total mass is the DM. We can see clearly, that the inferred ellipticity of the DM component is different from that of the gas distribution. The latter is rounder on cluster scales. However, the discrepancy between the two components changes towards the centre of the cluster as both ellipticities become more similar to that of the BCG. We also note that the asymmetry in direction to the NE clump is less prominent in the total mass than in the gas and that the overall mass distribution is mainly uni-modal. In particular, our finding does not support the hypothesis of \citet{Gomez2012} of a major merger event between two similar-sized clusters, like in the bullet cluster \citep{Clowe2006}. More likely, per our model, the NE clump was possibly a smaller object, such as a galaxy group infalling into AS1063. We assess this scenario in more detail with the different products and by-products of our new method in Sect.~\ref{sect:merging-event}. 

Notably, the addition of the B-spline surface deforms the elliptical symmetry of the main halo, as shown by the crimson dashed contours. This extends the mass towards the NE clumps, mimicking the effect of another dPIE as found by previous studies \citep{Caminha2016,Limousin2022}. The smooth mass components also present other deviations from the main halo symmetry, as shown by the fluctuation of the crimson contours around the main dPIE halo iso-masses. These fluctuations are only limited to the core due to the B-spline amplitudes fading away, as highlighted by the farthest contours from the cluster centre. Thus, it is hard to assess if these variations are linked to the mass distribution patterns on larger scales, such as local galaxy overdensities or mass haloes on the cluster outskirt. Further constraints from weak lensing analysis from the BUFFALO program will expand the area constrained by lensing and, thus, the surface where the B-spline can be defined.

\subsubsection{Cluster member masses}
\begin{figure}
    \centering
    \includegraphics[width=\linewidth]{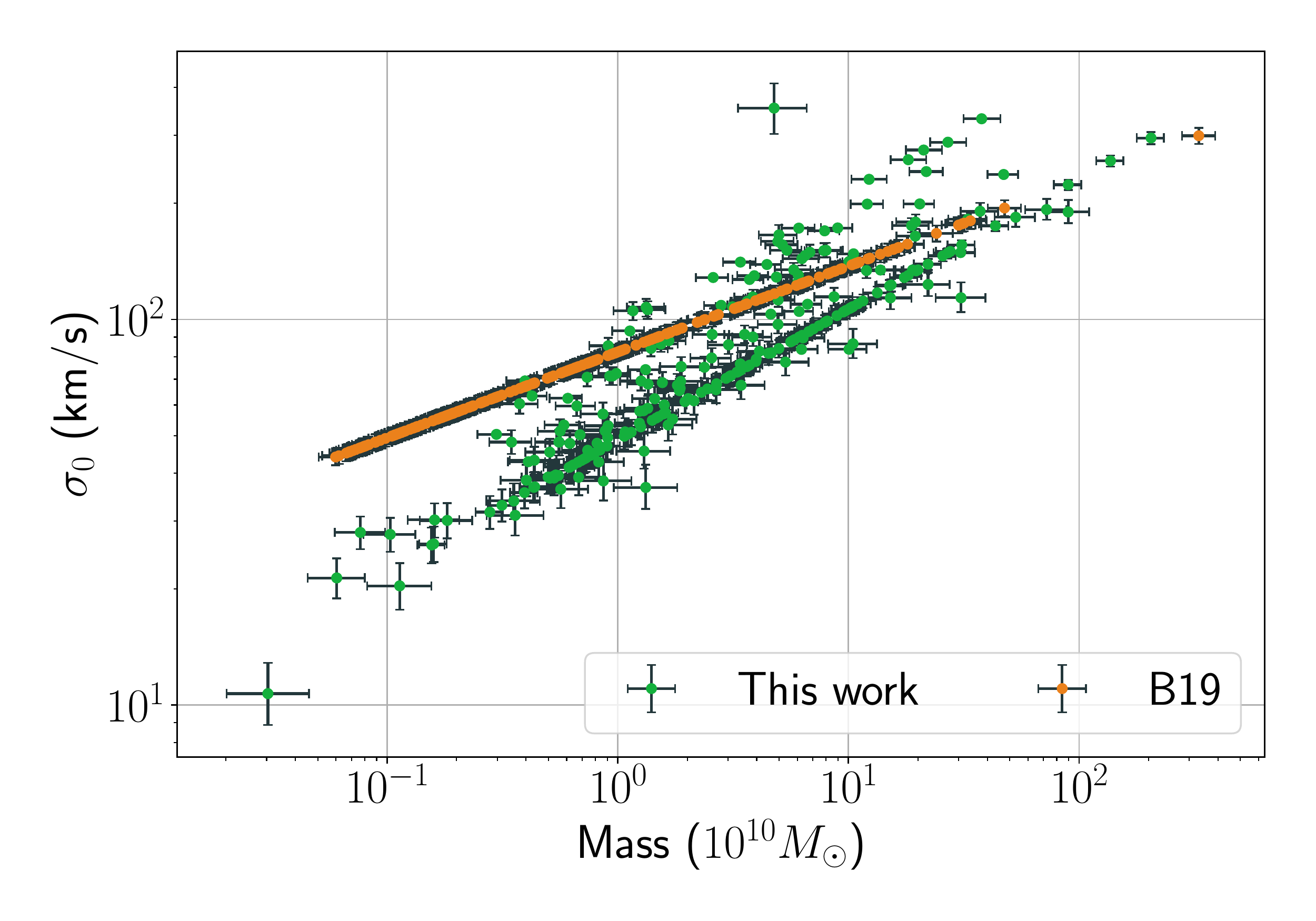}    
    \caption{Central velocity dispersion of cluster members as a function of their dPIE total masses from our model (green). We have also plotted in orange the model from B19 as comparison. The error bars represent the $"1\sigma"\, \rm CI$ of the posterior distribution. The slope obtained here is steeper than the one obtained in B19 (see Sect.\ref{sec:fundamental_plane_calibration}).}
    \label{fig:M_sigma_0}
\end{figure}
\begin{table}
\centering
\begin{tabular}{ccc}
    \hline
    Parameters & Median & $\sigma$\\
    \hline
    $a$&$-0.571$& $0.026$\\
    $b$&$0.302$& $0.018$ \\
    $c$&$1.478$& $0.075$ \\
    $\sigma_e^*$&$190.5$& $5.3$\\
    \hline
\end{tabular}
\caption{Parameters of the Faber \& Jackson scaling relation and the fundamental plane of elliptical galaxies obtained after the joint fit from X-ray and lensing constraints.}
\label{Tab:lens_fundamental_plane}
\end{table}

Figure~\ref{fig:M_sigma_0} presents the central velocity dispersion (i.e. as in the Lenstool code, $\sigma_{0}=\frac{2}{3}\sigma_{0\rm,th}$ from \citet{Limousin2005} definition) of the dPIE potentials as a function of their dPIE total mass from this work (green) and from B19 (orange). Thanks to our joint calibration process between the Faber \& Jackson and the fundamental plane, all of the modelled galaxies follow the same slope, providing consistency among all the cluster members, contrary to G22. Hence, our implementation has been successful at obtaining a scatter in the $\sigma_0$-$\text{Mass}$ relation without introducing biases for cluster members modelled with different observables. As expected from the calibration in Sect.\ref{sec:fundamental_plane_calibration}, our slope is steeper than the one obtained by B19, even after the addition of the lensing constraints. The parameter estimations obtained with the joint lensing and X-ray fit are listed in Table~\ref{Tab:lens_fundamental_plane}. As the median values are slightly changed, and the statistical errors have been reduced, the new medians are in agreement at $\sim1\sigma$ with the calibration. Thus, the lensing constraints have been successful in improving the accuracy of the fundamental plane estimation. The total mass contributed by cluster members without the BCG is $1.6\pm0.4$~$10^{12}M_{\odot}$ at $600$ kpc from the BCG.

\subsubsection{The gas distribution}
\begin{figure}
    \centering
    \includegraphics[width=\linewidth]{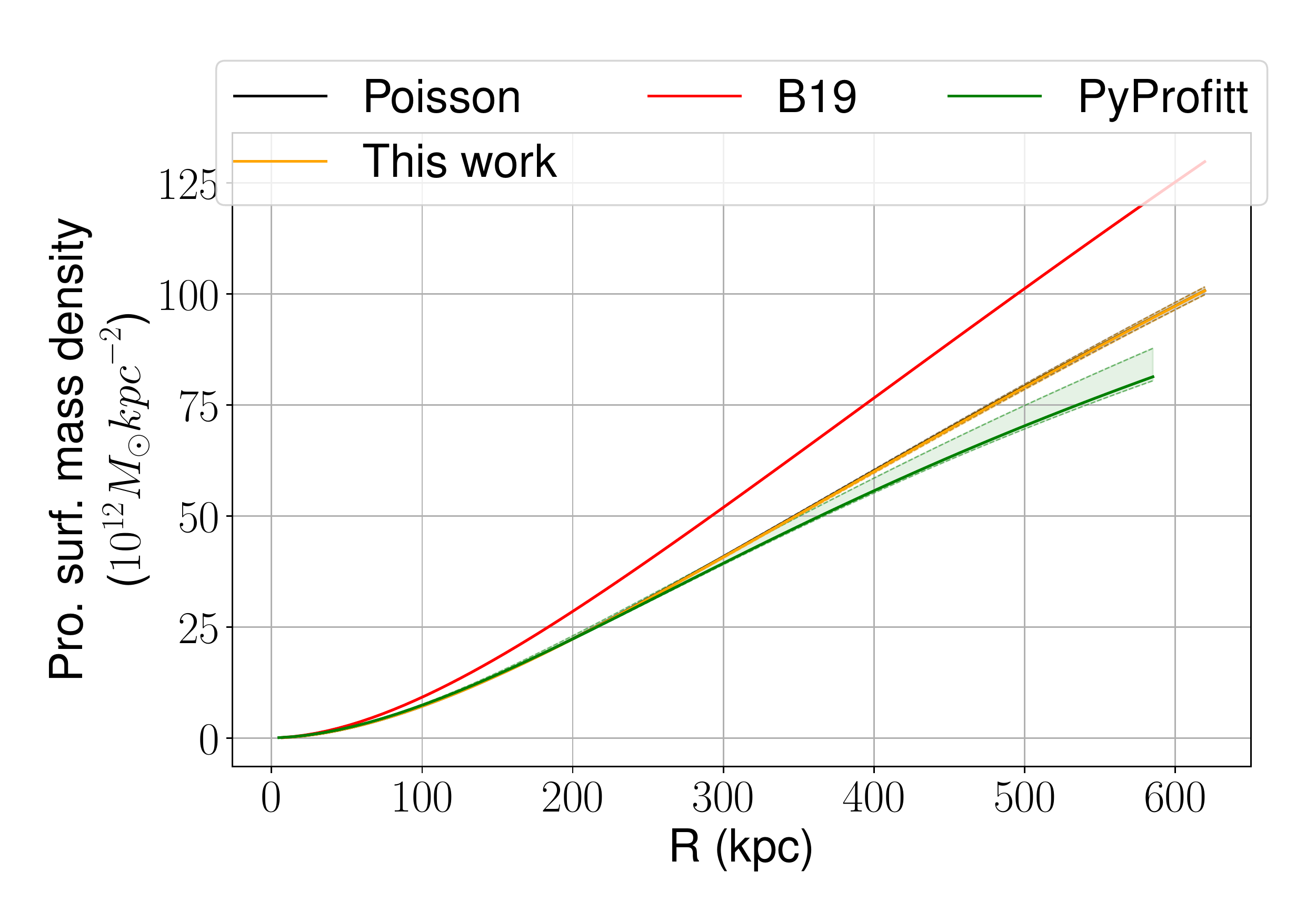}
    \caption{Projected gas mass density profiles integrated with a disk of increasing radius. The error represents the $"3\sigma"\, \rm CI$ of the posterior distribution except for B19 work, where only the best-fit model is publicly available. Our X-ray constraints and in B19 end at a radius of $\sim600$~kpc.}
    \label{fig:gas_prof}
\end{figure}

To compare the obtained gas distribution with other methods, we compute projected gas mass profiles shown in Fig.~\ref{fig:gas_prof} for several models. To assess the bias produced by the use of a Gamma-Poisson mixture instead of a Poisson statistic, we perform a fit of the gas using only the latter. As shown by the black and orange profiles, the use of the Poisson statistic has only a minimal impact on the overall fit as both curves are almost indiscernible from each other. Hence, it allows us to have a more satisfying statistical explanation of the considered X-ray constraints and provides the same gas distribution as the more standard Poisson likelihood. 

We also compare our mass profiles with the gas included in the B19 model (red solid line) as well as the profile obtained with the \textsc{PyProfitt} multiscale deprojection method \citep[green area]{Eckert2020}. The latter method is in good agreement with our results with a mass underestimate of only $3$ per cent, but the discrepancy is much larger with the B19 model. Indeed, their model shows an overestimation of $28$ per cent on average compared to our profiles. This model is the only one which has been optimised without the same set of X-ray constraints and parameters of the plasma emission model. Hence, this discrepancy can be partially explained by a difference when computing the effective area of the telescope, for example. We could not reproduce the analysis as \citet{Bonamigo2017,Bonamigo2018} because the energy at which this effective area has been computed is not publicly available. Hence, we can not assess if there are other reasons for the disagreement between our estimate and B19.

\section{Testing our reconstruction with the mass distribution of a mock cluster}

\label{sect:mock-mass-result}
To estimate the improvement in the mass reconstruction provided by our new method, we create a mock cluster based on our modelling of AS1063 without B-spline perturbations. We start from the best-fit model and remove all cluster members that are modelled according to the Faber \& Jackson law, assuming that the fundamental plane is a good representation of the actual cluster member masses. We use this mass model to predict X-ray counts with the associated Poisson-Gamma mixture while keeping other quantities fixed, such as the background count or the cooling functions. Regarding the lensing constraints, we use the barycentre of the source positions of the actual multiply-imaged systems of AS1063 and predict new multiple images with the mock mass. Thanks to this procedure, the realisation of the lensing constraints is fairly similar to reality. We then add a circular Gaussian error of $0.2$~$\rm arcsec$ on the positions of the images to account for line-of-sight perturbers and other systematics.

We create four different models to reproduce the mass of the mock cluster while assessing the improvements due to the inclusion of cluster members as well as in the large-scale distribution of the gas. We consider three models: one without the gas and galaxies with only the Faber \& Jackson scaling; one including cluster galaxies with the fundamental plane only and a third including only the X-ray emitting gas. And a fourth model that includes both modelling enhancements for cluster member galaxies and the X-ray emitting gas. To account for the combination of the fundamental plane and Faber \& Jackson relations, we split the considered galaxies between them to have the same proportions as in the real model. We created the two fundamental plane groups (i.e. galaxies with available $R_e$ or $\sigma_e$ measurements) by taking the most luminous galaxies from the two initial ones, and leaving the remaining elements to the Faber \& Jackson scaling. Indeed, inside the MUSE footprint and excluding contamination between cluster members, the faintest ones tend to be modelled with the latter relation as they do not have enough signal for morphological or spectroscopic measurements.

\begin{figure*}
    
    \includegraphics[width=\linewidth]{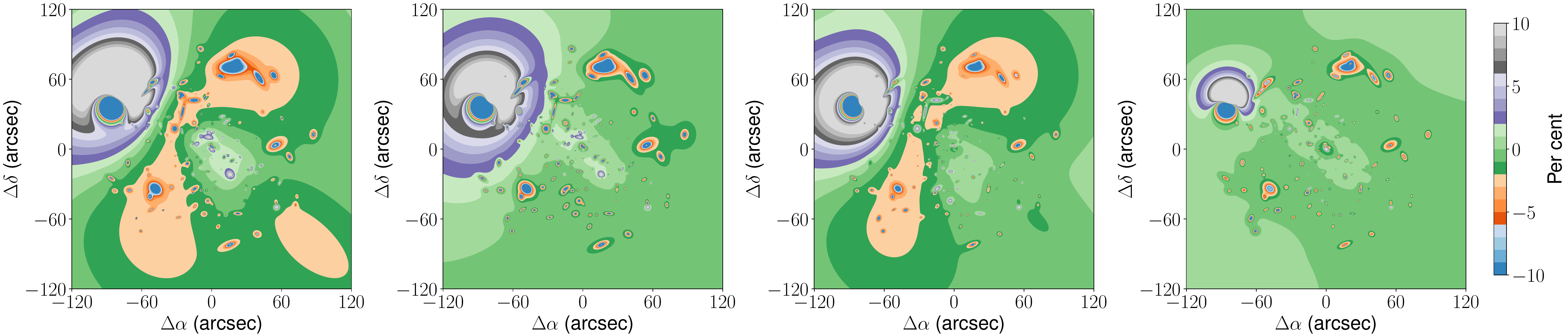}
    
    \caption{Maps of the relative differences between the projected surface mass density of each model tested on the mock cluster and the true one. From left to right, the models tested are the reference one, only the gas addition, only the fundamental plane one and finally, both of them. All maps have the same orientation, with the North in the upward direction and the East directed at the left.}
    \label{fig:fit-mass_tot_mock}
\end{figure*}

Figure~\ref{fig:fit-mass_tot_mock} shows the relative differences of the 2D mass distribution between the simulation and the best-fit of each model. From left to right, we have the reference model (i.e. no gas and uncalibrated Faber \& Jackson), only gas model, then the fundamental plane on top of the reference and finally, both enhancements. The best-fit parameters, as well as statistics from the posterior distributions, are available in Appendix~\ref{sect:dPIE-parameter} in Table~\ref{Tab:dPIE-parameter-mock}. If we do not consider the discrepancies around the NE halo that are due to the weaker effect of the strong lensing constraints, differences are mainly below $10$ per cent in absolute value in the core. As expected, the model with both improvements performs the best with errors mostly below $2$ per cent except for the NE halo and some cluster members that switched from the fundamental plane to the Faber \& Jackson relation. But the position of the NE halo is mainly constrained by its priors. The real NE position is included in the global posterior distribution of the two models using the fundamental plane scheme. Regarding the two other models, the relative declination of the NE halo is not recovered, with a maximum bias of $\sim9$~arcsec for the model without any improvements. We note that the imperfect modelling scheme of cluster members does not seem to have affected the overall distribution, but only locally, when the discrepancy between relations is higher. We discuss the differences due to the cluster member scaling relations in more detail in Sect.~\ref{sect:mock-cluster-member}.

As shown by the two models where the gas distribution is not considered separately, an underestimation of the mass is introduced where the gas distribution is locally modifying the main halo ellipticity. However, the gap is reasonable as it is below $4$ per cent, which indicates that models from previous studies are still providing a fairly accurate reproduction of the total mass. However, this could have an impact on analyses relying on an accurate estimation of the magnification, as a few per cent offset in mass can lead to significant differences in this quantity. It is also expected that in the case of a more perturbed cluster where the gas is not following the total mass, this underestimation would be greater thus biasing the resulting analysis.

\subsection{Gas distribution reproduction}

 \begin{figure}
    
    \includegraphics[width=\linewidth]{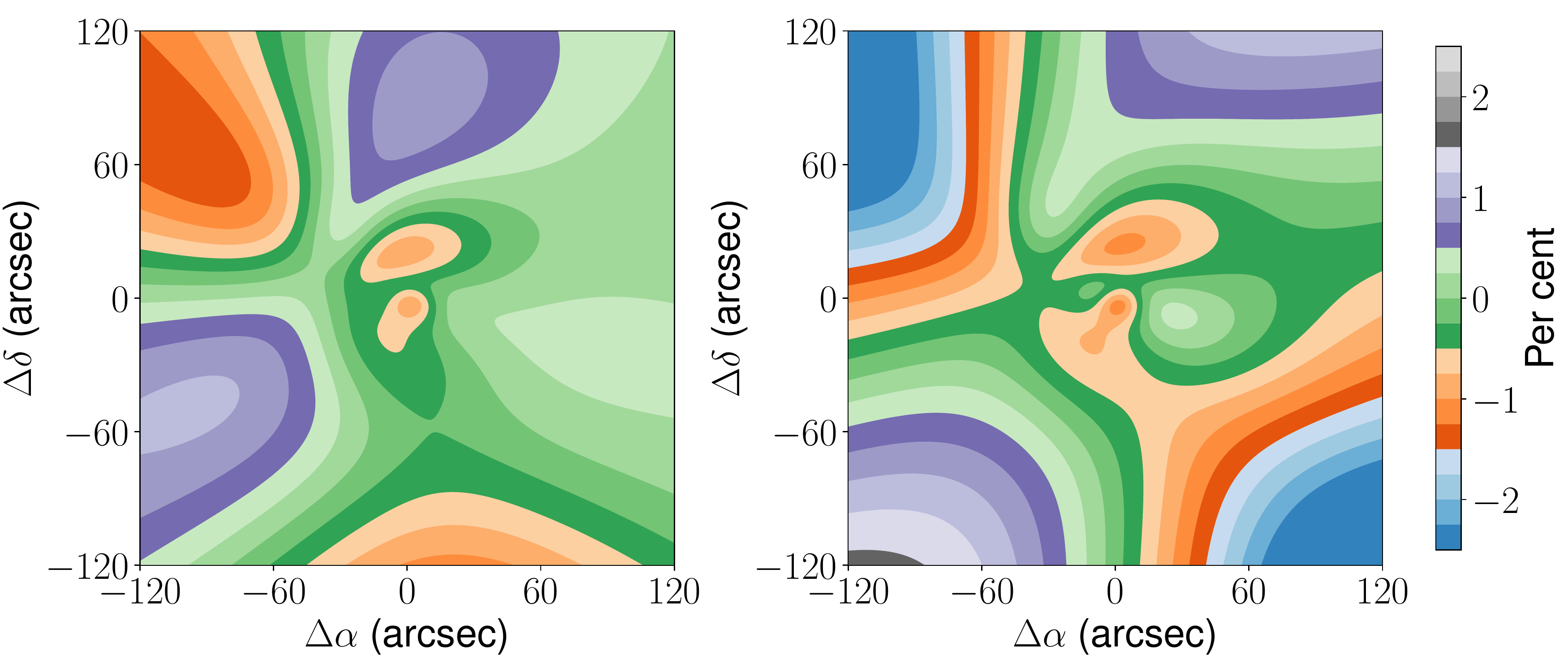}
  
    \caption{Maps of the relative differences between the projected gas mass distribution of the model that include only the gas and both the gas and the fundamental plane with the true distribution. The earlier and the latter are presented in the left and right panels, respectively}
    \label{fig:fit-mass_gas_mock}
\end{figure}

Figure~\ref{fig:fit-mass_gas_mock} shows the relative differences between the fitted gas distribution and the input one. As one can see, our method recovers it with excellent precision ($\leq2$ per cent) as the discrepancy is mostly $\leq1$ per cent for the two models. Regarding the X-ray intrinsic error, our models are able to recover the input values ($0.538$ count) with good agreement (i.e. included in the $"1\sigma"\,\rm CI$). Indeed, the model without the fundamental plane and the one with are presenting X-ray errors of $0.528^{+0.022}_{-0.023}$ and $0.529^{+0.024}_{-0.024}$ counts, respectively.

Regarding our goodness of fit procedure detailed in Sect.~\ref{sect:model-discri}, the models with and without the fundamental plane show  likelihoods for the best-fit model that are inside the $"1\sigma"\,\rm CI$ and the $"3\sigma"\,\rm CI$, respectively, of the expected likelihood distribution according to these count models. To assess with more confidence the previous results, we optimise $10$ gas-only models with new realisations of X-ray constraints based on the same initial count model. The best-fit models are showing $5$ over-fitting models, that are inside the $"1\sigma"\,\rm CI$, and $5$ under-fitting models split between the limits of the $"1\sigma"\,\rm CI$ and the $"3\sigma"\,\rm CI$. Hence, our two mock results made with a joint fit are consistent with these gas-only models. As we are here in an idealized case, assuming a larger $\rm CI$ of $"5\sigma"$ to define if a model reaches a good fit is a fair hypothesis according to the results on a mock cluster. We can also add that our goodness of fit score is biased by the non-consideration of the whole model posterior distribution for computation cost mitigation. We only use the best-fitting gas model to sample the expected likelihood distribution as it needs more than $10^{5}$ sample to provide reliable results on one model only, making percentile analysis among the whole distribution costly in terms of memory. Thus, we can expect that using the whole posterior distribution could provide different goodness of fit score than currently.

\subsection{Recovering of the cluster member masses}
\label{sect:mock-cluster-member}
\begin{figure*}
    \begin{minipage}{0.33\linewidth}
    \centering
    \includegraphics[width=\linewidth]{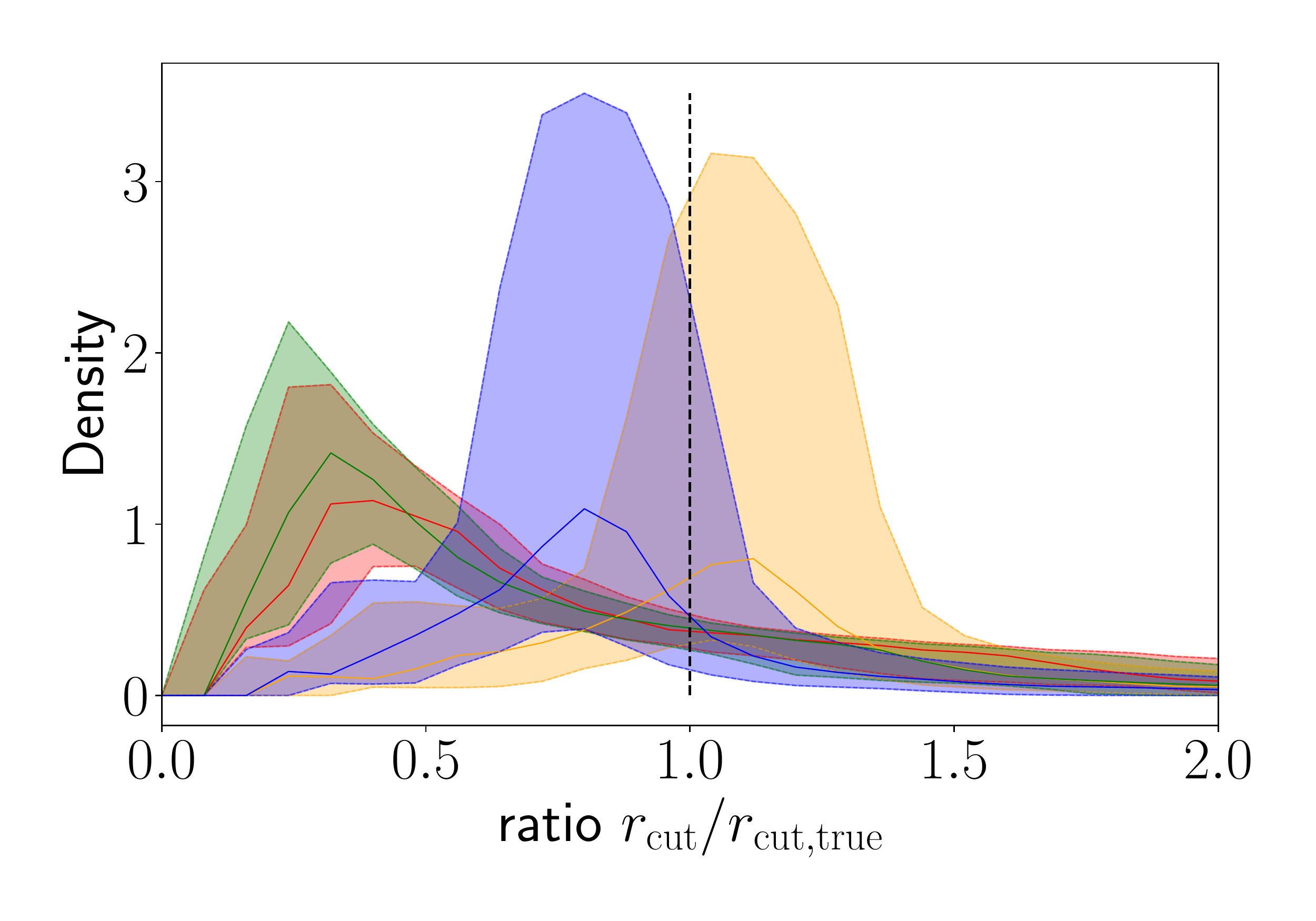}
    \end{minipage}
    \begin{minipage}{0.33\linewidth}
    \centering
    \includegraphics[width=\linewidth]{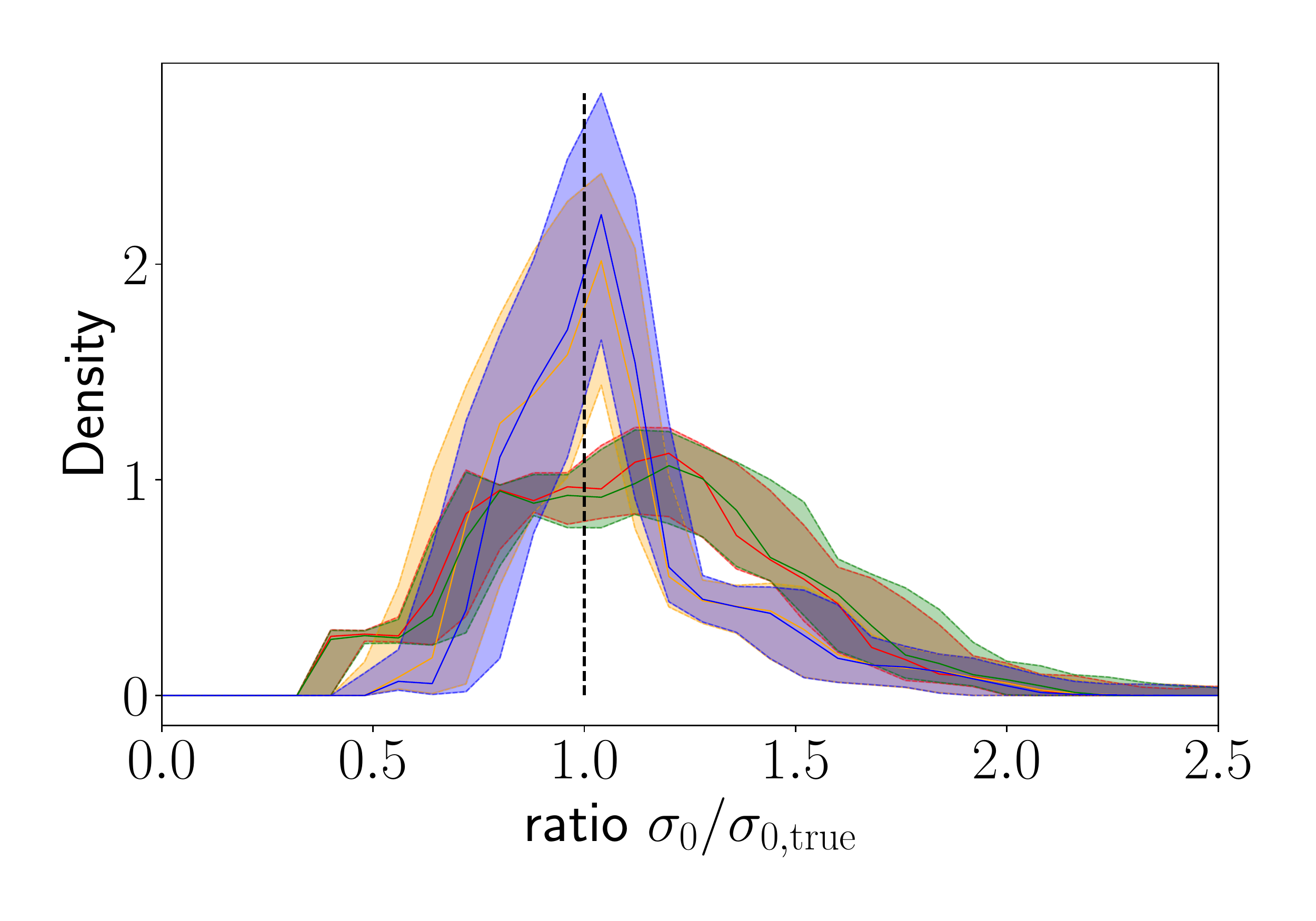}
    \end{minipage}
    \begin{minipage}{0.33\linewidth}
    \centering
    \includegraphics[width=\linewidth]{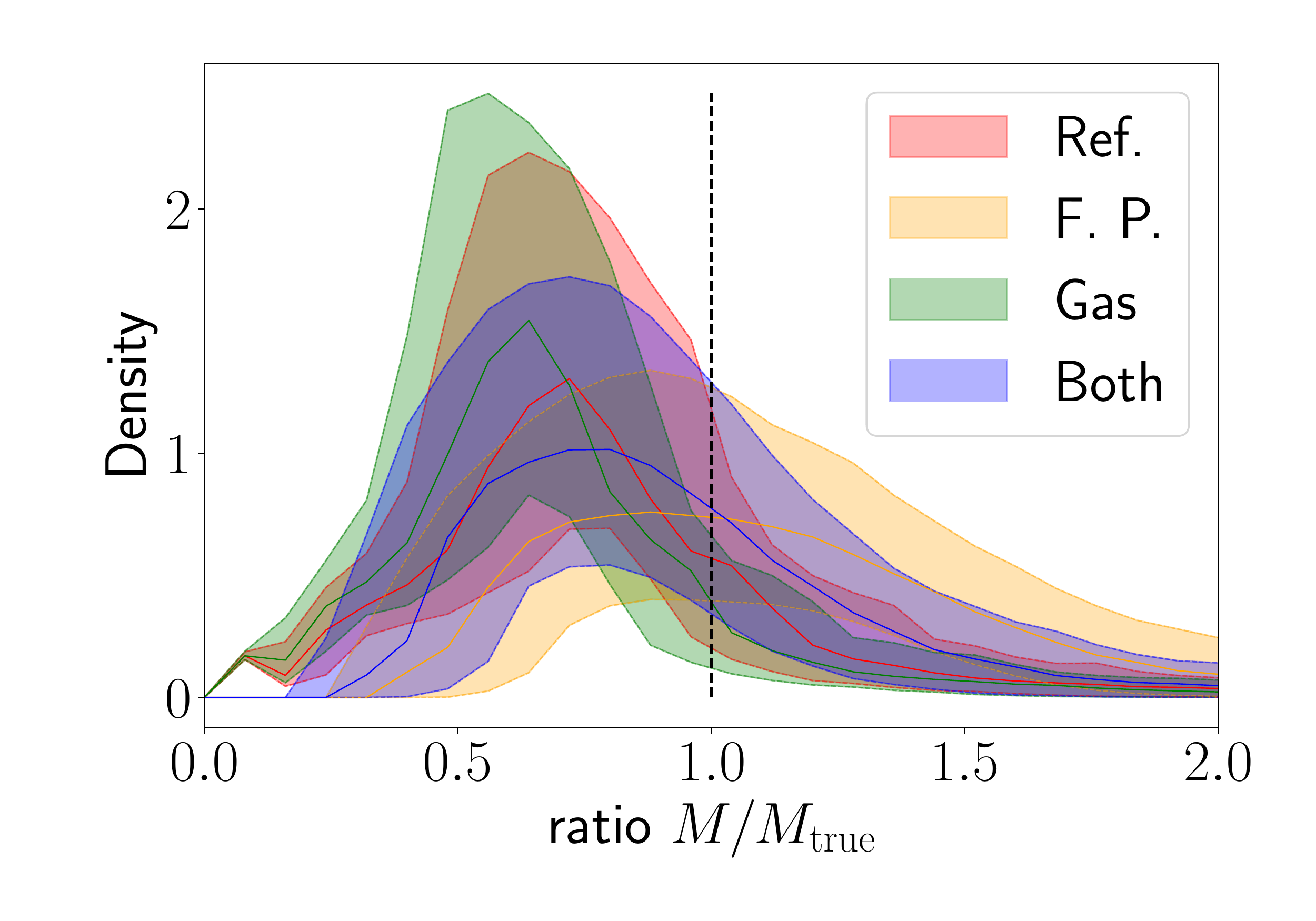}
    \end{minipage}
    \caption{Density distribution of the ratios between estimated dPIE parameters in the different models and their true values. Densities have been estimated for each model in the considered posterior distribution with the kernel density estimator implemented in \citet{Kumar2019} and a fixed bandwidth of $0.05$. The shaded areas and the plain lines represent the $"1\sigma"\, \rm CI$, and the median of the distribution of density. From left to right, the parameters considered are the cut radius, $r_{\rm cut}$, the central velocity dispersion, $\sigma_{0}$, and the total mass, $M$.}
    \label{fig:comparison_all_model}
\end{figure*}

To assess the efficiency of each model to estimate the mass of cluster members, we look at the recovered $r_{\rm cut}$, $\sigma_0$ and their total mass $M$. For each model in the posterior distribution, we build a distribution from the ratio between the considered parameter and its true value among all cluster members. We compute the density associated with the previous distribution with the kernel density estimator implemented in \citet{Kumar2019} with a fixed bandwidth of $0.05$. From the results on each model, we obtain the posterior distribution of this density. The plain lines and shaded areas in plots in Fig.~\ref{fig:comparison_all_model} show the median and the $"1\sigma"\, \rm CI$ of the density distribution for each parameter. As one can see, when using fundamental plane we recover the $r_{\rm cut}$ and $\sigma_0$ more accurately than for the two other models, that rely on the Faber \& Jackson solely. However, there are still offsets of $\sim20$ per cent for the $r_{\rm cut}$ when it is down to only a few per cent for the $\sigma_0$. In particular, this gap in the two parameters ($r_{\rm cut}$ and $\sigma_0$) can be explained by the goodness of fit of the scaling factor, $\nu$, between the light and the mass distribution. Indeed, the model with gas distribution and the one without have $\nu$ values of $1.87^{+0.35}_{-0.34}$ and $2.45^{+0.38}_{-0.36}$ for an input one of $2.14$, respectively. Hence, the biases on the posterior distribution are really similar to the ones on $r_{\rm cut}$ which is expected due to the fact that $r_{\rm cut}\propto\nu$ (see Appendix~\ref{app:c_p_explanation}) but also shows that the effect of $r^*_{\rm core}/r^*_{\rm cut}$ have a minor influence on the final estimation.

Regarding the total mass, $M$, shown on the right plot in Fig.~\ref{fig:comparison_all_model}, the differences when using of the fundamental plane with respect to the other models are less striking. Indeed, the results from the two models using the plane present similar biases than the models relying on the simpler scaling law. The main differences are in the precision of the estimation, which is lower for the two models using the plane, encompassing with a slightly higher probability density the areas around $1$. As the $r_{\rm core}$ estimation can not explain these biases (i.e. models with the fundamental plane are overestimating it), the only explanation is that good estimations of $r_{\rm cut}$ or $\sigma_0$ are not obtained for the same model in the posterior frequently. Hence, estimating each parameter independently is more reliable than combining them, according to our results.

\begin{figure*}
    \begin{minipage}{0.33\linewidth}
    \centering
    \includegraphics[width=\linewidth]{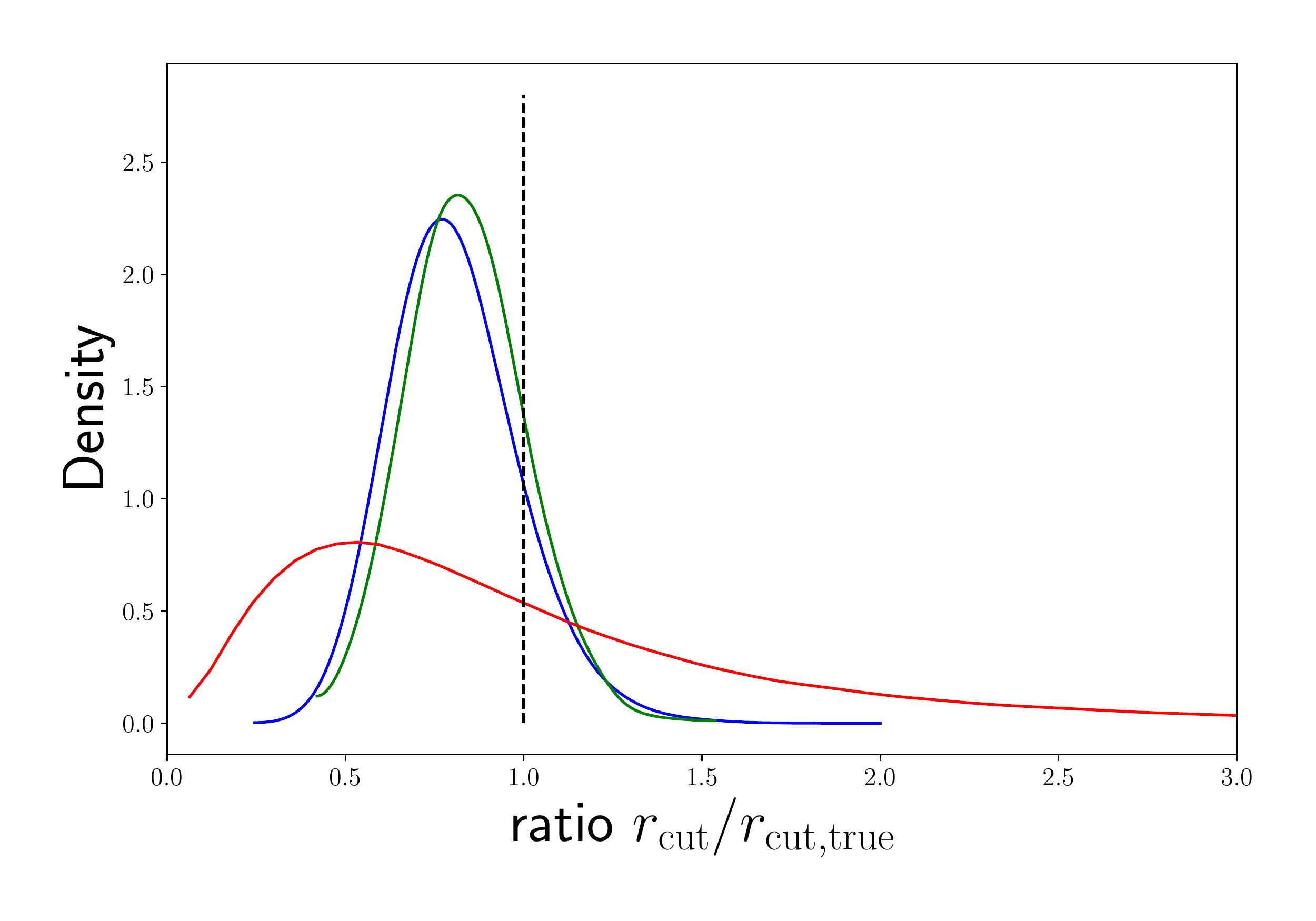}
    \end{minipage}
    \begin{minipage}{0.33\linewidth}
    \centering
    \includegraphics[width=\linewidth]{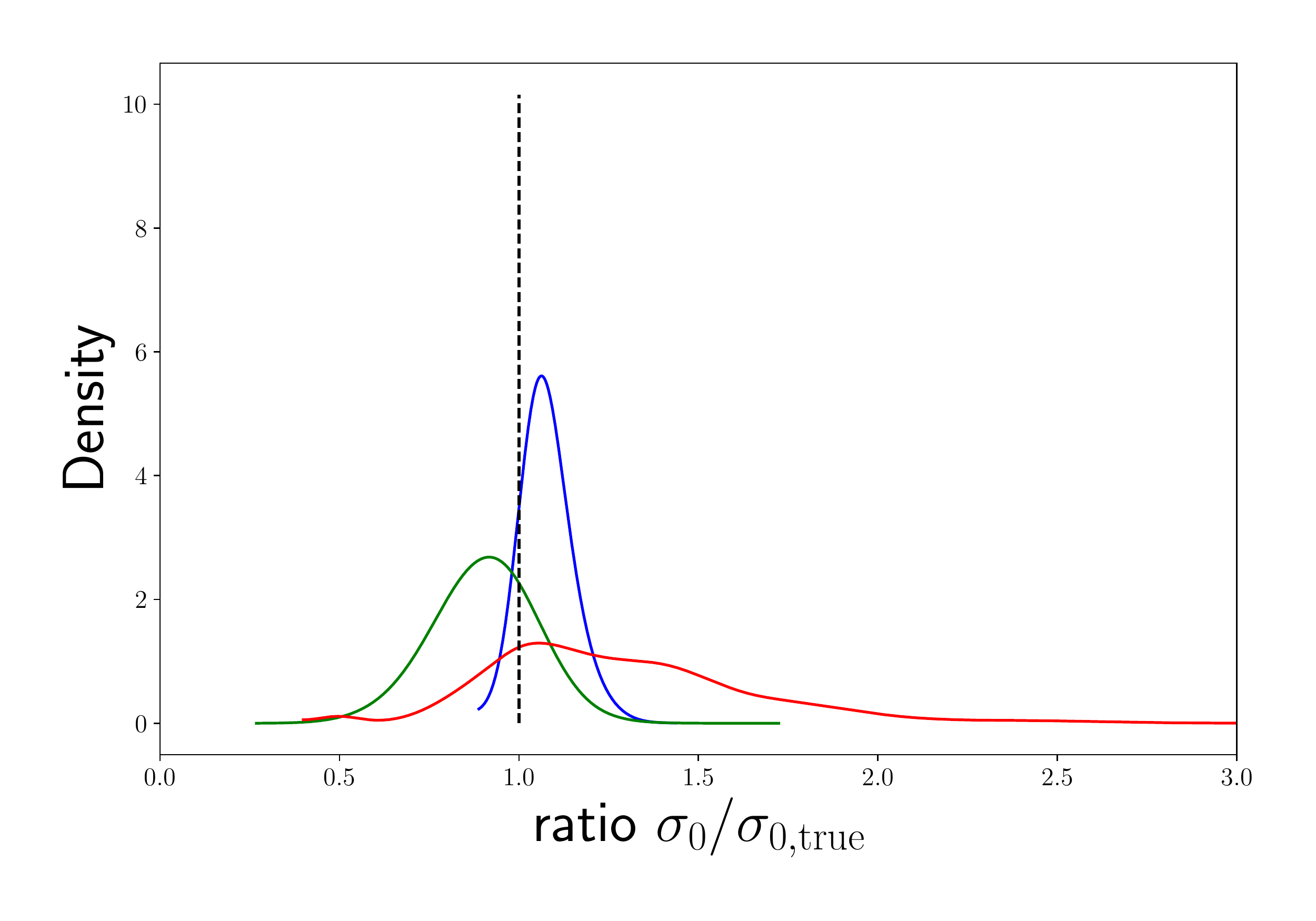}
    \end{minipage}
    \begin{minipage}{0.33\linewidth}
    \centering
    \includegraphics[width=\linewidth]{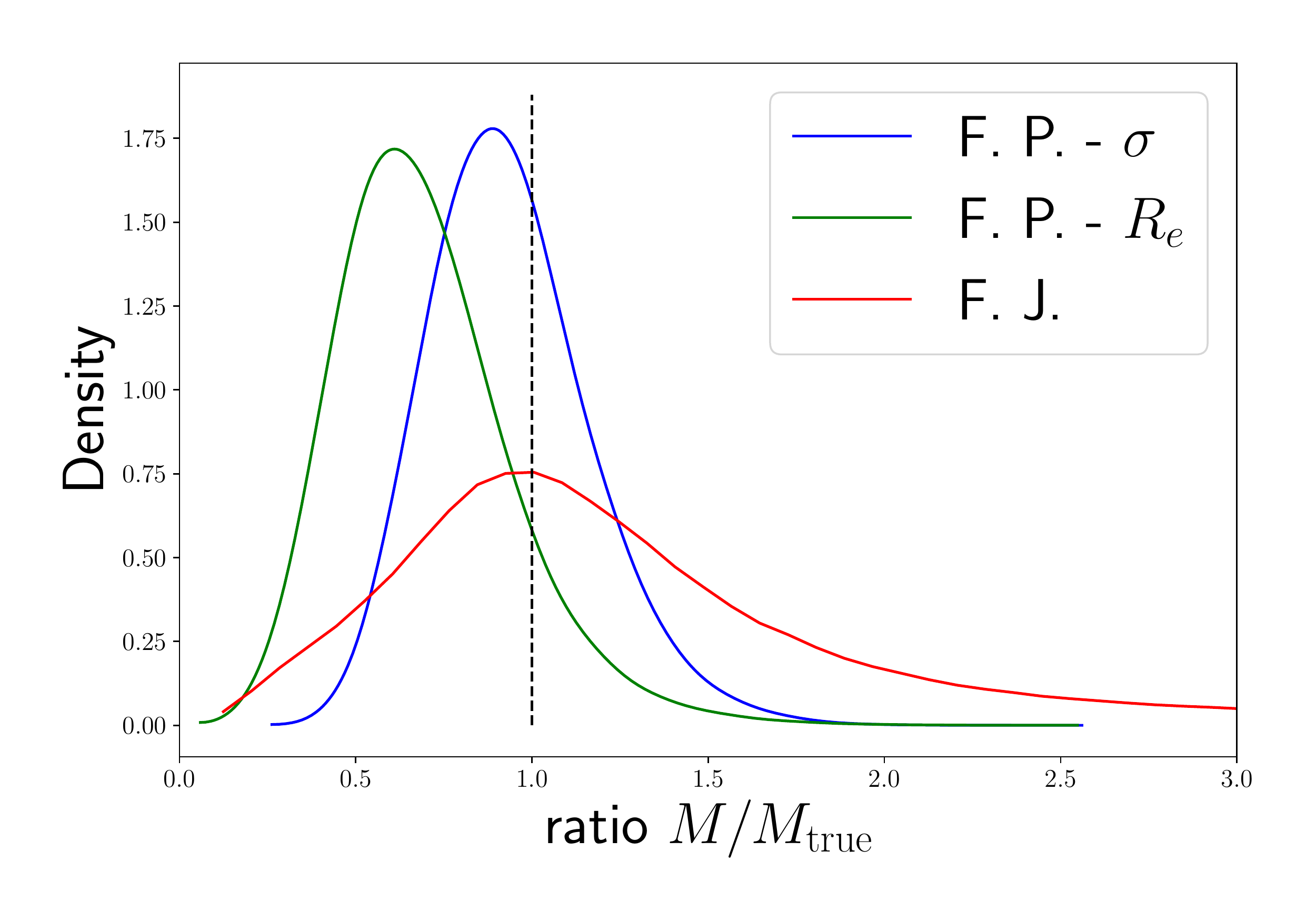}
    \end{minipage}
    \caption{Density distribution of the ratios between estimated dPIE parameters of each cluster member sub-groups and their true values. The density has been estimated on the concatenated distribution of ratio among all posterior models. The kernel density estimator, as well as the order of the parameters, are the same as in Fig.~\ref{fig:comparison_all_model}.}
    \label{fig:detail_model}
\end{figure*}

The method presented in this work used three different modelling schemes for the cluster members. We assess their efficiency in recovering the associated profile parameters in Fig.~\ref{fig:detail_model} with the model using the full method. Again, we build the distribution of the ratios presented previously for the three groups but we have to adapt the presentation scheme as the distribution for one model of the $r_{\rm cut}$ or $\sigma_0$ ratio leads to a Dirac distribution for the fundamental plane schemes. It is due to the proportional relations between the true values and the fitted ones for these parameters. Hence, we concatenate the distribution of each ratio from all models in the posterior, and we then estimate their density through the same method. Similarly to before, the model that use the fundamental plane provides a better estimation of $r_{\rm cut}$ and $\sigma_0$. One can see that we have the same bias for both schemes on $r_{\rm cut}$ estimation dominated by the $\nu$ estimation. Regarding $\sigma_0$ estimation, the two schemes show two opposite biases, with an underestimation when using $R_e$ as input and an overestimation when it is $\sigma_0$. In fact, the overestimation is easily explained from the behaviour of the projection factor $c_p$ (see appendix~\ref{app:c_p_explanation}) that scales up or down with $\nu$ at fixed $r^*_{\rm core}/r^*_{\rm cut}$. Hence, as $\sigma_0=\sigma_e/c_p$, if $\nu$ is lower than the true value, we will overestimate $\sigma_0$. It is more difficult to judge the pattern of the other scheme as also $\nu$ influences  $\sigma_{e, \rm FP}$. It scales exponentially with it according to the fundamental plane equation. Thus, we can deduce that the decrease in $\sigma_{e, \rm FP}$ with $\nu$ is faster than the $c_p$ ones, leading to the global underestimation observed. 

Regarding the total mass, $M$, estimation, we can see that the fundamental plane scheme with $\sigma_e$ as input is the best, as expected from our pre-modelling analysis in Sect.~\ref{sect:FP_prediction_accuracy}. It is followed by the Faber \& Jackson scaling relation and, finally, the second fundamental plane. Interestingly, the latter scheme is the only one not to overestimate $\sigma_0$. Thus, it cannot compensate for $r_{cut}$ underestimation as the two others. Both fundamental plane schemes are not equally stable according to a random realisation of constraints, as in their case, we have not modified the input observational parameters besides the multiple images. As shown by the previous analysis, the critical parameter in the wrong estimation pattern is $\nu$. Hence, physically motivated priors could solve this problem as it is linked to the mass-to-light ratio in our assumptions. 

In comparison with the reference model that does not have a calibration of the cluster member scaling relation, we can see here that the Faber \& Jackson scheme has a density centred on $1$ for $M$ estimation. Hence, as seen in Fig.~\ref{fig:M_sigma_0}, the calibration helps to have a homogeneity between the different relations.

We assume that representing the real mass distribution of cluster members with the fundamental plane of elliptical galaxies is a fair hypothesis. Hence, the previous results show that incorporating knowledge of members kinematics as well as using a more complex relation than the Faber\&Jackson scaling one allow us to recover more accurately and robustly the parameters of the cluster members mass profiles. This method effectively reduces the biases in the parameter estimation, as well as it improves the consistency of the mass model with existing observational information.

\section{Discussion}
\label{sect:discussion}

Our new scheme for the modelling of cluster member masses outlined in Sect.~\ref{sect:clus-memb-dist} makes several assumptions that we discuss in Sect.~\ref{sect:cluster-member-hypot-consistency} in regard to the obtained mass models. Thanks to the addition of X-ray constraints, we obtain some important insights into the cluster gas thermodynamics. Finally, thanks to this new method, we discuss the hypothesis that AS1063 has undergone a merging event in Sect.~\ref{sect:merging-event}, and its gas fraction mapping in Sect.~\ref{sect:gas_fraction}.

\subsection{Cluster member mass hypotheses}
\label{sect:cluster-member-hypot-consistency}
\begin{figure*}
    \begin{minipage}{0.49\linewidth}
    \centering
    \includegraphics[width=\linewidth]{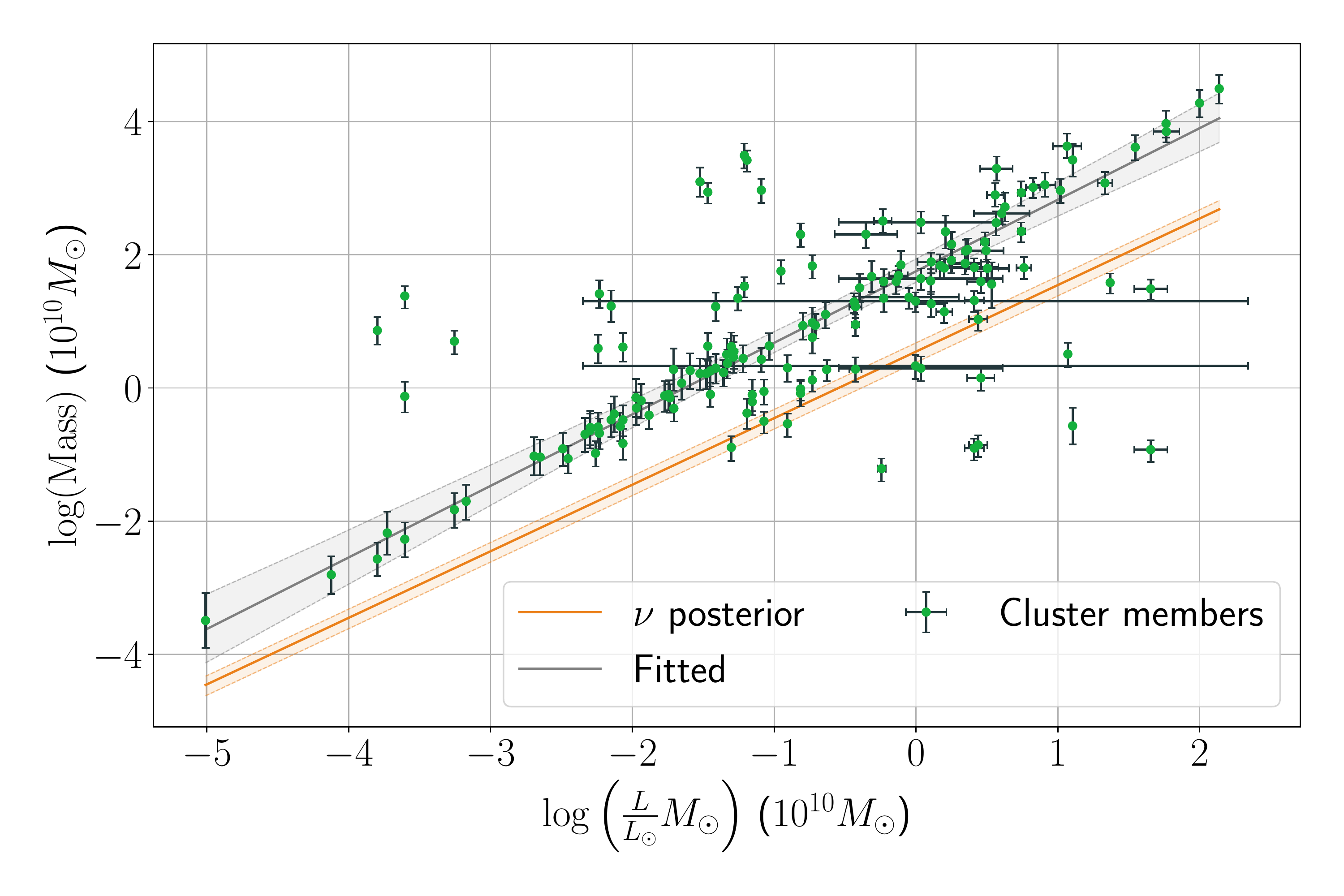}
    \end{minipage}
    \hfill
    \begin{minipage}{0.49\linewidth}
    \centering
    \includegraphics[width=\linewidth]{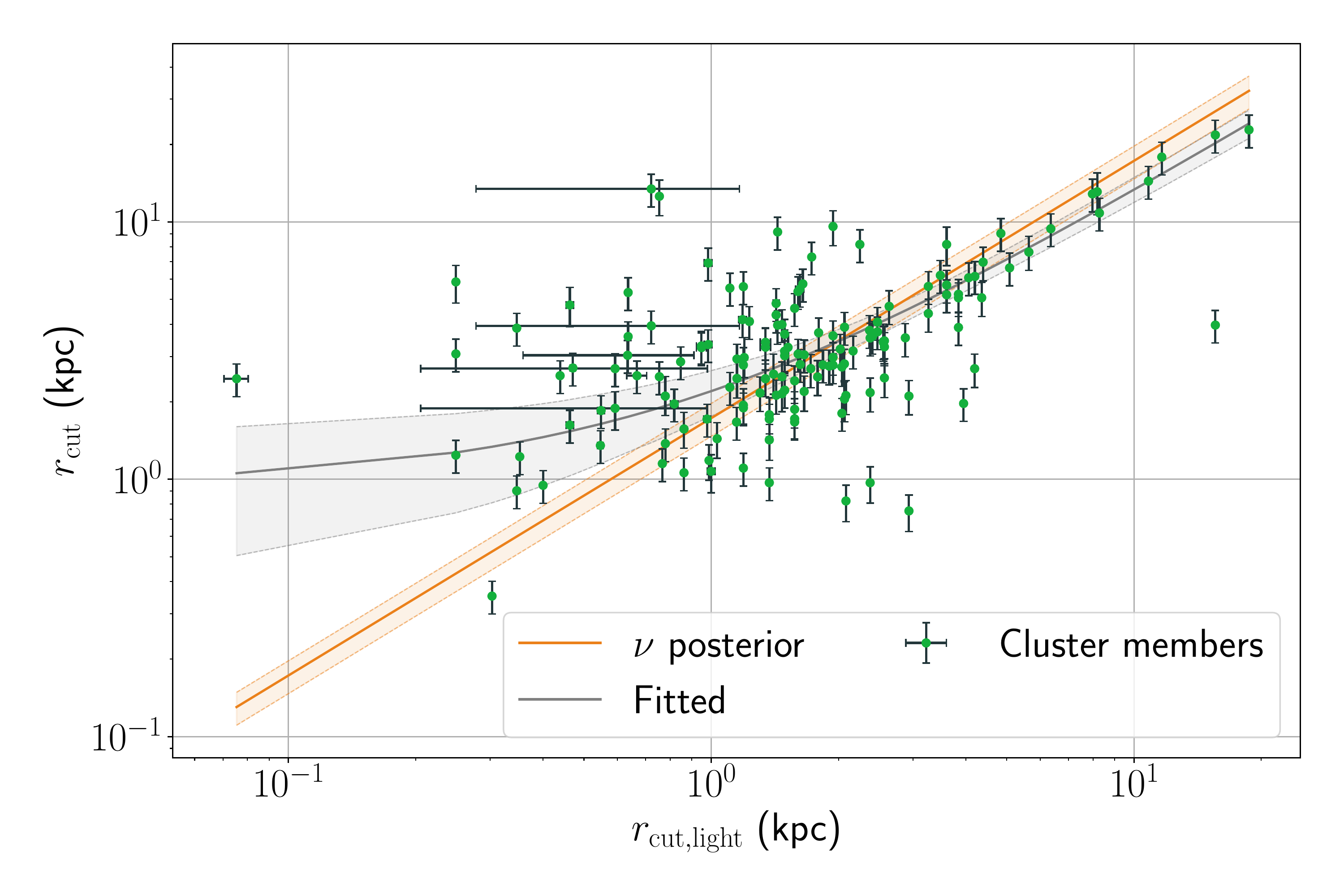}
    \end{minipage}
    \caption{\textit{Left panel:} Plot of the cluster member dPIE total mass as a function of their luminosity. Error bars represent the standard deviation of each quantity. \textit{Right panel:} Plot of the cluster member dPIE $r_{\rm cut}$ as a function of the $r_{\rm cut, light}$ estimated on their Sersic light profiles. For both panels, the grey area represents the actual relations fitted on the presented data, and the orange one represents the relations assumed in our modelling hypothesis, and its estimation through the joined X-rays and lensing fit.}
    \label{fig:nu_interpretation}
\end{figure*}

\begin{table}
\centering
\begin{tabular}{cc}
    \hline
    Parameters & Median$^{84 \text{ per cent limit}}_{16 \text{ per cent limit}}$\\
    \hline
    $\beta$&$2.43^{+0.25}_{-0.21}$\\
    $\delta$&$0.074^{+0.073}_{-0.069}$\\
    $\zeta$&$1.24^{+0.11}_{-0.12}$\\
    $\eta$&$3.85^{+0.26}_{-0.25}$\\
    \hline
\end{tabular}
\caption{Parameters of the more realistic relations linking $M$ and $L$ and also $r_{\rm cut}$ and $r_{\rm cut,light}$.}
\label{Tab:cluster-member-relations}
\end{table}

As detailed in Appendix~\ref{app:c_p_explanation}, our 3D model of the cluster member mass and light distribution assumes a constant mass-to-light ratio that has the same value as the scale factor, $\nu$, between the light and mass profiles. In particular, these assumptions allow us to link $R_e$ to $r_{cut}$, and to compute $c_p$. Still, there is no input information in the modelling neither on the total luminosity, $L$, nor the cut radius of the light distribution, $r_{\rm cut,light}$. As the previous quantities are available, we address the consistency of our hypotheses in Fig.~\ref{fig:nu_interpretation} presenting the dPIE total mass as a function of the cluster member luminosity in the left panel, and the scaling between $r_{\rm cut}$ and $r_{\rm cut,light}$ in the right one. The assumed relations between these quantities and their estimations from our joint X-rays and lensing optimisation are presented in orange, where the plain line and the shaded area represent the median and the $"1\sigma"\, \rm CI$ of the posterior distribution, respectively. As the assumed relations are simplifications of the known reality (e.g. a constant mass-to-light ratio), we sample the more realistic ones, $M=\beta L^{1+\delta}$, and, $r_{\rm cut}=\zeta r_{\rm cut,light}+\eta$, on the data plotted. The obtained posterior distribution is represented in grey with the same scheme as previously. The parameter statistics for both relations are given in Table~\ref{Tab:cluster-member-relations}.

Our model provides $"2\sigma"\,\rm CI$ for $\nu$ which overlap with the same $CI$ for $\beta$ and $\zeta$. We note that our result lies between the estimate of $\beta$ and $\zeta$, indicating that our simplified model will need to be more complex to allow for such discrepancy. Indeed, sampling on both $\beta$ and $\zeta$ shows two disjoints $"5\sigma"\,\rm CI$. As expected, $\delta>0$ is favoured, but our results prefer way lower values than the $~0.35$ estimated by \citet{Bolton2008} and \citet{Koopmans2009} from galaxy-galaxy lensing analysis; this value is actually the upper bound of the $"5\sigma"\,\rm CI$ for $\delta$. It is possible that $\delta$ measurements are biased by the galaxy mass estimation, which has been performed assuming the homology resulting in the observed differences with previous works. As seen on the right panel in Fig.~\ref{fig:nu_interpretation}, the relation between $r_{\rm cut}$ and $r_{\rm cut,light}$ seems to present a non-zero offset with strong confidence indicating a turn for small $r_{\rm cut,light}$. It is difficult to assess the meaning of this relation as the use of a dPIE for the light distribution is already a proxy of the real light distribution that simplifies the $c_p$ computation. Indeed, a dPIE can approximate a Sérsic profile \citep{Suyu2014} partially, but it cannot reproduce its central and asymptotic behaviours.

Nevertheless, $\nu$ is neither providing a perfect representation nor a robust estimation of the previous relations. It is, however, a fair proxy to both of them as the measured relations are mainly taken into account in the wide $\nu$ posterior, and the modelling hypotheses are thus consistent with the actual results. Considering that cluster members are only accounting for a small percentage of the total mass and any increase in the model complexity will worsen the already high computational cost, this implementation of the fundamental plane provides more benefits than drawbacks. We also do not intend to propose a final description as it is only the second iteration of strong lensing mass modelling that includes the fundamental planes, with G22 being the first one. Hence, new schemes that would model the previous relations more closely to reality will allow us to improve the robustness of cluster members mass estimation.

Some of the comments addressed here could also be relevant in the case of G22-like modelling as our scheme using $\sigma_{0,\rm FP}$ is quite similar to theirs. Indeed, we also have $R_e$ proportional to $r_{\rm cut}$ through admittedly a different parametrization, but a translation is possible. The other differences are the association of the light and mass distributions, and the computation of $c_p$ if taken into account. In any case, if a value of $\nu$ or $c_p$ is assumed (e.g. $\nu=1$ for B19), a parameter transformation between approaches can be put in place.

\subsection{Merging events}
\label{sect:merging-event}
As previously mentioned in Sect.\ref{sect:constraints_repro}, it is suspected that AS1063 is undergoing a merging event. It was first proposed by \citet{Gomez2012} in the form of a major merger (1:4 mass ratio) happening perpendicular to the LOS with arguments based on the kinematics of cluster members, a high intra-cluster gas temperature and the asymmetry of the X-ray emission. Since then, further studies have brought up new arguments in favour of this scenario. \citet{deOliveira2021} analysed the intra-cluster light distribution which shows evidence for a cluster merger. \citet{Xie2020} and \citet{Rahaman2021} reported a giant radio halo and a gradient in the temperature distribution with a bow shape, respectively. Both are indicators of a merging event. The spectral analysis of the cluster members made by \citet{Mercurio2021} revealed a bimodal redshift distribution among the $~1200$ objects measured. Their results on the kinematics properties of the galaxies near the NE halo are also in favour of a different origin than the rest of the cluster members. In that context, we want to show in this section the ability of our method products and by-products to combine pieces of evidence from both lensing and X-ray analyses to help confirm such hypotheses. 

\begin{figure*}
    \begin{minipage}{0.48\linewidth}
    \centering
    \includegraphics[width=\linewidth]{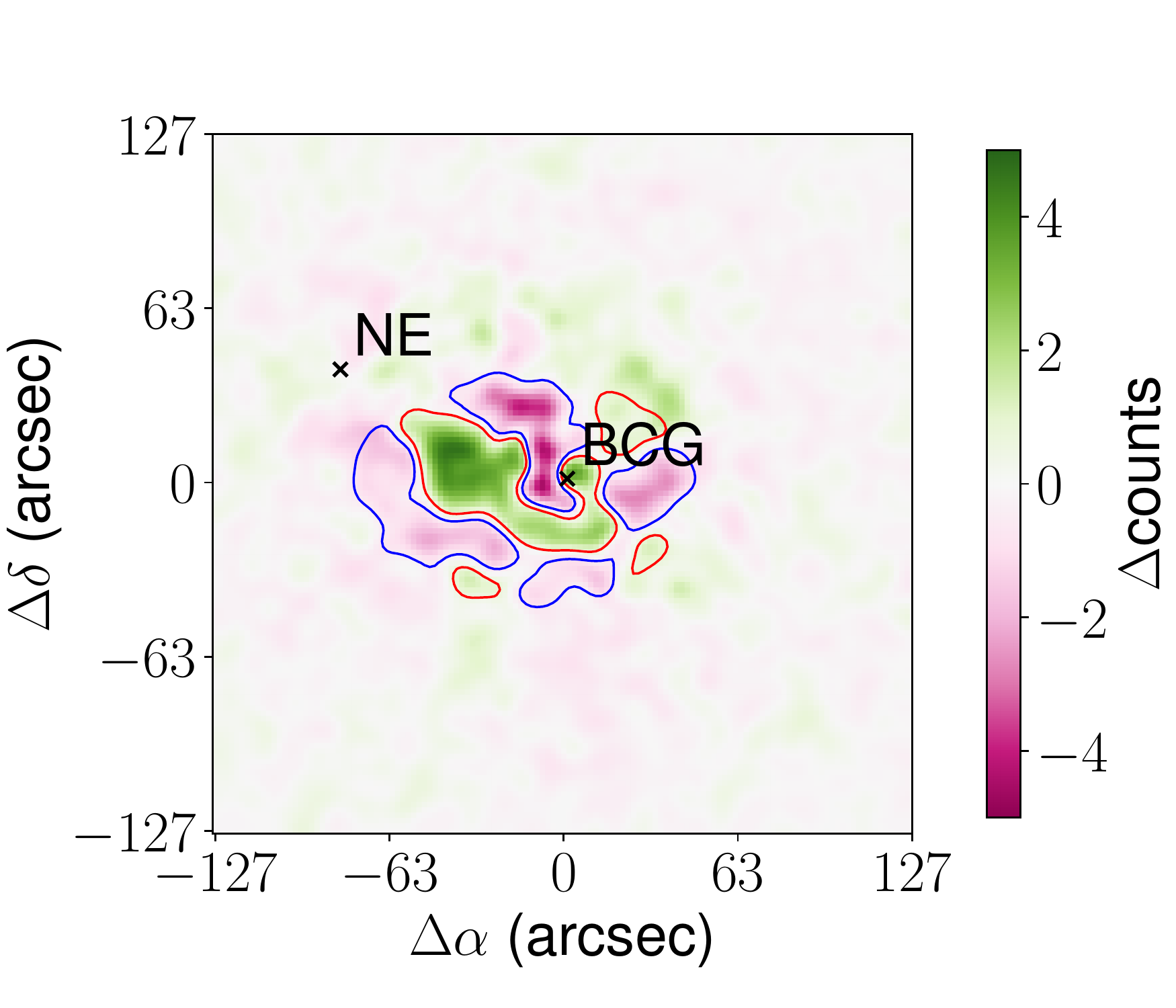}
    \end{minipage}
    \begin{minipage}{0.48\linewidth}
    \centering
    \includegraphics[width=\linewidth]{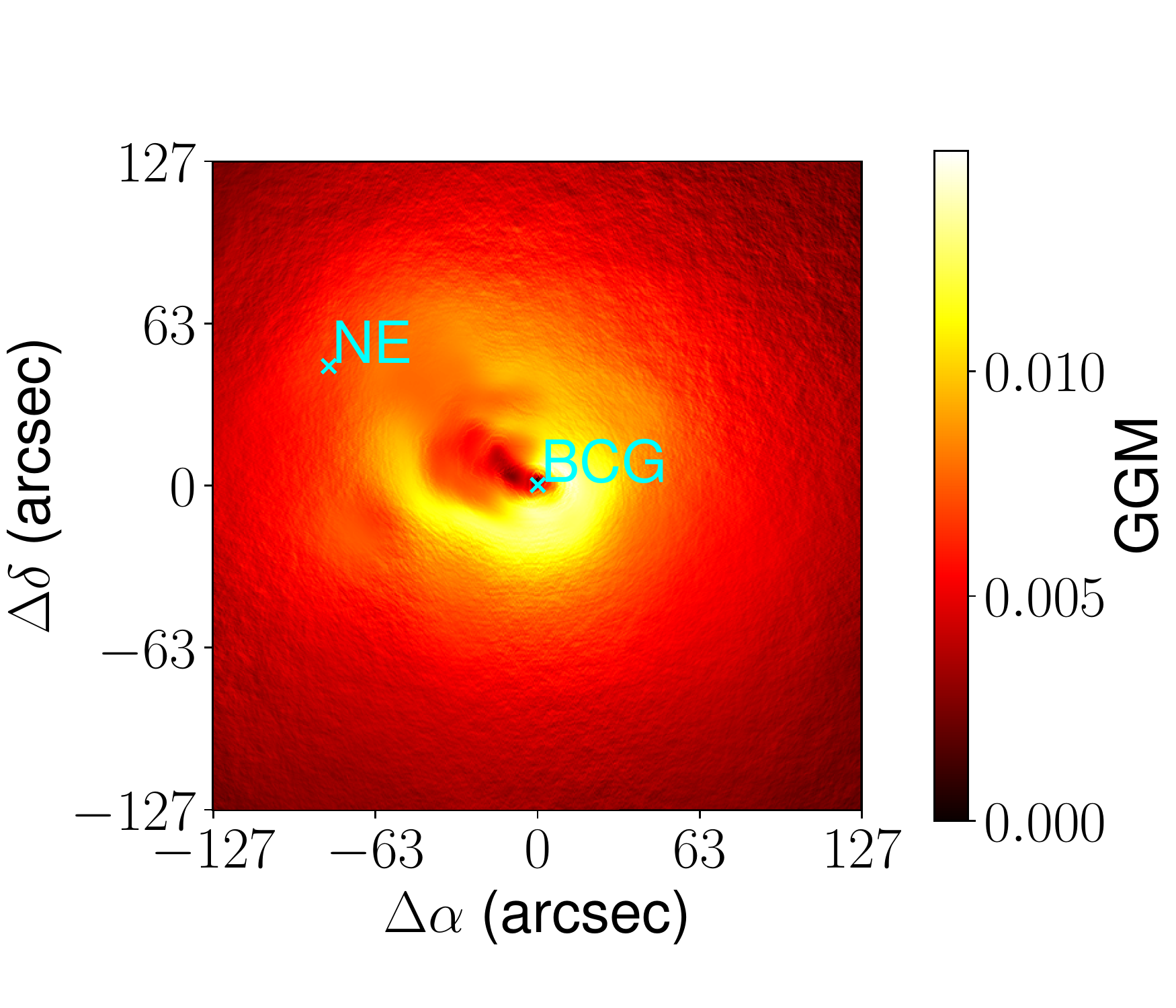}
    \end{minipage}
    \caption{\textit{Left panel:} Map of the smoothed residual (Gaussian kernel of $2$ pixel size) between the observation and the best-fit model. Blue and red contours highlight the region of negative and positive excess, respectively. These regions are used to probe a possible gas sloshing quantitatively with ICM properties. \textit{Right panel:} Adaptively smoothed count map between $0.5$-$7.0$ keV with a Gaussian gradient magnitude method \citep{Sanders2021}.}
    \label{fig:merging_event-1}
\end{figure*}

Our modelling of the X-ray surface brightness is only able to fit details at the cluster scale; thus, signs of gas sloshing from an infalling object can be visible in the map of the residual counts. Indeed, the results presented in Sect.\ref{sect:constraints_repro} show a shallow pattern of overestimation and underestimation of the observed counts. Figure~\ref{fig:merging_event-1} presents the residual map smoothed with a Gaussian kernel of $2$ pixel size, as well as the count map smoothed with an adaptive Gaussian gradient method\footnote{\url{https://github.com/jeremysanders/ggm}} \citep{Sanders2021}. The earlier shows more clearly the pattern previously mentioned, which could be the result of the initial object containing the NE halo that has fallen into AS1063 following the SW to NE direction. The axis of this trajectory is the same as indicated in previous works \citep{Gomez2012,Rahaman2021}. The latter map and its "U" shaped pattern are an indication of ram pressure stripping, in particular, that an object which seems to match with the BCG position is moving in the Intra-Cluster Medium (ICM). If we invert the reference frame, it can be interpreted as the ICM moving around the BCG as a consequence of an infalling object passing in the cluster. 

\begin{figure*}
   
    \includegraphics[width=\linewidth]{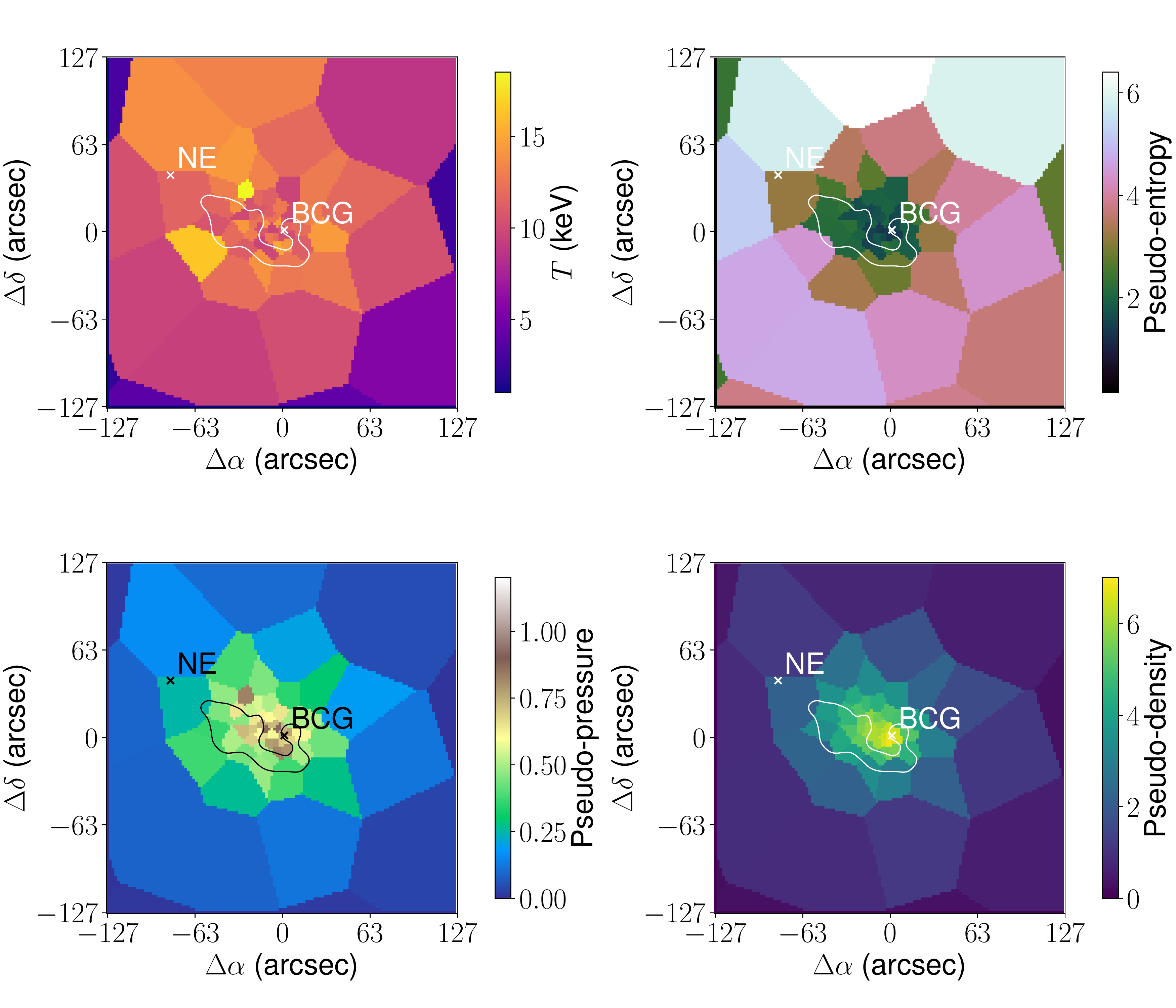}
    \caption{Maps of the ICM properties obtained from the X-ray emission, with the temperature (\textit{top left}), pseudo-entropy (\textit{top right}), pseudo-pressure (\textit{bottom left}) and  pseudo-density (\textit{bottom right}). The white or black contour highlights the "spiral" pattern seen in the X-ray fit residual.}
    \label{fig:merging_event-2}
\end{figure*}

In addition to the information related to the X-ray surface brightness and its fit, we have in hand maps of the ICM properties that are direct products of our method or which can easily be deduced from them. Indeed, to define our plasma emission model, we need a map of the cluster temperature that is produced along with the Norm of the APEC model. From these two quantities, we deduce a pseudo-entropy, a pseudo-pressure and a deprojected gas density. The three latter quantities, as well as the temperature for a SN threshold of $6$, are shown in Fig.\ref{fig:merging_event-2}, with the "spiral" pattern highlighted with white contours. Globally, these maps show a unimodal distribution with an asymmetry towards the  NE, but they are clearly in disagreement of two distinguishable structures as seen in the case of a major merger with a close mass ratio such as the Bullet cluster \citep{Million2009}. Indeed, there only seems to be one low entropy area associated with the AS1063 centre. Thus there is no trace of a remnant low entropy gas of a second cluster core, and the pseudo-pressure is also not showing a complex structure, as expected in the case of a major merger event. However, AS1063 does not look totally relaxed, as the temperature map presents some gradients in agreement with a perturbed state. From this analysis and the mass ratio of $~300$ between the two dPIEs haloes representing the main mass component and the NE one, the remaining plausible scenario is the fall of a much less massive object, such as a galaxy group, into AS1063. Such events will produce some sloshing of the gas, as previously proposed, which can be constrained quantitatively from the ICM properties in the under/over estimation area of the X-ray surface brightness \citep{Ueda2021}. In particular, it is expected that the entropy and temperature are higher in the negative excess region than in the positive one, with a difference higher for the first quantity. It is the opposite for the density, and the pressure should be unchanged. Due to maps binning being on a scale similar to the considered area, they are not the best tools to measure the differences between these regions.

\begin{figure}
   \includegraphics[width=\linewidth]{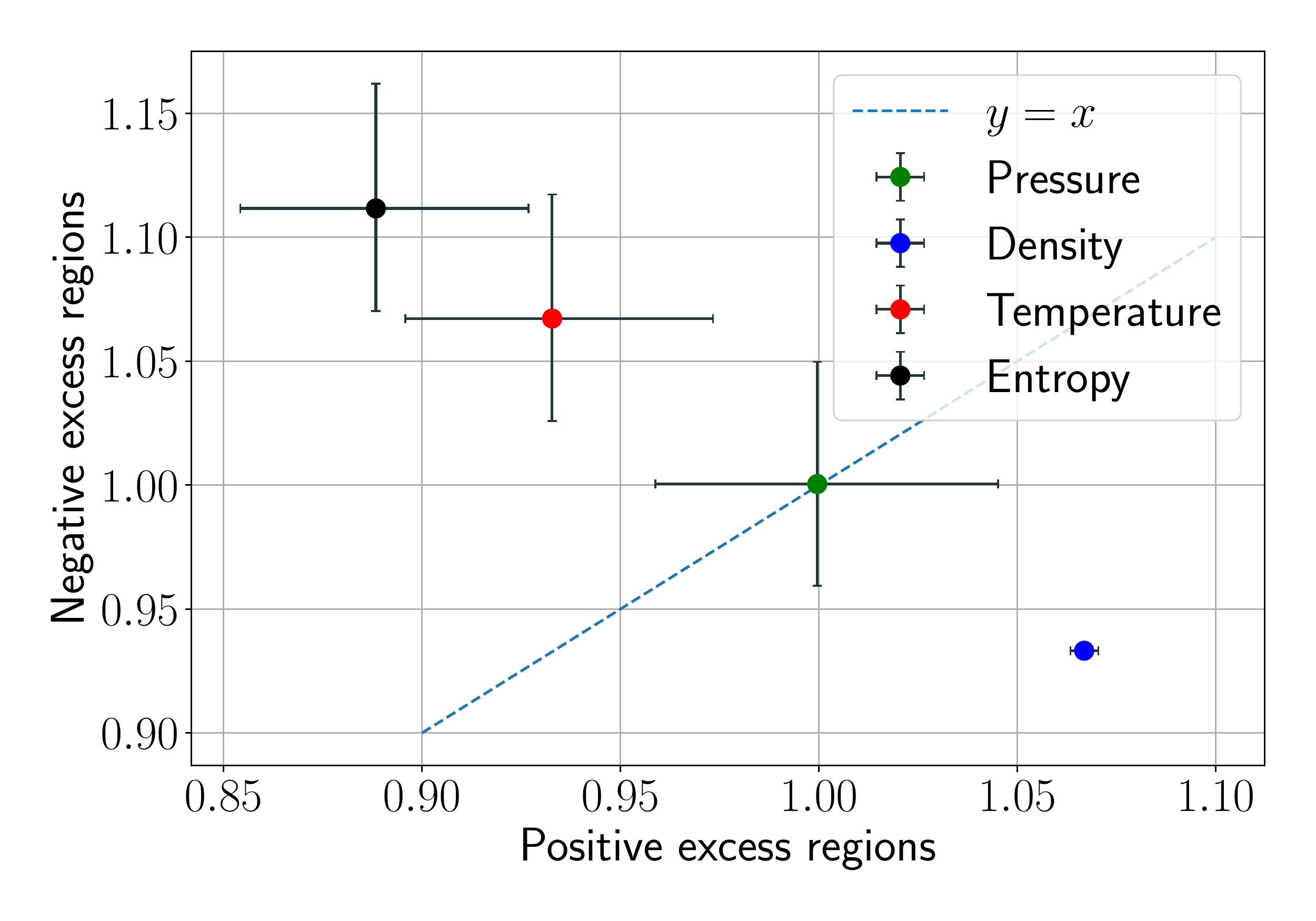}
    \caption{Comparison of the ICM properties between the positive and negative excess regions, which are presented in Fig.\ref{fig:merging_event-1} with the red and blue contour, respectively. The error bars represent the $"1\sigma"\, \rm CI$ of the posterior distribution.}
    \label{fig:merging_event-3}
\end{figure}

To obtain a quantitative answer on the presence of gas sloshing, we extract two spectra based on the blue (negative excess) and red (positive excess) contours shown in Fig.\ref{fig:merging_event-1}, and fit them with the same method as described in Sect.~\ref{sect:X-ray-data-analysis}. We present in Fig.\ref{fig:merging_event-3} the temperature, entropy, pressure and density measurements with a cosmetic normalization to plot them next to each other. As one can see, it shows the expected pattern for gas sloshing. If we look more precisely in detail at the posterior distribution of all measurements, the $"5\sigma"\,\rm CI$ of the entropy and density agree with that phenomenon, and it is $"3\sigma"\,\rm CI$ for the temperature. Hence, we can assess that we do detect the existence of gas sloshing in AS1063. Such a process is in agreement with a minor merger event between the cluster and a much smaller infalling object like a galaxy group. Our method and the tools we developed are able to give us important insights into the AS1063 dynamical state without requiring a disjoint analysis, thus showing an example of the possible study that will benefit from this kind of modelling.

\subsection{Hydrostatic equilibrium and gas fraction}
\label{sect:gas_fraction}
As we model with separate haloes, the intra-cluster gas and the rest of the matter, we do not assume the hydrostatic equilibrium. Thus, our method is well suited to analyse the hydrostatic bias by comparing its results with another method on the same observables. To perform this comparison with the associated hydrostatic mass of AS1063, we fit a Navarro–Frenk–White profile \citep{Navarro1991} on the X-ray observations with the \textsc{Hydromass}\footnote{\url{https://hydromass.readthedocs.io/en/latest/intro.html}} package based on \citet{Ettori2010} method. We use the same count, background and exposure maps as well as the plasma emission model, but we use a wider FoV than the one used for the \textsc{Lenstool} optimisation. Indeed, we fit the X-ray surface brightness up to almost $1$ Mpc as we are less dominated by the background in the circular binning. To obtain a temperature profile, we use the same fitting method presented in Sect.~\ref{sect:X-ray-data-analysis}, but instead of using a Voronoï tessellation, we make circular binning with $3000$ counts each in the $4$-$7$ keV band.

\begin{figure*}
    \begin{minipage}{0.49\linewidth}
    \centering
    \includegraphics[width=\linewidth]{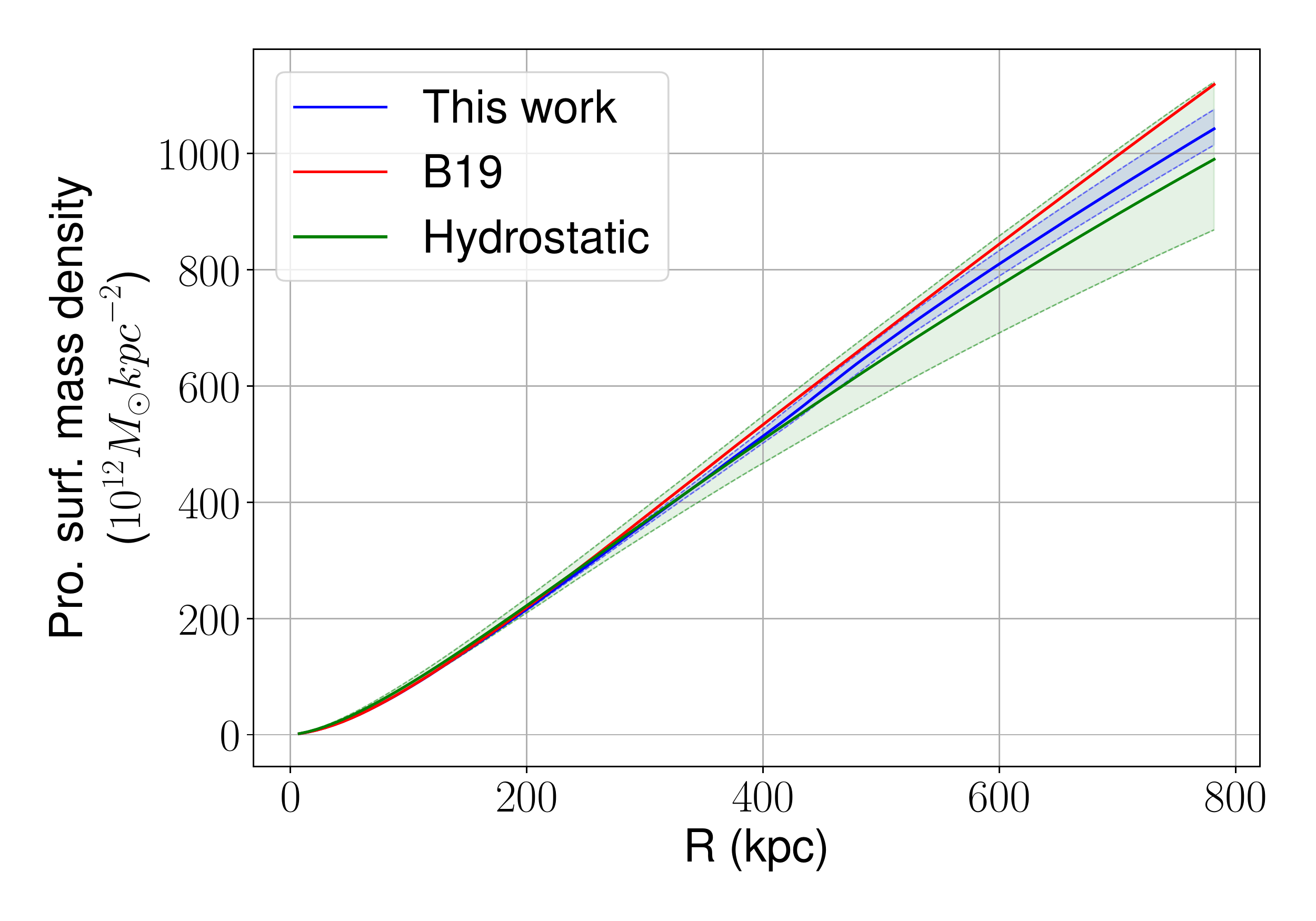}
    \end{minipage}
    \begin{minipage}{0.49\linewidth}
    \centering
    \includegraphics[width=\linewidth]{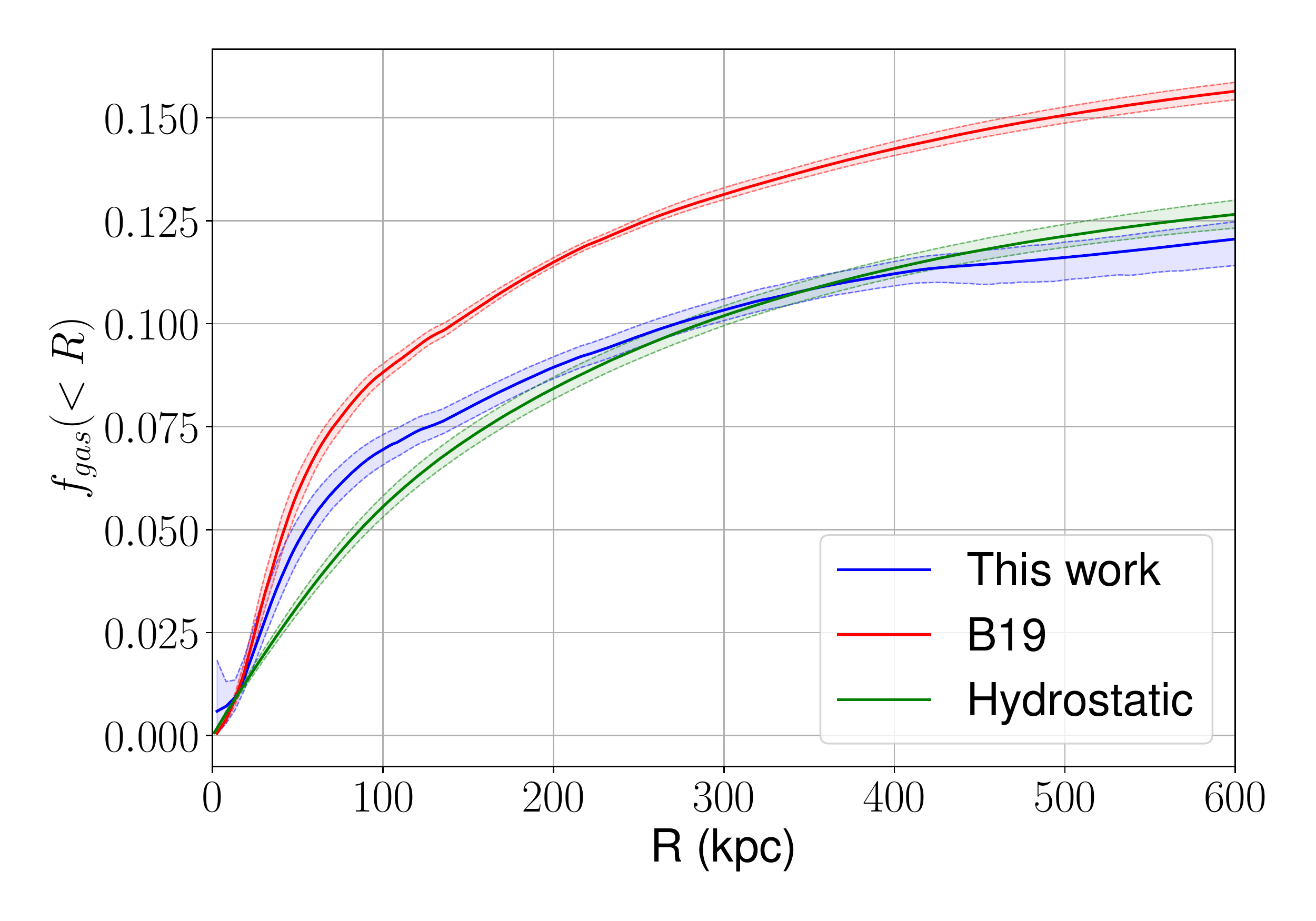}
    \end{minipage}
    \caption{Integrated profiles of the 2D total mass density (\textit{left panel}), and gas fraction (\textit{right panel}) integrated in spheres of increasing radii. For both panels, plain lines and shaded regions represent the median and the $"3\sigma"\, \rm CI$ of the profiles distribution among the $4000$ models of the original posterior distributions.}
    \label{fig:Profile_hydromass}
\end{figure*}

Figure~\ref{fig:Profile_hydromass} presents the integrated projected mass profiles and the deprojected gas fraction of the hydrostatic reconstruction, B19 and our models. Plain lines show the median profiles among $4000$ models from the posterior distribution when the shaded areas highlight $"3\sigma"\, \rm CI$ of the distribution. As one can see, the hydrostatic model is in quite good agreement with both lensing models, in particular beyond $\sim60$ kpc, where they are both included in the $"3\sigma"\,\rm CI$ of the hydrostatic mass. Below this radius, lensing mass profiles are almost the same and, on average less than $15$ per cent away from each other. The closest multiple image from the profile centre (i.e. BCG) is at a radius of $75$ kpc, thus, the lensing reconstructions are more determined by modelling assumptions than the actual constraints in the central area. It is then hard to assess if this discrepancy is meaningful. If we only consider the profiles beyond this radius, the hydrostatic mass bias is then inferior to $1$ per cent, which is expected for a slightly perturbed cluster \citep{Ansarifard2020}. This shows that AS1063 is close to being in hydrostatic equilibrium, which would not be the case if it were undergoing a major merger event. It is consistent with the scenario of a minor merger presented previously, and already discussed by \citet{Sartoris2020}. Regarding the gas fraction, our model provides results similar to the hydrostatic equilibrium beyond $\sim220$ kpc as both $"3\sigma"\,\rm CI$ overlapped. Overall, the average of the absolute differences between both is of $10$ per cent below $600$ kpc, mainly dominated by the area of disagreement close to the BCG, which shows a gas fraction $17$ per cent lower than ours. We note that this X-ray fit provides a different result on the gas mass than the multi-scale deprojection presented in Sect.\ref{sect:res-mass}. In that case, our model presents a slightly lower deprojected gas mass than the NFW fit. Hence, both the total mass discrepancy and the previous one lead to this disagreement on the gas fraction, even if both of them are in quite close agreement. There is a significant difference with the B19 model, which is a direct consequence of their gas mass estimation which is higher than the two others, as seen in Fig.~\ref{fig:gas_prof}.

\begin{figure*}
   
    \includegraphics[width=\linewidth]{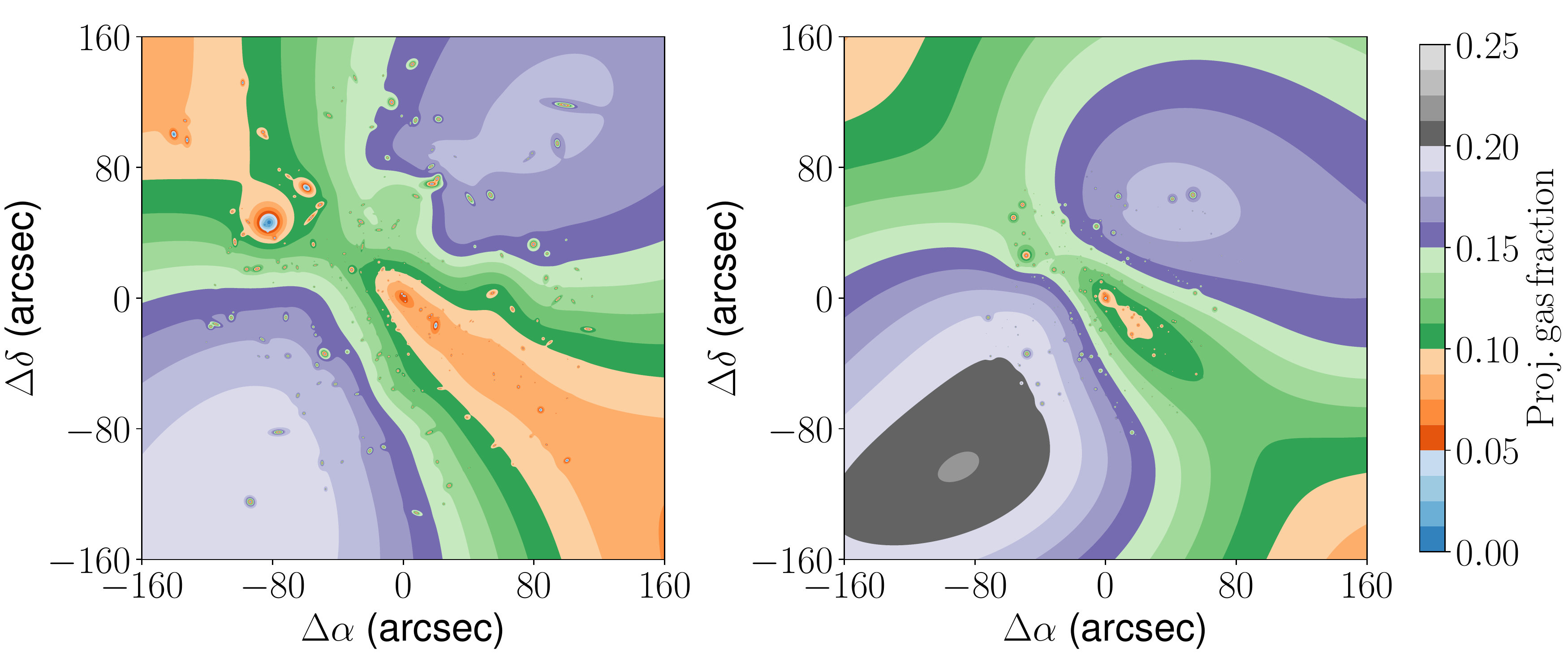}
    
    \caption{Maps of the gas fraction from our best-fit model and B19 one in the left and right panel, respectively.}
    \label{fig:map_gas_fraction}
\end{figure*}

We present the 2D distribution of the projected gas fraction in Fig.\ref{fig:map_gas_fraction} for B19 and our best-fit model. If, globally, the observed gas fraction is similar with a median of the relative difference of $11$ per cent between both maps, there are some differences in the morphology. Indeed, the ellipticity discrepancies between the DM and the gas can be clearly seen in our model when it is less evident in B19. In particular, the lobes of a lower or higher gas fraction have different shapes between both models on areas that are mainly outside the constrained one. It can be explained by the differences in the modelling and the assumptions on the parameters, such as dPIEs' cut radius, $r_{cut}$. In addition, more complexity can be observed in our 2D distribution that has been added by the B-spline surfaces in the area mentioned in Sect.\ref{sect:res-mass}. We note that the 2D analysis of the gas fraction is possible with our modelling method without the need for any extra-tools, thus allowing simple access to the position and shape of the different cluster components.

In recent work, \cite{Cerini2022}, derived and demonstrated the utility of power spectrum analysis of fluctuations in the lensing maps and X-ray surface brightness maps to assess the dynamical state of the gas in clusters. They were able to demonstrate that departure from hydrostatic equilibrium and the existence of small-scale turbulent processes, like shocks, gas sloshing etc, could be extracted from combined analysis using these new metrics. Similarly, our work also amply demonstrates the gain from the combination of multi-wavelength data to characterise the dynamical state of clusters. What is powerful about such methods is the leveraging a set of observables that are independent of the dynamical state (lensing data) along with other observables (X-ray data) to derive new insights into the dynamics and assembly of galaxy clusters, specifically the role of on-going mergers and their signatures. 

\section{Summary and conclusion}
\label{sect:conclusion}

In this paper, we present a method to reconstruct the mass of galaxy clusters from the combination of strong lensing, X-ray emission and cluster member morphology and kinematics data. We incorporate all these observables into a single process by performing a joint optimisation of multiple image positions and X-ray photon counts. Measurements on galaxies are added in the form of physically motivated priors on the three galaxy scaling relations. This new method has been implemented in the software \textsc{Lenstool}, and test it on the well-observed cluster AS1063. Both software and mass models are released publicly to the community. 

We find that our method is able to recover the strong lensing constraints like other available state-of-the-art methods. Furthermore, our model of the gas captures the asymmetric shape of AS1063 X-ray emission efficiently. Thus, our model is able to increase the complexity of the reconstruction by deviating from the forced elliptical symmetry of the main mass haloes with the addition of the gas distribution. The presented treatment of cluster members takes advantage of the fundamental plane of elliptical galaxies, which allows the addition of a scatter in the $\sigma_0$-$\text{Mass}$ plane in comparison to previous treatments. Thanks to our combined calibration process, we propose a coherent modelling of the cluster member mass through the fundamental plane and the Faber \& Jackson scaling relation.

We use a mock cluster based on AS1063 to assess the robustness of our method to the assumed uncertainties on the constraints as well as the benefit of the enhancement presented in this work. We find that we recover the gas mapping up to less than $3$ per cent of error, as well as the intrinsic error added to the gas emission models which is inside the $"1\sigma"\,\rm CI$. In the case of AS1063, the non-inclusion of the gas-only leads to a maximum of a few per cent error in the total mass. In particular, it is not possible to capture the deviation from the main elliptical symmetry with a solely parametric treatment. We expect that this might be more significant for perturbed clusters where the gas distribution is less well correlated with the DM content. We compare the ability of our current cluster member model to previous work and report that using the combination of the fundamental plane and Faber \& Jackson scaling relations with a calibration from actual measurements allows us to reduce the bias on the estimation of dPIE's $\sigma_0$ and $r_{cut}$. However, the improvement on the posterior distribution of the cluster member masses propagated from these two  quantities is less important. 

We assess the consistency of our cluster member modelling hypotheses with available measurements to probe the connection between mass and light as well as the assumed scalings of their light and mass profiles. We find that with our simplification, we provide a fair but not robust, estimation of both as they present similar behaviour but with different scaling slopes. Future modelling improvements could aim at finding new schemes that would avoid our assumptions while still leveraging the empirically found correlations in the fundamental plane.

Thanks to our combination of X-ray and lensing, we obtain a much more exhaustive view of cluster lenses in comparison to a disjoint analysis. In particular, from the residual of the X-ray emission fit to the gas thermodynamics combined with an accurate representation of the mass distribution, we provide new evidence for a recent merging event in AS1063, rejecting the major merger possibility. Indeed, we find proof for gas sloshing thanks to the tools we put in place to map the intra-cluster gas temperature. In addition, as the cluster components are disentangled, we can observe the differences between the baryonic and non-baryonic components by mapping in 2D the gas fraction compared to earlier work \citet{Bonamigo2018}, as our total mass reconstruction is more flexible thanks to the B-spline surface. We find that AS1063 is nearly at the hydrostatic equilibrium, in agreement with the analysis by \citet{Sartoris2020}.

The new method presented here includes baryonic components explicitly and self-consistently and demonstrates its capabilities. The next step would be to extend this work to a model of a sample of cluster lenses, such as the complete set of BUFFALO clusters or other newly observed cluster lenses with the recently launched \textit{James Webb Space Telescope} (\textit{JWST}). These will be well suited as some of the most extreme cluster lenses in these samples consist of highly perturbed clusters that have undergone or are actively undergoing major merger events. Thus, the inclusion of the modelling of the gas will improve the total mass distribution and help assess the dynamical state of the cluster. 
Performing the analysis we present here on a larger sample of clusters, we would be able to measure offsets between the different cluster elements and provide constraints on a possible self-interaction of the DM. Thanks to the more exhaustive picture of the dynamical state of clusters, the comparison with results from simulated merging events would be easier and more complete.

Such a sample will offer a homogeneous window into the high-redshift universe that will strongly benefit from the enhanced consistency achieved using the multi-wavelength observations of clusters. Hence, reducing biases on lensed galaxy measurements where a local bad estimation of the total mass can have here more impact than on analyses focused on cluster physics. A natural improvement to our method would be the inclusion of the weak lensing constraints to reduce the induced bias on the core and extend the scale on which the mass distribution is reconstructed, for example, a merging with \textit{Hybrid}-\textsc{Lenstool} \citep{Niemiec2020} method. The codes to use weak lensing measurements are already available, but the structures at the outskirts have to be identified beforehand with a free-form method. Modelling of the gas would benefit from taking inspiration from these methods with a multi-scale reconstruction, losing some properties valuable to analyse the cluster state but providing a highly accurate mapping of the associated mass in exchange.

As demonstrated by \citet{Mahler2022}, in the modelling of the \textit{JWST} data of the first lensing cluster to be observed, SMACS J0723.3-7327, the flexibility to include an additional mass component that corresponds to the baryons locked into the ICL is likely to prove invaluable. Such addition could be done with a similar methodological extension as we presented here for the X-ray data in a self-consistent way. The prospects for comprehensive modelling of the multiple mass components of a cluster, robustly, to include all the baryonic contributors: stellar component of cluster members, X-ray gas and the ICL, in addition to the dominant DM component, is becoming feasible.

Finally, we aim to provide one of the first steps of a new cluster mass modelling standard, which is based on a self-consistent and simultaneous inclusion of multi-wavelength data and multiple observables. This new methodology, we believe, provides the appropriate enhancement in the sophistication of cluster mass modelling aligned with the improvement in the availability of observational data as we transit from the \textit{HST} to the \textit{JWST} era of cluster and high-redshift studies.

\section*{Acknowledgements}
The authors are grateful to the referee for carefully reading the manuscript and for valuable suggestions and comments, which helped improve and clarify it. This work used data and catalogue products from HFF-DeepSpace, funded by the National Science Foundation and Space Telescope Science Institute (operated by the Association of Universities for Research in Astronomy, Inc., under NASA contract NAS5-26555). This research has been made possible by observations made with the NASA/ESA Hubble Space Telescope obtained from the Space Telescope Science Institute. These observations are associated with program GO-15117. We also made use of observations collected at the European Southern Observatory under ESO programme(s) 60.A-9345(A), 095.A-0653(A) and 186.A-0798. M.L. acknowledges the Centre National de la Recherche Scientifique (CNRS) and the Centre National des Etudes Spatiale (CNES) for their support. MM acknowledges the project PCI2021-122072-2B, financed by MICIN/AEI/10.13039/501100011033, and the European Union “NextGenerationEU”/RTRP. DJL is supported by STFC grant ST/T000244/1. DJL and MJ are supported by the United Kingdom Research and Innovation (UKRI) Future Leaders Fellowship `Using Cosmic Beasts to uncover the Nature of Dark Matter' (grant number MR/S017216/1).

\section*{Data Availability}
The \textsc{Lenstool} output files describing the mass model presented in this article are available in GitHub at \href{https://github.com/njzifjoiez/AS1063-model-Beauchesne-2023}{https://github.com/njzifjoiez/AS1063-model-Beauchesne-2023}. These files can be interpreted without the need for the latest release of \textsc{Lenstool}. In particular, there is no need for the extensions presented in this paper, such as the fundamental plane relation or the X-ray modelling. We also released publicly the redshift catalogue produced with the new analysis of the MUSE datacube at the following link: \href{https://cral-perso.univ-lyon1.fr/labo/perso/johan.richard/MUSE_data_release/}{https://cral-perso.univ-lyon1.fr/labo/perso/johan.richard/MUSE\_data\_release/}



\bibliographystyle{mnras}
\bibliography{example} 




\appendix
\section{B-spline modelling of the X-ray background}
\label{app:X-ray-background-model}
The approach outlined here is empirical and aimed at modelling the number of counts directly (i.e. it implies a unitary instrument response) with B-spline functions of degree $3$. We used it on \textit{Chandra} observation, but it is a general method that can be applied to other X-ray observatories as long as they provide the same kind of data. The whole python code that we developed is publicly available in the \textsc{Lenstool} repository and takes benefit of the \textsc{Sherpa} fitting environment as well as the B-spline routines \textit{splev} and \textit{splrep} \citep{Dierckx1993} through their \textsc{Scipy} wrappers \citep{2020SciPy-NMeth}.

\begin{figure*}
    \begin{minipage}{.24\linewidth}
    \centering
    \includegraphics[width=\linewidth]{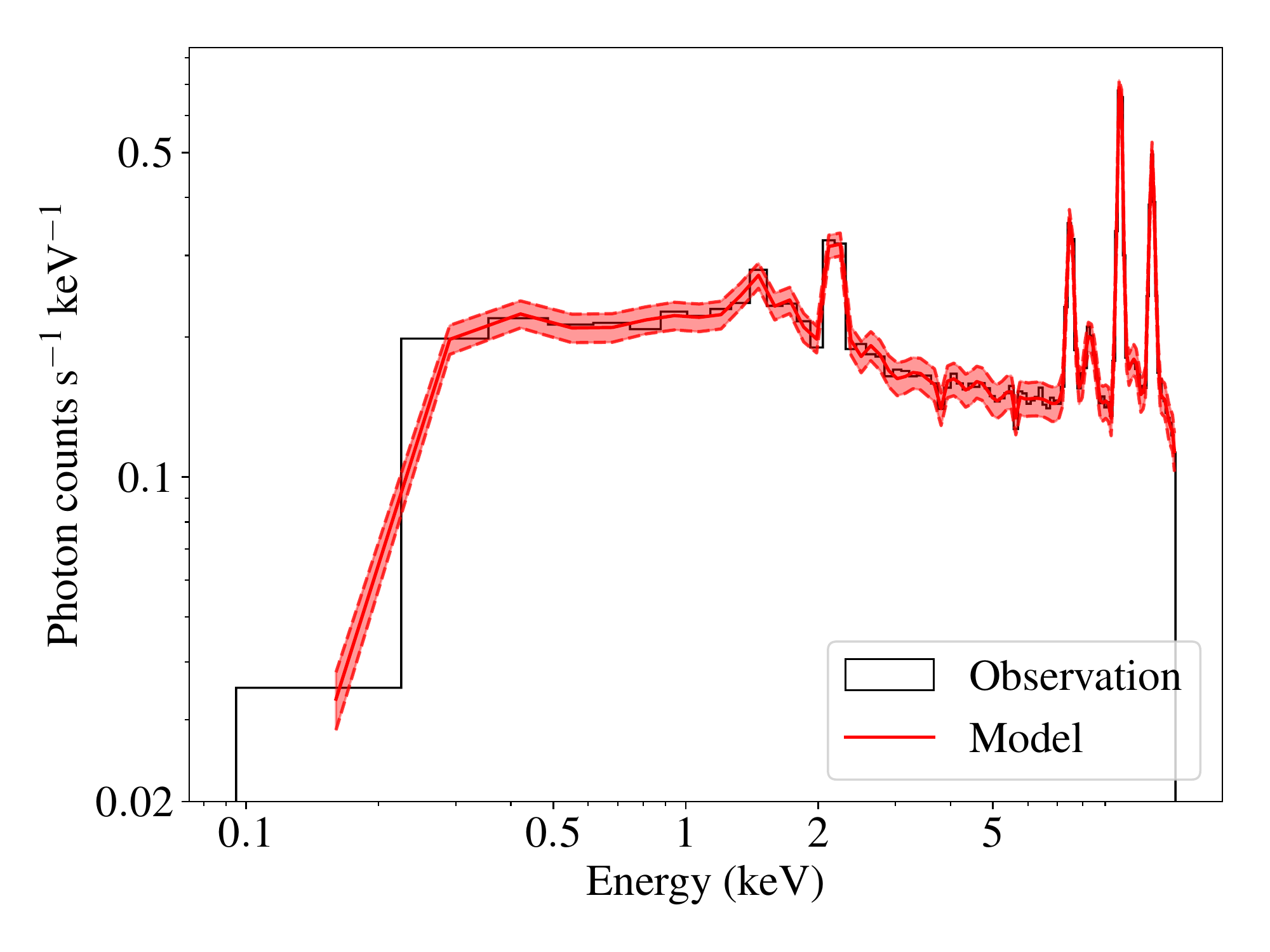}
    \end{minipage}
    \begin{minipage}{.24\linewidth}
    \centering
    \includegraphics[width=\linewidth]{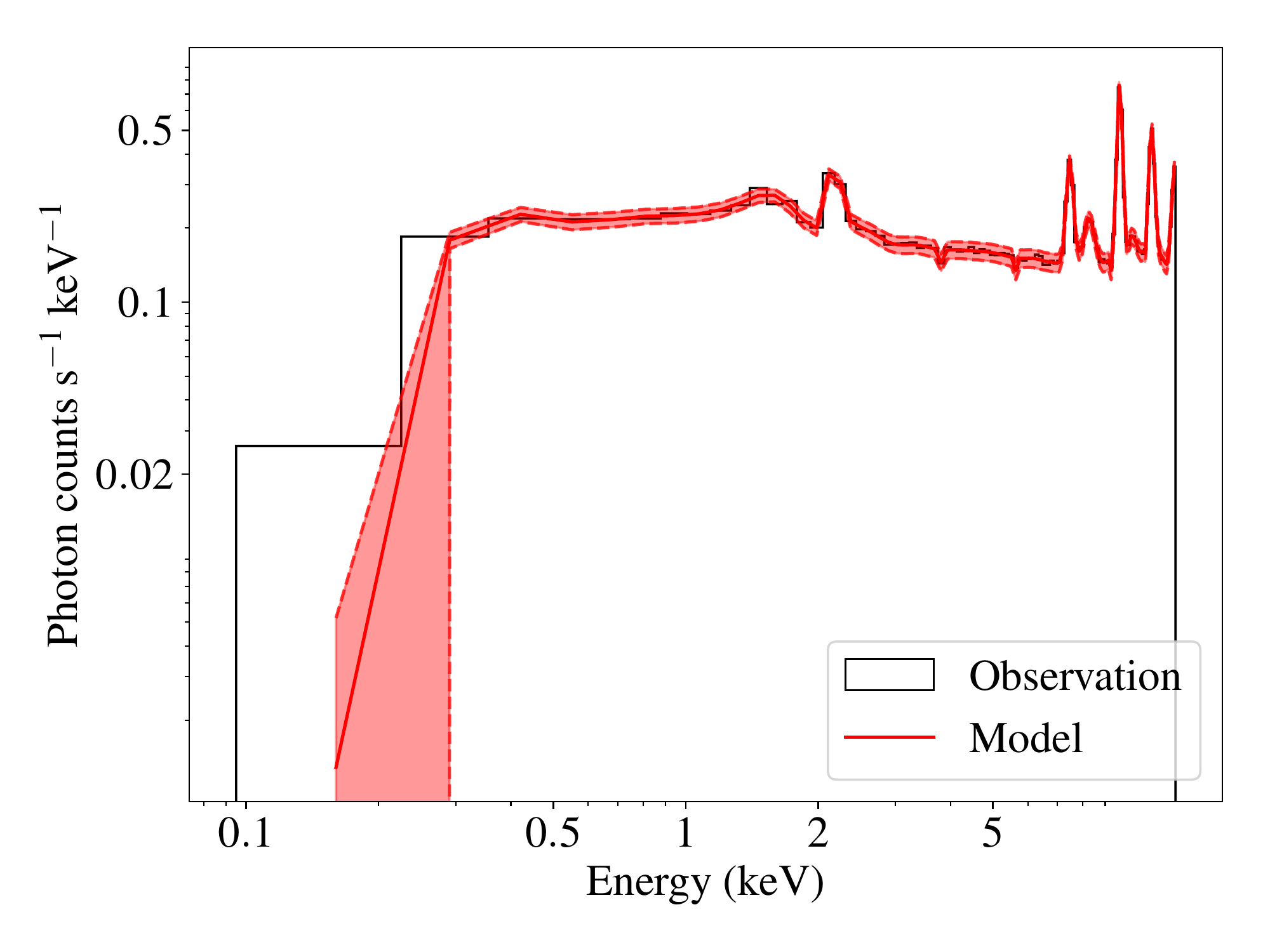}
    \end{minipage}
    \begin{minipage}{.24\linewidth}
    \centering
    \includegraphics[width=\linewidth]{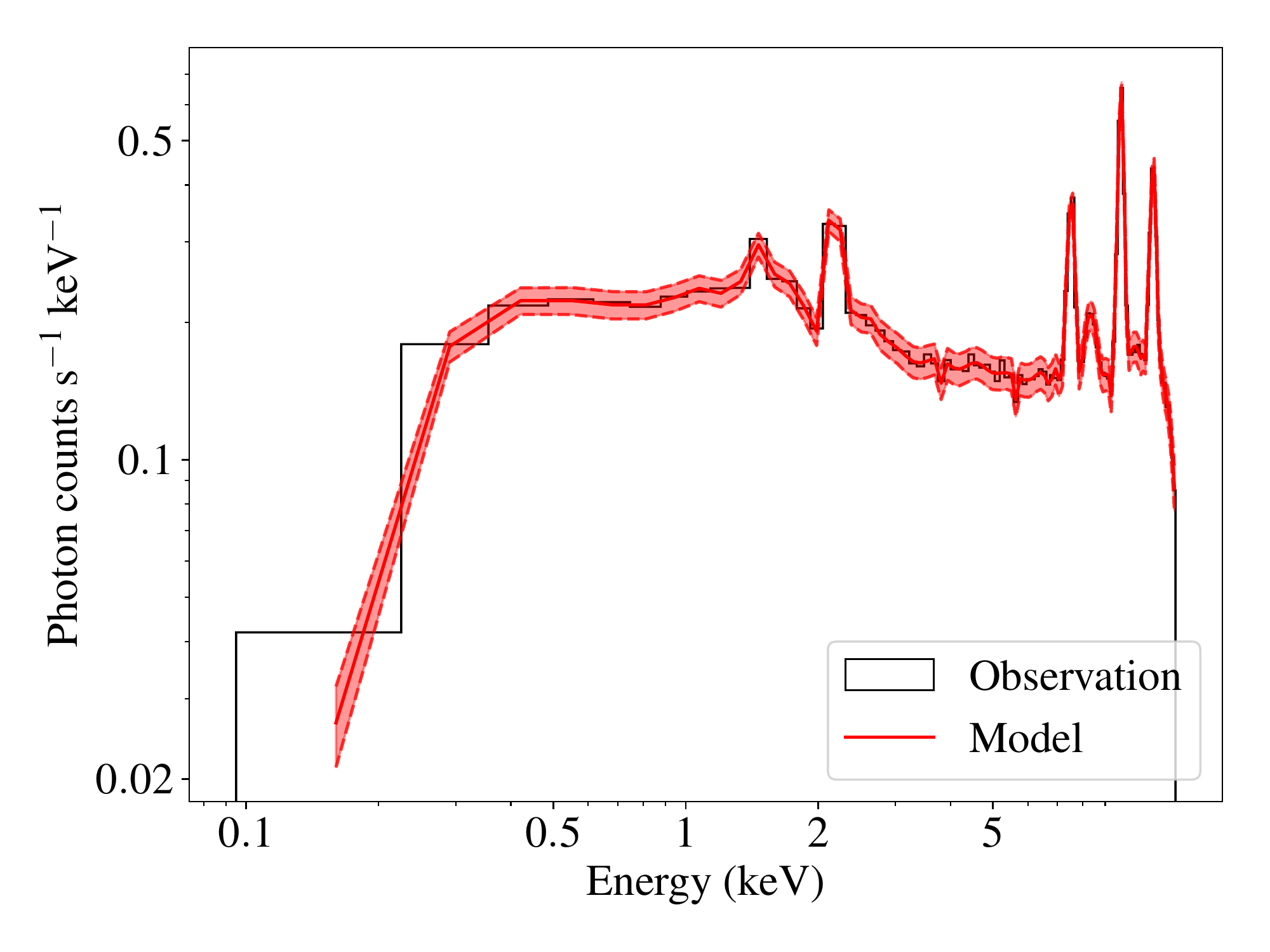}
    \end{minipage}
    \begin{minipage}{.24\linewidth}
    \centering
    \includegraphics[width=\linewidth]{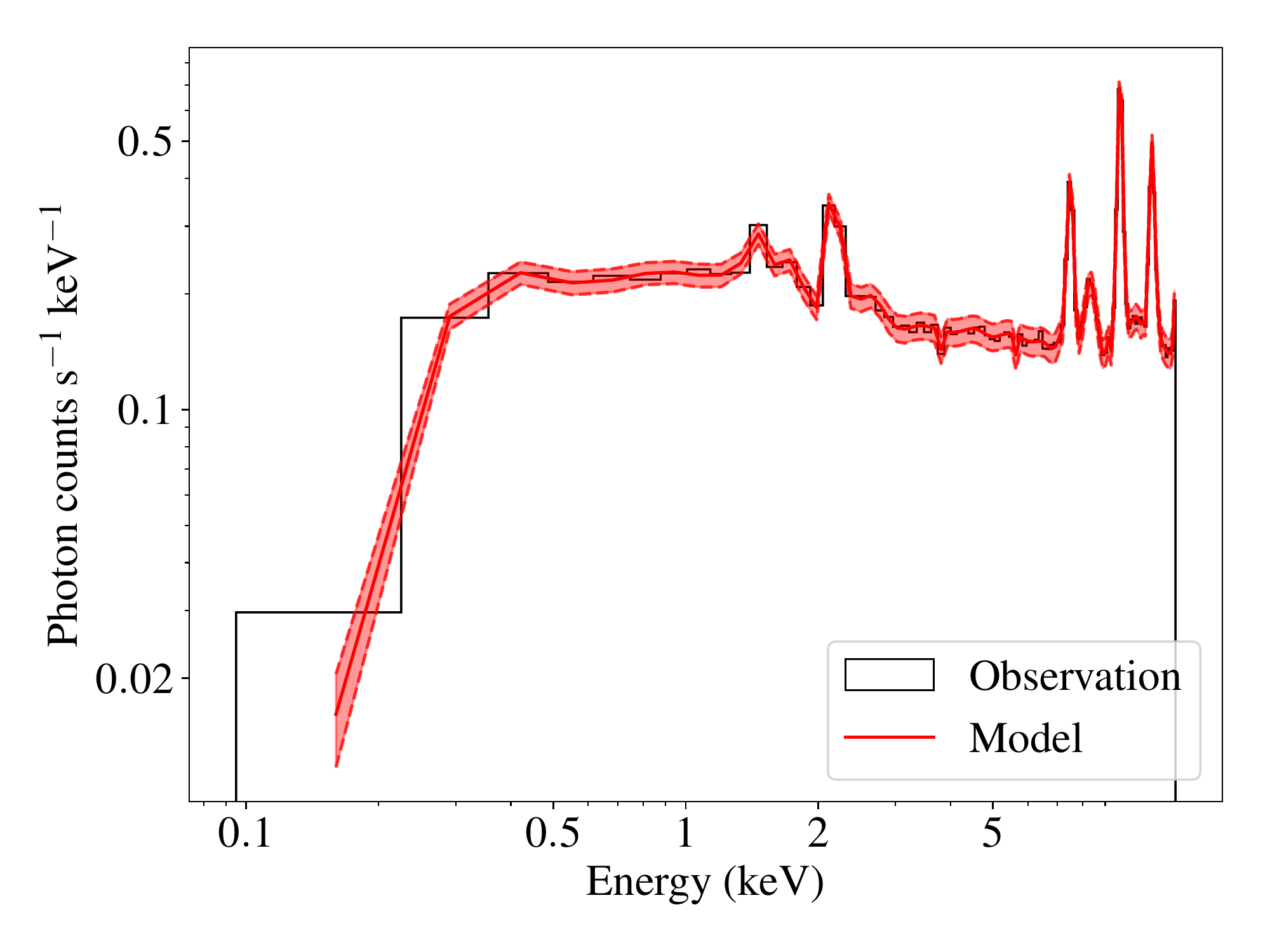}
    \end{minipage}
    \caption{Spectra of blank-sky observations for each of the four CCD in the $\left[0,13\right]$ keV energy band associated with observations $18611$. Black histograms represent the photon counts, and plain red lines highlight the median of the background model. Shaded red regions represent the uncertainty on the parameter estimations}
    \label{fig:fit-per-ccd}
\end{figure*}

We start our modelling by fitting each CCD spectra on the $\left[0,13\right]$ keV energy band on each observation. This range has been chosen to ensure that the background model will have the right value on the edge of the $\left[0.5,12\right]$ keV one. Each fit is made in two steps; the first is an optimisation with a $\chi^2$ statistic with the \citet{Gehrels1986} approximation implemented in \textsc{Sherpa} through a Levenberg-Marquardt method. During this optimisation, only the B-spline coefficients are not fixed, while the knots positions and their numbers are obtained by the \textit{splrep} routine. The coefficient values given by the preceding routine are used to define starting points for the $\chi^2$ minimisation. Finally, the second step is a simplex optimisation from the results of the last one with the \textsc{Xspec} \textit{Cstat}. Initially, we wanted only to include this last step as a Poisson likelihood is more adapted to the considered datasets, but the optimisation was not converging to satisfying models in some cases. The resulting spectra models for the observation of $18611$ on each CCD are presented in Fig.~\ref{fig:fit-per-ccd}.

\begin{figure*}
    \begin{minipage}{.48\linewidth}
    \centering
    \includegraphics[width=\linewidth]{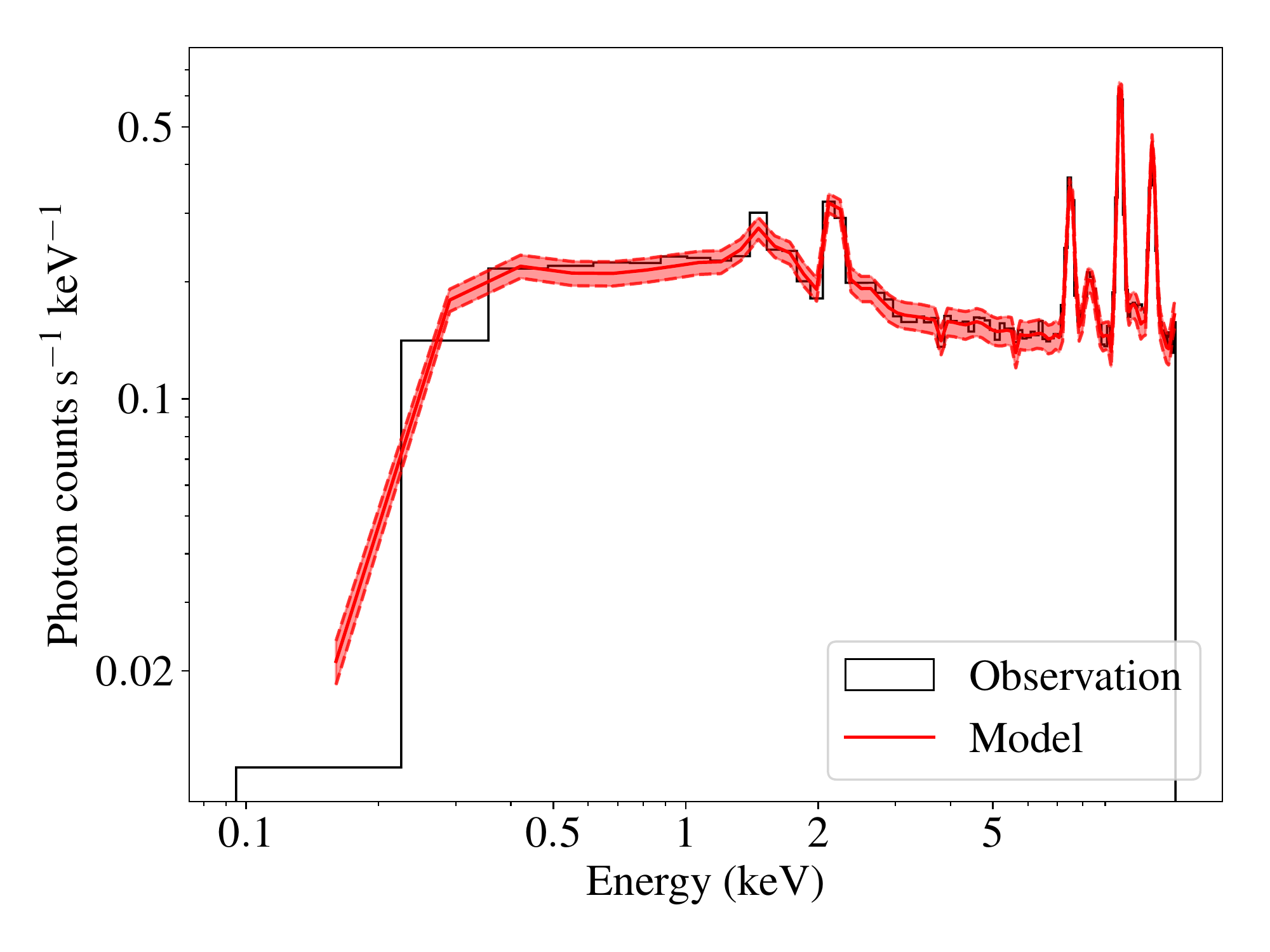}
    \end{minipage}
    \begin{minipage}{.48\linewidth}
    \centering
    \includegraphics[width=\linewidth]{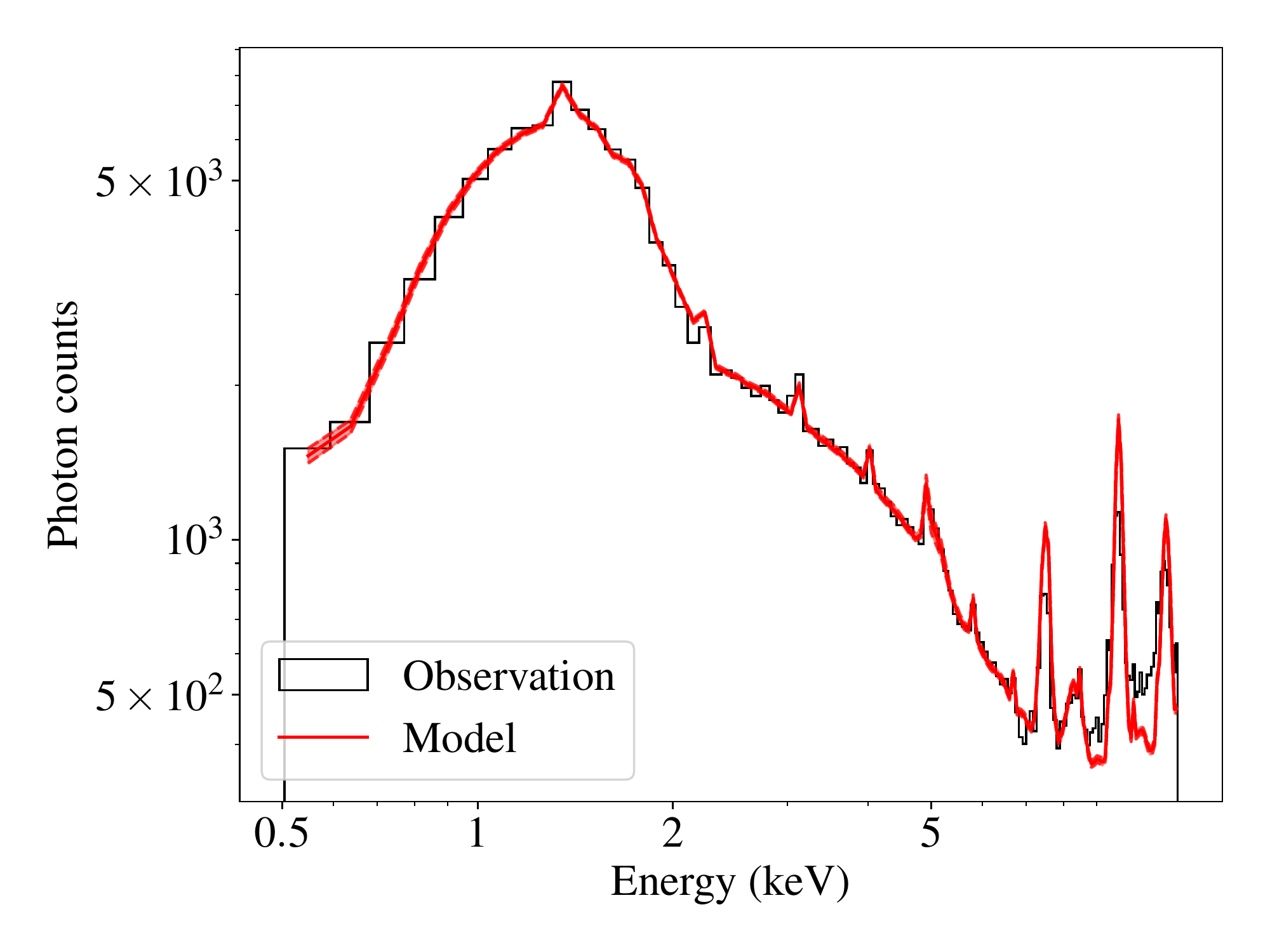}
    \end{minipage}
    \caption{\textit{Left panel:} Black histogram presents the spectrum of the blank-sky background associated with the observation $18611$ on the core of AS1063. The red line and shaded region represent the background model and the uncertainties in the parameter estimation. The extraction region is a square of $8.3$ arcmin around the cluster centre and is overlapped on different CCDs. \textit{Right panel:} Spectrum of AS1063 in the previously defined extraction regions combining the three observations considered for that work. The colour scheme is the same as before except for the model where red plain lines and shaded region represent the median and the $"1\sigma"\, \rm CI$ of the posterior distribution of the spectrum model.}
    \label{fig:fit-whole-cluter}
\end{figure*}

To adapt the previous spectra model to the region where is extracted the source spectrum, we extract the one from the same area in the \textit{blank-sky} background. We then fit it with a linear combination of the CCD model that contributes using a simplex method with the \textit{Cstat} statistic. By allowing the coefficient of the linear combination to deviate from the ratio of the overlapping area of the CCD on the region of interest on the total region area, we aimed to take into account spatial differences on a unique CCD. Fig.~\ref{fig:fit-whole-cluter} shows the model obtained for the whole cluster fit of Sect.~\ref{sect:X-ray-data-analysis} on the observation $18611$ in the left panel. Before using this background model, we re-scale the coefficient of the linear combination with the ratio of counts on the $\left[9.5,12\right]$ keV range between the observation and the \textit{blank-sky} per CCD. In this energy band, it mostly accounts for the particle background as the effective area of \textit{Chandra} is almost null. We make this re-scaling to take into account the differences per CCD due to the discrepancy in exposure times. Finally, the spectrum model is used in the nested sampling method with a normalisation per observation.

The right panel of Fig.~\ref{fig:fit-whole-cluter} presents the result of the fitting procedure on the whole cluster. It highlights one of the limitations of this approach which is the fluctuation of the amplitude of the fluorescent emission lines among a CCD for high energies. Hence, to improve this modelling, we could allow them to vary, and a second enhancement would be the disentanglement of the instrumental and sky background. However, as we are mainly interested in the core of clusters where we have a high SN and the energy range of interest for the source is $\left[0.5,7\right]$ keV, both of these issues should not affect our results.

\section{Bias on the measurement of the LOSVD}
\label{app:LOSVD-bias}

\begin{table}
\centering
\begin{tabular}{cc}
    \hline
    Parameters & Distribution\\
    \hline
    $V$& $\mathcal{U}(-250,250)$\\
    $\sigma_{\rm True}$& $\mathcal{U}(10,400)$\\
    $h_1$& $\mathcal{U}(-0.15,0.15)$\\
    $h_2$& $\mathcal{U}(-0.15,0.15)$\\
    \hline
\end{tabular}
\caption{Prior distributions on the LOSVD parameters used to generate a sample of simulated spectra from AS1063 cluster members.}
\label{Tab:dist-sampling-LOSVD}
\end{table}

Following \citet{Bergamini2019}, we created mock spectra to test and correct our estimation of the LOSVD measurements with \textsc{pPXF} and \textsc{pyMultiNest} libraries. To obtain spectra that will represent the galaxy population of AS1063, we use the linear combination of template spectra with weights obtained from the best-fit solution for each galaxy considered. We then use \textsc{pPXF} to convolve them with different LOSVD parameters. In particular, the velocity $V$, the velocity dispersion $\sigma$ and the two Hermites moments $h_1$ and $h_2$ are sampled randomly from the distribution shown in Table~\ref{Tab:dist-sampling-LOSVD} and the Legendre polynomial coefficients are taken from the best-fit solution. For each galaxy (i.e. $107$ elements), we sampled $100$ mock spectra that we transformed to MUSE-like measurements by degrading the spectral resolution and FWHM to match the real ones. We finally add noise based on the real spectra variance that we adjusted in amplitude to obtain a signal-to-noise sampled with a uniform probability between $3$ and $50$. In that way, we have mock errors taking into account specific variations, such as the subtraction of skylines in the real extracted spectra.

\begin{figure*}
    \begin{minipage}{.33\linewidth}
    \centering
    \includegraphics[width=\linewidth]{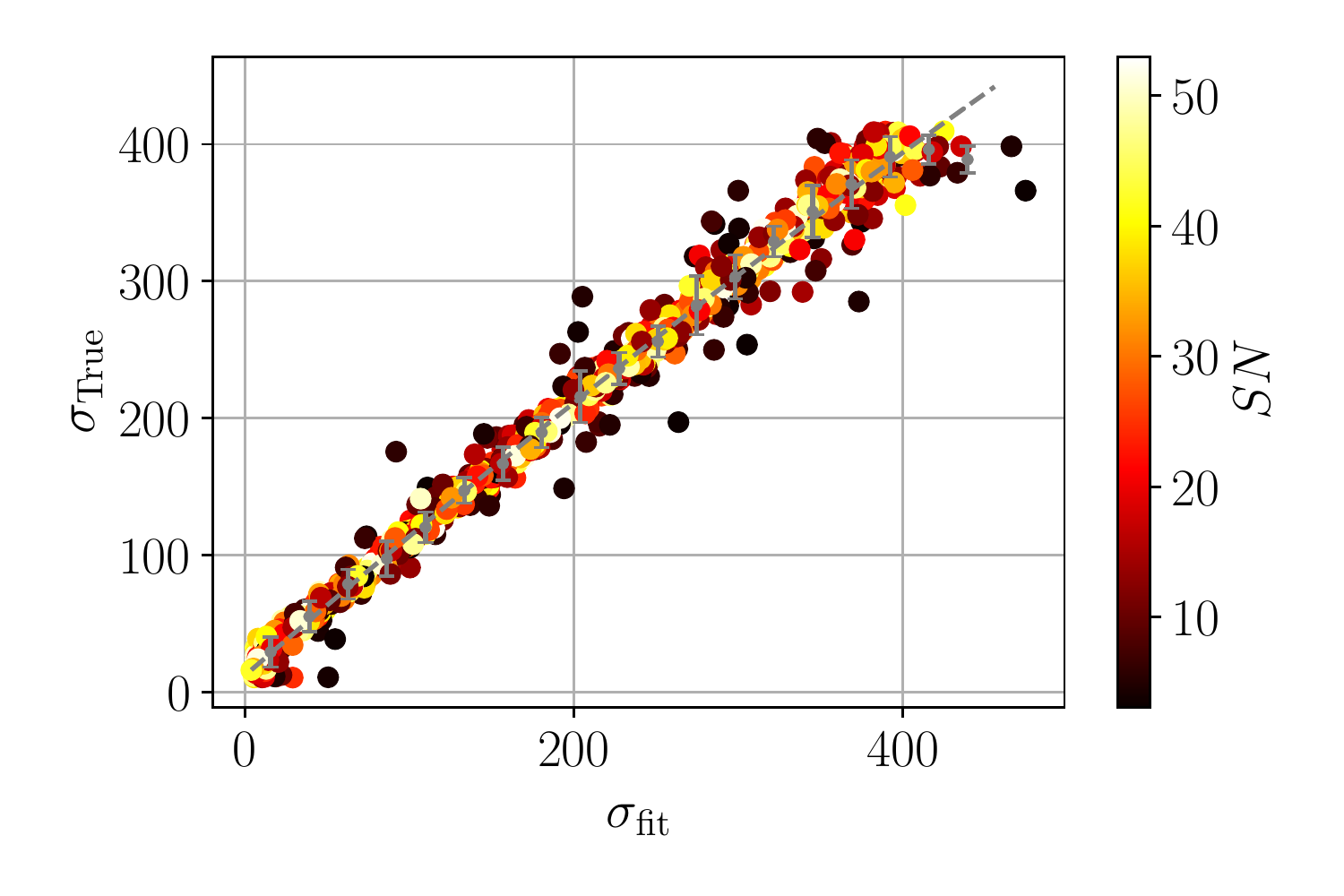}
    \end{minipage}
    \begin{minipage}{.33\linewidth}
    \centering
    \includegraphics[width=\linewidth]{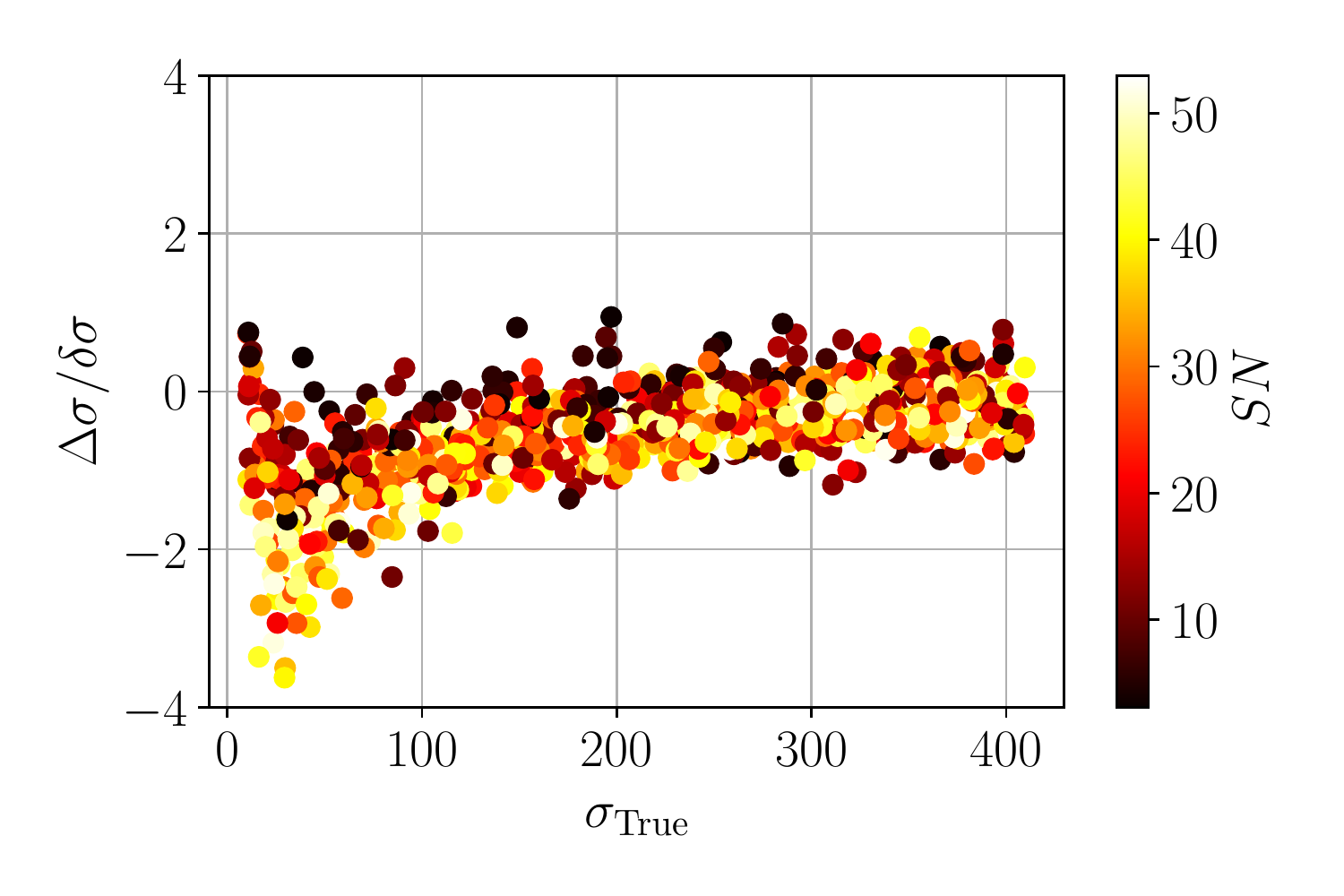}
    \end{minipage}
    \begin{minipage}{.33\linewidth}
    \centering
    \includegraphics[width=\linewidth]{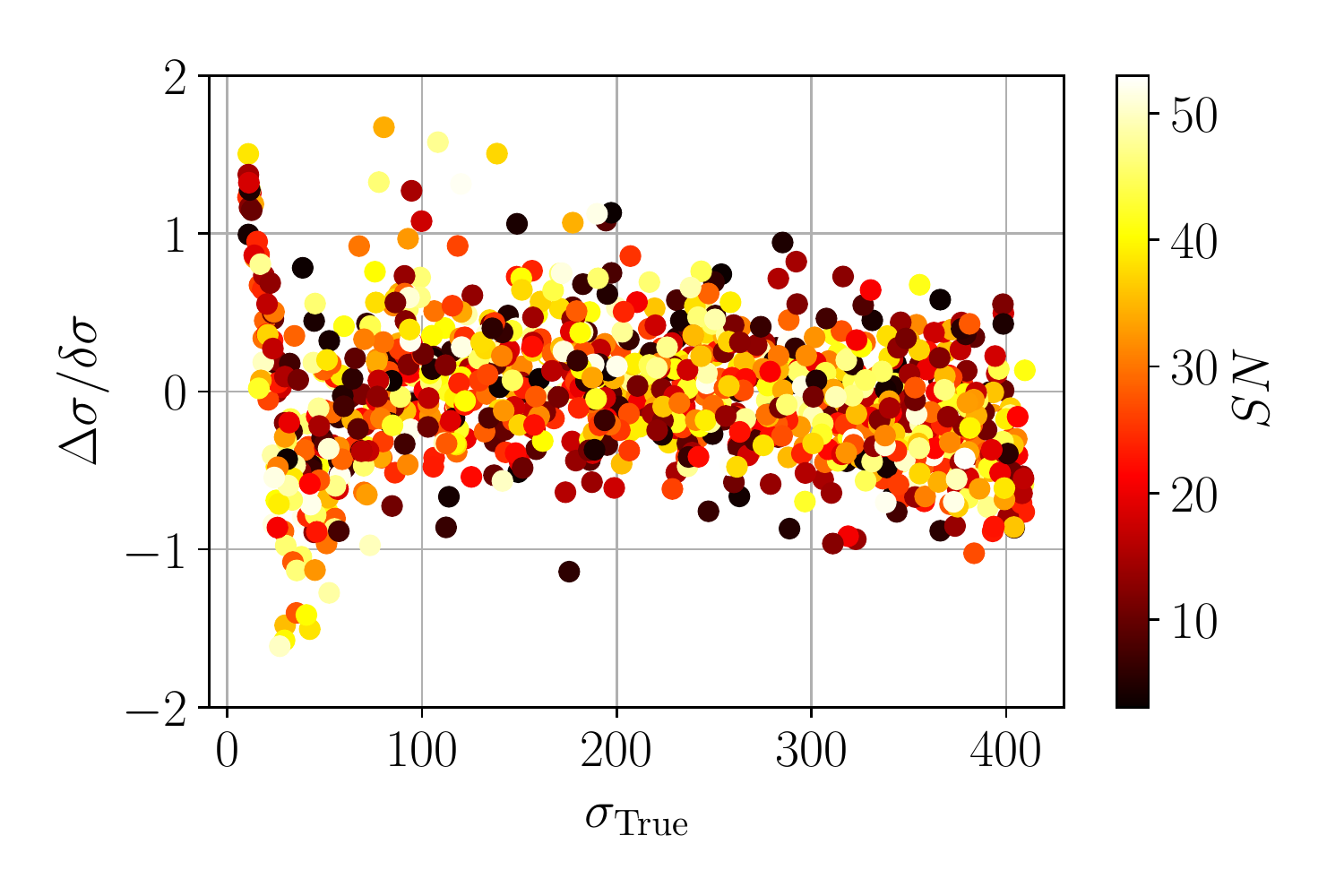}
    \end{minipage}
    \caption{\textit{Left Panel:} Plot of the true velocity dispersion, $\sigma_{\rm True}$, and the best-fit one, $\sigma_{\rm fit}$. \textit{Middle Panel:} Plot of the ratio between the $\Delta\sigma$ (i.e. $\sigma_{\rm True}-\sigma_{\rm fit}$) and the statistical uncertainty measured on the posterior distribution in function of $\sigma_{\rm True}$. \textit{Right Panel:} Same as middle panel but $\sigma_{\rm fit}$ has been corrected with relations given in Eq.~\ref{Eq:sigma_poly}.}
    \label{fig:correction_err_sigma}
\end{figure*}

The mock spectra have been fitted with the exact same methodology as outlined in Sect.~\ref{sect:MUSE-data-analysis} and $\sigma_{\rm fit}$ which is the median of obtained posterior distribution on $\sigma$ is plotted against the true value $\sigma_{\rm True}$ in Fig.\ref{fig:correction_err_sigma}. Our fit globally follows the input LOSVD parameters with a small bias that we estimate by fitting a polynomial of degree $2$ on binned values of $\sigma_{\rm fit}$. The binned data and the fitted polynomial are shown by the cyan scatter and lines in the previous graph and the polynomial expression is the following:
\begin{equation}
    \sigma_{\rm fit,Corr}=-2.029\times10^{-4}\sigma_{\rm fit}^2+1.035\sigma_{\rm fit}+12.22
    \label{Eq:sigma_poly}
\end{equation}
Thus, $\sigma_{\rm fit}$ is suffering a small underestimation of around $3$ per cent that is increased at low $\sigma_{\rm True}$ value (i.e. $\sigma_{\rm True}<100$ $\rm km/s$). It is well visible in the central plot in Fig.\ref{fig:correction_err_sigma} that shows the ratio between $\Delta\sigma=\sigma_{\rm fit}-\sigma_{\rm True}$ and the statistical errors from the posterior distribution $\delta\sigma$ in function of $\sigma_{\rm True}$. However, after applying Eq.~\ref{Eq:sigma_poly} on all the velocity dispersion posterior distribution, we obtain a $\sigma_{\rm fit,Corr}$ and $\delta\sigma_{\rm corr}$ that provides a better estimation of $\sigma_{\rm True}$ as we can see on the plot on the right column of Fig.\ref{fig:correction_err_sigma}. Indeed, the ratios between the true and the estimation errors are mostly contained in the $[-1,1]$ intervals. 

In comparison, results from \citet{Bergamini2019} on mock spectra present a few per cent of overestimation on $\sigma_{\rm fit}$ as well as an underestimation of the real errors. Thus, if we also obtain similar biases on the best-fit solution of the velocity dispersion, our $\delta\sigma_{\rm corr}$ are presenting a correct view of the measurement errors, which supports our use of a nested sampling method instead of the default least square from the \textsc{pPXF} package.

\section{From $\sigma_e$ to $\sigma_0$}
\label{app:c_p_explanation}

$\sigma_0$ is usually considered as a representation of the central velocity dispersion of the associated dPIE potential, but its link with the actual measurement of the LOSVD parameters is not straightforward, and we need to make some assumptions about the mass and light distribution of the considered galaxies. Following \citet{Bergamini2019,Newman2013}, the LOS velocity dispersion $\sigma_{\rm LOS}$ is given by:
\begin{equation}
    \sigma^2_{\rm LOS}(R)=\frac{2G}{I(R)}\int^{\infty}_{R} \nu(r) M_{\rm 3D}(r) \mathcal{F}(r)\, r^{2\beta_{\rm aniso}-2}{\rm d}r
\end{equation}
Where $I$ and $\nu$ are the surface density and density of the luminosity distribution, respectively. $M_{\rm 3D}$ is the total mass enclosed in a sphere of radius $r$. The two lasting terms, $\mathcal{F}$ and $\beta_{\rm aniso}$ are linked to the orbit of stars considered that in the case of isotropic orbits are expressed as follows \citep{Cappellari2008}:
\begin{equation}
    \mathcal{F}(r)=\sqrt{r^2-R^2} \text{   and   } \beta_{\rm  aniso}=0
\end{equation}
The actual measurements $\sigma_{\rm ap}$ are $\sigma_{\rm LOS}$ values averaged in an circular aperture of radius $R'$ that is defined as follows:
\begin{equation}
   \label{Eq:sigma_formula_compact}
   \sigma^2_{\rm ap}(R')=\frac{2\pi}{L(R')}\int_0^{R'} \sigma^2_{\rm LOS}(R) I(R) R {\rm d}R
\end{equation}
With $L(R')$ the luminosity enclosed in a circle of radius $R'$. In the case of a dPIE potential, we have the following relations for the different terms involved in the $\sigma_{\rm ap}$ expression:
\begin{subequations}
\begin{align}
    \nu\left(r\right)&=\frac{\sigma_0^2(r_{\rm core}+r_{\rm cut})}{2 \Upsilon_{\odot} \pi Gr_{\rm core}^2 r_{\rm cut}}\frac{1}{\left(1+\left(\frac{r}{r_{\rm core}}\right)^2\right)\left(1+\left(\frac{r}{r_{\rm cut}}\right)^2\right)}\label{Eq:dPIE_formula_sigma-1}\\
    B&=\frac{2\sigma_0^2 r_{\rm cut}}{G\left(r_{\rm cut}-r_{\rm core}\right)}\label{Eq:dPIE_formula_sigma-2}\\
    I\left(r\right)&=\frac{B}{\Upsilon_{\odot}}\left(\frac{1}{\sqrt{r^2+r_{\rm core}^2}}-\frac{1}{\sqrt{r^2+r_{\rm cut}^2}}\right)\label{Eq:dPIE_formula_sigma-3}\\
    M_{\rm 3D}\left(r\right)&=B\left(r_{\rm cut} \arctan\left(\frac{r}{r_{\rm cut}}\right)-r_{\rm core} \arctan\left(\frac{r}{r_{\rm core}}\right)\right)\label{Eq:dPIE_formula_sigma-4}\\
    L\left(r\right)&=\frac{B}{\Upsilon_{\odot}}\left(\sqrt{r^2+r_{\rm core}^2}+r_{\rm cut}-\sqrt{r^2+r_{\rm cut}^2}-r_{\rm core}\right)\label{Eq:dPIE_formula_sigma-5}
\end{align}
\end{subequations}
Where we define the pre-factor $B$ for clarity. To properly match $\sigma_e$ to $\sigma_0$, we have to make assumptions that fit the ones taken in Sect.~\ref{sect:clus-memb-dist}. In particular, we assumed $\nu R_e=R_{e,\rm dPIE}$ (condition $1$). For the convenience of the fundamental plan implementation, we would like a relation of that type between the velocities dispersion $\sigma_e=c_p\sigma_0$ (condition $2$). Thus, we make the following assumption on the light distribution and its relations to the mass one:
\begin{itemize}
    \item The light distribution is well represented by a dPIE profile with parameters $\sigma_{0\rm,light}$, $r_{\rm core, light}$ and $r_{\rm cut, light}$.
    \item dPIEs associated with the light and the mass shared the same centre coordinates, ellipticity and position angle.
    \item Relations between parameters from the mass and light potentials:
    \begin{itemize}
        \item $\sigma_{0\rm,light}=\sigma_{0}$ (condition $2$)
        \item $r_{\rm core, light}=\frac{r_{\rm core}}{\nu}$ (condition $1$)
        \item $r_{\rm cut, light}=\frac{r_{\rm cut}}{\nu}$ (condition $1$)
    \end{itemize}
\end{itemize}
Where, $\sigma_{0}$, $r_{\rm core}$ and $r_{\rm cut}$ are the dPIE parameters associated with the mass distribution. We can now define the projection factor $c_p$ and gives a physical interpretation to $\nu$. Injecting Eq.~\ref{Eq:dPIE_formula_sigma-1}, \ref{Eq:dPIE_formula_sigma-3}, \ref{Eq:dPIE_formula_sigma-4} and \ref{Eq:dPIE_formula_sigma-5} into Eq.~\ref{Eq:sigma_formula_compact} gives the following expression for the former term:
\begin{equation}
\begin{aligned}
    &c_p^2(R)=\frac{6}{\pi}\frac{(r_{\rm core}+r_{\rm cut})\nu^2}{r_{\rm core}^2r_{\rm cut}}\\
    &\times\left(\sqrt{\left(\frac{r_{\rm core}}{\nu}\right)^2+R^2} -\frac{r_{\rm core}}{\nu}
    -\sqrt{\left(\frac{r_{\rm cut}}{\nu}\right)^2+R^2} +\frac{r_{\rm cut}}{\nu}\right)^{-1}\\
    &\times\int_0^R R'dR'\int_{R'}^{\infty}\frac{r_{\rm cut}\tan^{-1} \left(\frac{r}{r_{\rm cut}}\right)-r_{\rm core}\tan^{-1} \left(\frac{r}{r_{\rm core}}\right)}{\left(1+\left(\frac{r\nu}{r_{\rm cut}}\right)^2\right)\left(1+\left(\frac{r\nu}{r_{\rm core}}\right)^2\right)}\frac{\sqrt{r^2-R'^2}dr}{r^2}
    \label{Eq:projection factor}
\end{aligned}
\end{equation}
This expression is similar to the one presented by \citet{Bergamini2019} in Equation~C16, and we only have a scaling between the light and total mass tracer in the form of $\nu$ in addition. Thanks to the assumptions on $c_p$, $\nu$ becomes more than simply a spatial scaling between light and mass distribution, and it is linked to the ratio of the luminosity $L$ to the total mass $M_{\rm tot}$ by the following expression:
\begin{align}
    L&=\frac{\pi \sigma_{0\rm,light}^2}{G}\frac{r_{\rm cut, light}^2}{r_{\rm cut, light}-r_{\rm core, light}}\\
    &=\nu^{-1} M_{\rm tot}
\end{align}
\section{New multiply-imaged systems}
\label{app:New_im_sys}
\begin{figure*}
    \begin{minipage}{0.32\linewidth}
    \centering
    \includegraphics[width=\linewidth]{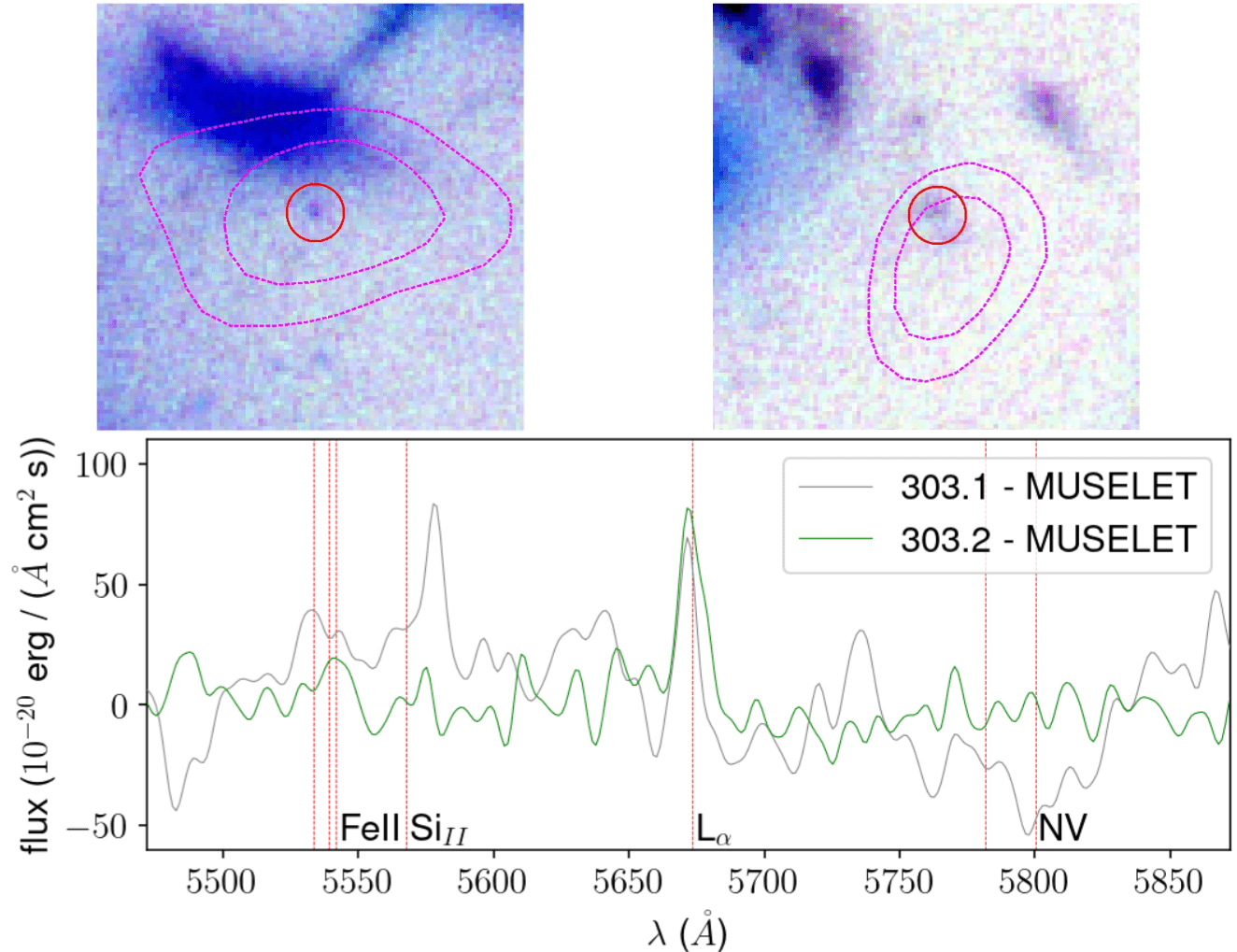}
    \end{minipage}
    \hfill
    \begin{minipage}{0.32\linewidth}
    \centering
    \includegraphics[width=\linewidth]{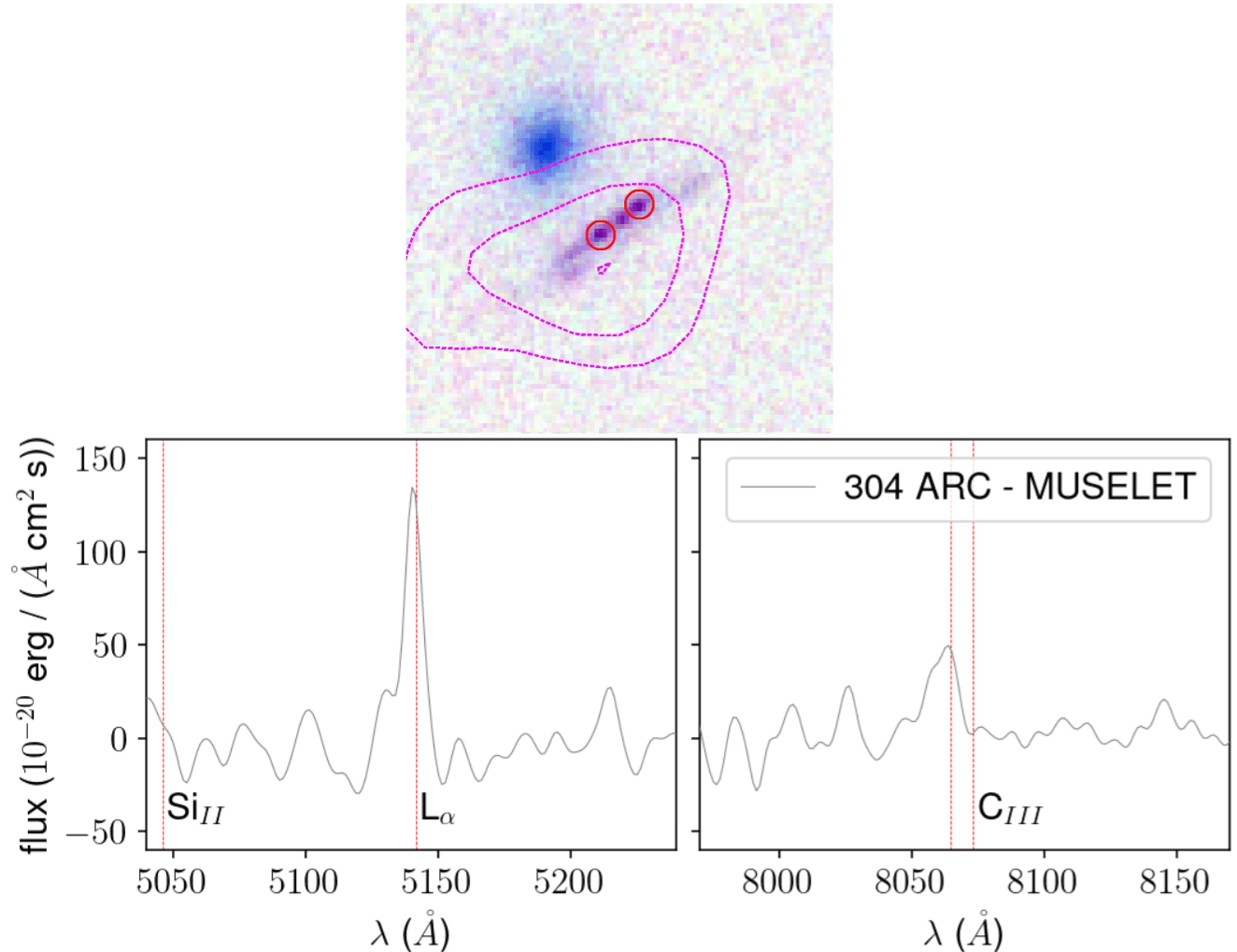}
    \end{minipage}
    \hfill
    \begin{minipage}{0.32\linewidth}
    \centering
    \includegraphics[width=\linewidth]{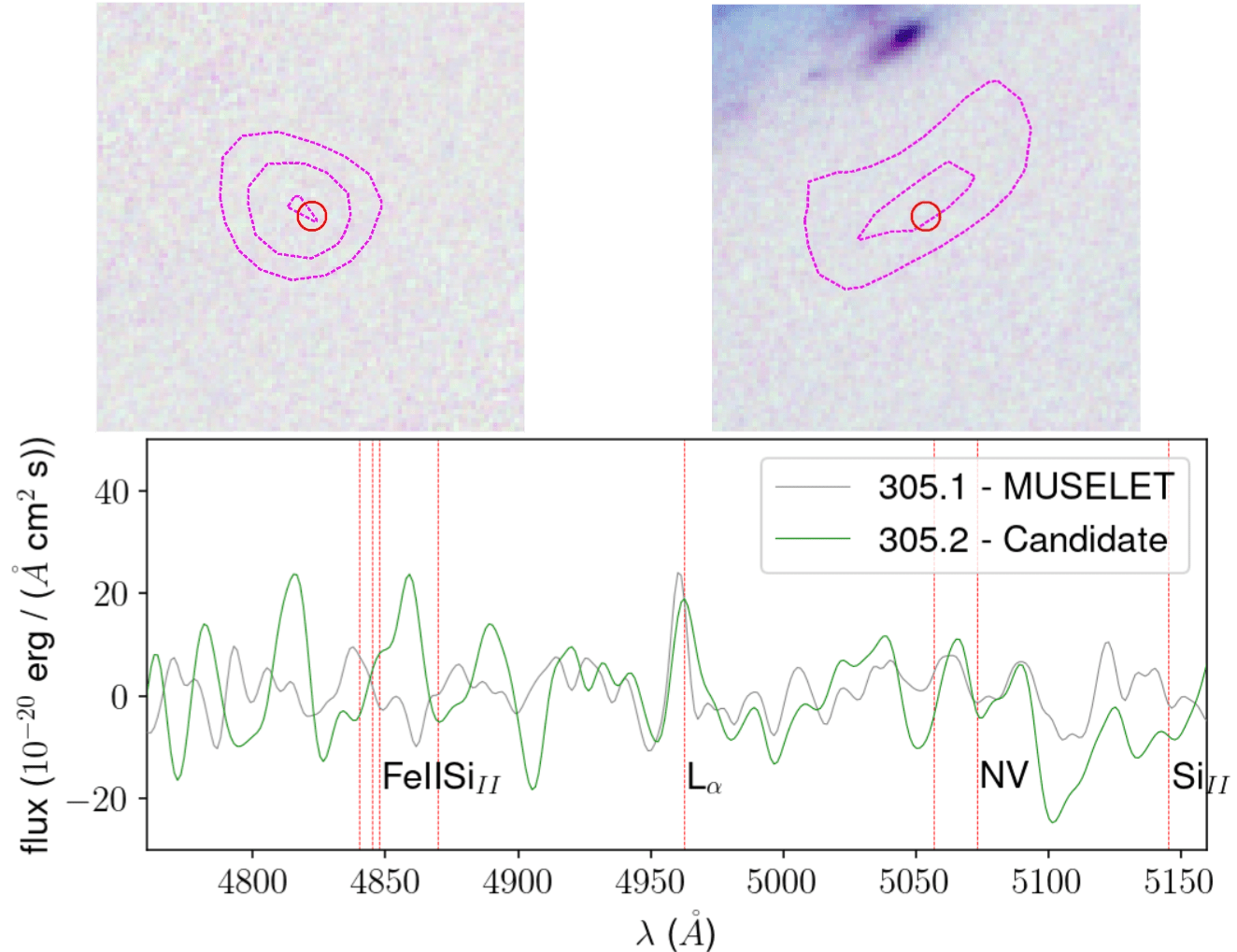}
    \end{minipage}
    \caption{Images and spectra of the newly identified multiply-imaged systems. The top row shows inverted coloured images from \textit{HST} bands F814W, F606W and F435W with magenta contours highlighting the MUSE narrow band associated with the Lyman~$\alpha$ emission. The images are ordered by increasing the index of the multiple images (i.e. \textit{XXX.y} where \textit{XXX} denotes the multiply-imaged system index and \textit{y} the image number in the system), and spectra are shown in the bottom row, with possible emission/absorption lines highlighted. They are centred on emission lines used to define the redshift of the system. In the label, MUSELET indicates a direct detection in the MUSE dataset when Candidate is for an indirect one.}
    \label{fig:sys_mul}
\end{figure*}

\begin{table}
\centering
\begin{tabular}{cccc}
    \hline
    ID & $\alpha(\rm{J2000})$&$\delta(\rm{J2000})$& $z_{\rm system}\pm\sigma$\\
    \hline
    $303.1$&$342.175814$&$-44.536371$&$3.666\pm0.003$\\
    $303.2$&$342.187884$&$-44.540421$&$3.666\pm0.003$\\
    $304.1$&$342.173801$&$-44.541178$&$3.230\pm0.004$\\
    $304.2$&$342.173909$&$-44.541235$&$3.230\pm0.004$\\
    $305.1$&$342.180303$&$-44.524639$&$3.082\pm0.002$\\
    $305.2$&$342.183869$&$-44.525756$&$3.082\pm0.002$\\
    \hline
\end{tabular}
\caption{Positions and redshifts of the newly identified multiply-imaged systems used in this work.}
\label{Tab:pos-sys-multiple}
\end{table}

Fig.~\ref{fig:sys_mul} present the \textit{HST} images as well as the MUSE Lyman~$\alpha$ narrow band and line of the new multiply-imaged systems used in the strong lensing analysis. Most of these images have been detected with the automated emission lines and redshift estimator MUSE Line Emission Tracker (\textsc{MUSELET}) provided in the  \textsc{mpdaf}\footnote{\url{https://mpdaf.readthedocs.io/en/latest/muselet.html}} python package. The previous objects are denoted by the \textsc{MUSELET} label when the Candidate label is for indirect detection based on strong lensing model prediction and narrow-band images of the main emission lines (see \citet{Richard2021} for more details). Notably, the system $305$ is the only one that has no \textit{HST} counterpart as previously seen in similar clusters \citep{Lagattuta2017}. The positions and the measured MUSE redshift of these new multiply-imaged systems are presented in Table~\ref{Tab:pos-sys-multiple}.

\section{SN threshold impact of the X-ray fit}
\label{app:SN_threshold}
\begin{figure*}
    \includegraphics[width=\linewidth]{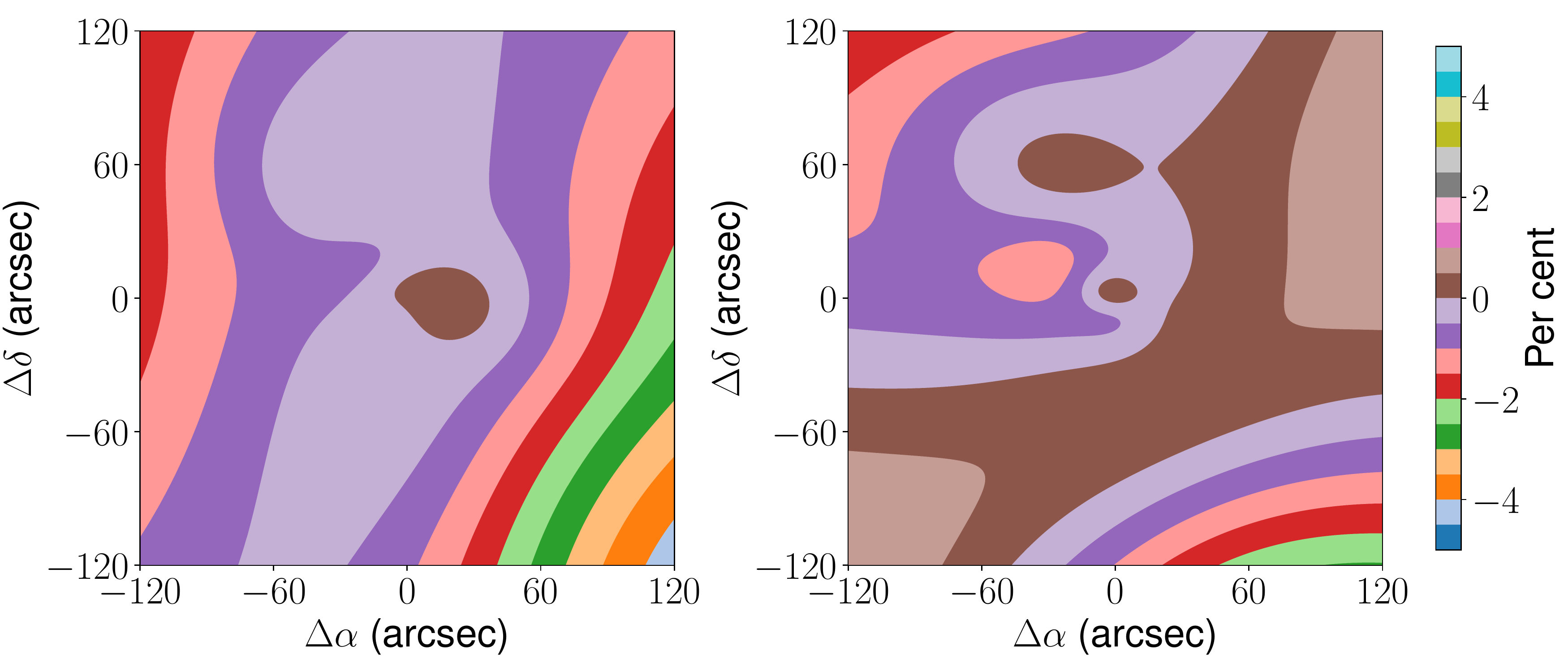}
    
    \caption{Relative differences between the best fit gas mass model made with a SN threshold of $14$ and $10$ (\textit{left panel}), and with a SN of $6$ and $10$ (\textit{right panel}).}
    \label{fig:relative_diff_mass_SN}
\end{figure*}

To assess the possible biases behind our arbitrary choice of SN threshold used in the binning of the cluster field detailed in Sect.~\ref{sect:X-ray-data-analysis}, we fit the gas distribution with only the X-ray constraints, which were made with two other SN thresholds: $6$, $14$. These results are then compared to the X-ray-only run used in the model discrimination of the gas-only part explained in Sect.~\ref{sect:model-discri}. In particular, Fig.~\ref{fig:relative_diff_mass_SN} presents the relative differences between the best-fit model made with an SN of $10$ (i.e. SN used in this article) and the best-fit model associated with an SN of $6$ (right panel) and an SN of $14$ (left panel). As we can see, the differences are quite limited in the case of AS1063 as the differences in the 2D mapping of the gas are less than $4$ per cent on the constrained area and mainly below $2$ per cent if we except regions on the right bottom corner of the plot. We can see that the biases on the reconstruction of the gas distribution are pretty low and are not affecting the results of our analyses.

However, if the reconstruction of the projected mass distribution is negligibly affected, it has more influence on the discrimination model of the gas distribution. Indeed, the likelihood values have changed for the best fit model: -33219 ($SN=14$), -33204 ($SN=10$) and -33179 ($SN=6$). The discrepancies are expected as we used different constraints, but it also affected our goodness of fit procedure as all the percentiles of the expected likelihood distribution have changed. With a SN of $14$; the best fit model likelihood is outside the $"5\sigma"\,\rm CI$ and would not be considered good enough to model the data. In the case of a SN of $6$, the likelihood is inside the $"3\sigma"\,\rm CI$, which is better than what we obtained for the SN of $10$ (i.e. between the bounds of the $"3\sigma"\,\rm CI$ and $"5\sigma"\,\rm CI$).

We showed that the mass measurements provided by our techniques are only affected negligibly by the choice of SN threshold, as the constraints given by the count map are way stronger than the fluctuation of the cooling functions for different binning. These results are dependent of the considered cluster, as the amplitude of the cooling function variations will depend on the temperature gradient between two bins. The only impacting aspect is the bias in the discrimination process, as we could have needed to use more dPIEs to model the gas if we used a SN of $14$. But we can balance this problem by the improvement in the likelihood when we reduce the SN threshold, which increases our confidence in our temperature measurement method. It may also be possible to choose an optimal SN by using the one minimizing the likelihood at given gas mass models.

\section{Mass model dPIE parameters}
\label{sect:dPIE-parameter}
\subsection{Mass model of AS1063}
\begin{table*}
\centering
\begin{tabular}{cccccccc}
    \hline
    ID & $\Delta\alpha$&$\Delta\delta$&$\phi$ &$\epsilon$&$r_{\rm core}$&$r_{\rm cut}$&$\sigma_{0\rm,lt}$\\
    & $\rm{arcsec}$&$\rm{arcsec}$&$\rm{degree}$&&$\rm{kpc}$&$\rm{kpc}$&$\rm{km/s}$\\
    \hline
    DM-BCG&$[0.0]$&$[0.0]$&$-38.4^{+0.3}_{-0.3}(-37.7)$&$0.657^{+0.013}_{-0.013}(0.682)$&$112.9^{+4.6}_{-4.1}(118.8)$&$[3000]$&$1200^{+10}_{-9}(1195)$\\
    DM-NE&$-83.8^{+5.7}_{-5.8}(-81.9)$&$41.8^{+3.8}_{-3.4}(46.6)$&$[0.0]$&$[0.0]$&$[2.5]$&$86^{+42}_{-32}(61)$&$496^{+98}_{-77}(459)$\\
    BCG&$[0.0]$&$[0.0]$&$[-36.29]$&$[0.431]$&$7.5^{+3.5}_{-3.1}(12.5)$&$355^{+128}_{-110}(547)$&$363^{+26}_{-28}(399)$\\
    Gas-1&$9.0^{+3.3}_{-3.1}(8.8)$&$-35.1^{+5.1}_{-6.3}(-47.3)$&$93.0^{+16.8}_{-19.1}(81.7)$&$0.091^{+0.046}_{-0.044}(0.082)$&$662.3^{+18.1}_{-19.5}(689.2)$&$[1250]$&$316^{+9}_{-8}(312)$\\
    Gas-2&$0.3^{+0.2}_{-0.2}(0.4)$&$-1.1^{+0.1}_{-0.1}(-1.1)$&$-12.2^{+1.2}_{-1.3}(-11.1)$&$0.322^{+0.016}_{-0.017}(0.336)$&$66.5^{+2.0}_{-1.9}(63.7)$&$[1250]$&$212^{+5}_{-5}(204)$\\
    Gas-3&$-20.7^{+0.5}_{-0.5}(-19.7)$&$16.6^{+0.5}_{-0.4}(16.1)$&$-31.1^{+1.5}_{-1.6}(-31.9)$&$0.147^{+0.011}_{-0.011}(0.162)$&$169.3^{+3.1}_{-2.9}(173.0)$&$[1250]$&$372^{+6}_{-6}(385)$\\
    \hline
\end{tabular}
\caption{\textit{From left to right:} dPIE's ID, relative right ascension ($\Delta\alpha$), relative declination ($\Delta\delta$), position angle ($\phi$), Ellipticity ($\epsilon$), core radius ($r_{\rm core}$), cut radius ($r_{\rm cut}$) and velocity dispersion ($\sigma_{0\rm,lt}$). Parameters in brackets are fixed a priori, and sampled parameter distributions are presented as $\text{median}^{84 \text{ per cent limit}}_{16 \text{ per cent limit}}(\text{Best-fit})$. The relative coordinates are calculated from the BCG centre at $\alpha({\rm J2000})=342.183213$ and $\delta({\rm J2000})=-44.530897$. We add the $\rm lt$ label to the velocity dispersion as we refer to its definition in the \textsc{lenstool} implementation.}
\label{Tab:dPIE-parameter}
\end{table*}

Table~\ref{Tab:dPIE-parameter} presents statistics from the sampling of the dPIE parameters associated with each mass halo defined independently. There are two main DM haloes, one centred on the BCG positions (DM-BCG) and one on the NE clump (DM-NE). The parameters of the BCG dPIE are presented on the BCG label, while the three potentials used for the gas distribution have the Gas prefix. 

\subsection{Mass models of the mock cluster}
\begin{table*}
\hspace*{-1.cm}
\begin{tabular}{ccccccccc}
    \hline
    Model&ID & $\Delta\alpha$&$\Delta\delta$&$\phi$ &$\epsilon$&$r_{\rm core}$&$r_{\rm cut}$&$\sigma_{0\rm,lt}$\\
    && $\rm{arcsec}$&$\rm{arcsec}$&$\rm{degree}$&&$\rm{kpc}$&$\rm{kpc}$&$\rm{km/s}$\\
    \hline
    Mock&DM-BCG&$[0.0]$&$[0.0]$&$-38.5$&$0.615$&$111.6$&$[3000]$&$1196$\\
    &DM-NE&$-84.5$&$37.7$&$[0.0]$&$[0.0]$&$[2.5]$&$37$&$736$\\
    &BCG&$[0.0]$&$[0.0]$&$[-36.29]$&$[0.431]$&$9.9$&$273$&$367$\\
    &Gas-1&$4.1$&$-12.1$&$119.9$&$0.212$&$519.3$&$[1250]$&$336$\\
    &Gas-2&$0.5$&$-1.1$&$-8.3$&$0.292$&$65.3$&$[1250]$&$215$\\
    &Gas-3&$-22.8$&$17.1$&$-27.9$&$0.133$&$154.0$&$[1250]$&$338$\\
    &&&&&&&&\\
    &&&&&&&&\\
    Ref.&DM-BCG&$[0.0]$&$[0.0]$&$-38.3^{+0.1}_{-0.1}(-38.0)$&$0.573^{+0.008}_{-0.008}(0.569)$&$105.1^{+4.1}_{-3.9}(98.2)$&$[3000]$&$1256^{+9}_{-14}(1259)$\\
    &DM-NE&$-80.2^{+1.8}_{-1.9}(-77.9)$&$46.3^{+1.4}_{-1.4}(48.0)$&$[0.0]$&$[0.0]$&$[2.5]$&$274^{+18}_{-36}(291)$&$459^{+23}_{-21}(445)$\\
    &BCG&$[0.0]$&$[0.0]$&$[-36.29]$&$[0.431]$&$9.2^{+5.5}_{-3.9}(7.2)$&$162^{+171}_{-89}(64)$&$363^{+22}_{-22}(340)$\\
    &&&&&&&&\\
    &&&&&&&&\\
    Gas&DM-BCG&$[0.0]$&$[0.0]$&$-38.5^{+0.1}_{-0.1}(-38.6)$&$0.613^{+0.008}_{-0.008}(0.616)$&$107.1^{+3.9}_{-4.5}(106.3)$&$[3000]$&$1200^{+10}_{-13}(1190)$\\
    only&DM-NE&$-81.8^{+1.8}_{-2.0}(-79.8)$&$43.1^{+1.6}_{-1.6}(42.6)$&$[0.0]$&$[0.0]$&$[2.5]$&$224^{+50}_{-74}(239)$&$396^{+46}_{-30}(386)$\\
    &BCG&$[0.0]$&$[0.0]$&$[-36.29]$&$[0.431]$&$9.7^{+5.1}_{-4.1}(9.1)$&$192^{+126}_{-103}(260)$&$364^{+21}_{-20}(352)$\\
    &Gas-1&$0.7^{+2.3}_{-2.4}(2.9)$&$-8.3^{+2.8}_{-2.9}(-9.5)$&$125.7^{+5.5}_{-5.8}(126.5)$&$0.204^{+0.035}_{-0.038}(0.217)$&$467.0^{+16.0}_{-19.2}(487.2)$&$[1250]$&$345^{+9}_{-9}(348)$\\
    &Gas-2&$0.3^{+0.2}_{-0.3}(0.2)$&$-0.9^{+0.1}_{-0.1}(-0.8)$&$-8.7^{+1.5}_{-1.5}(-8.8)$&$0.307^{+0.017}_{-0.017}(0.301)$&$68.1^{+2.2}_{-2.3}(69.6)$&$[1250]$&$226^{+6}_{-6}(229)$\\
    &Gas-3&$-24.3^{+0.7}_{-0.7}(-24.6)$&$19.0^{+0.7}_{-0.7}(18.9)$&$-26.0^{+6.7}_{-5.4}(-25.9)$&$0.084^{+0.022}_{-0.023}(0.067)$&$147.4^{+4.7}_{-4.8}(146.8)$&$[1250]$&$314^{+11}_{-11}(314)$\\
    &&&&&&&&\\
    &&&&&&&&\\
    FP&DM-BCG&$[0.0]$&$[0.0]$&$-38.3^{+0.1}_{-0.1}(-38.3)$&$0.572^{+0.008}_{-0.008}(0.575)$&$109.1^{+3.6}_{-3.4}(105.2)$&$[3000]$&$1252^{+12}_{-10}(1265)$\\
    only&DM-NE&$-82.8^{+1.9}_{-1.8}(-79.6)$&$41.6^{+1.5}_{-1.5}(39.9)$&$[0.0]$&$[0.0]$&$[2.5]$&$230^{+50}_{-76}(271)$&$454^{+51}_{-29}(409)$\\
    &BCG&$[0.0]$&$[0.0]$&$[-36.29]$&$[0.431]$&$12.2^{+5.9}_{-4.4}(9.3)$&$245^{+102}_{-118}(107)$&$369^{+21}_{-21}(343)$\\
    &&&&&&&&\\
    &&&&&&&&\\
    All&DM-BCG&$[0.0]$&$[0.0]$&$-38.5^{+0.1}_{-0.1}(-38.5)$&$0.609^{+0.009}_{-0.008}(0.611)$&$107.7^{+3.8}_{-3.8}(104.9)$&$[3000]$&$1194^{+11}_{-10}(1195)$\\
    &DM-NE&$-83.0^{+1.7}_{-1.7}(-83.7)$&$40.1^{+1.4}_{-1.6}(40.8)$&$[0.0]$&$[0.0]$&$[2.5]$&$97^{+69}_{-44}(54)$&$512^{+146}_{-89}(662)$\\
    &BCG&$[0.0]$&$[0.0]$&$[-36.29]$&$[0.431]$&$11.6^{+5.7}_{-4.7}(4.6)$&$238^{+110}_{-113}(267)$&$362^{+21}_{-20}(323)$\\
    &Gas-1&$1.0^{+2.2}_{-2.3}(3.2)$&$-8.3^{+2.9}_{-2.7}(-11.3)$&$125.1^{+5.3}_{-6.0}(115.9)$&$0.21^{+0.034}_{-0.037}(0.171)$&$462.6^{+15.2}_{-17.1}(494.1)$&$[1250]$&$344^{+9}_{-9}(340)$\\
    &Gas-2&$0.2^{+0.3}_{-0.3}(0.1)$&$-0.9^{+0.1}_{-0.2}(-0.8)$&$-8.8^{+1.6}_{-1.6}(-9.6)$&$0.306^{+0.017}_{-0.016}(0.313)$&$68.1^{+2.2}_{-2.2}(70.4)$&$[1250]$&$226^{+6}_{-6}(230)$\\
    &Gas-3&$-24.4^{+0.7}_{-0.7}(-24.3)$&$19.0^{+0.7}_{-0.7}(19.0)$&$-24.9^{+6.6}_{-5.3}(-31.4)$&$0.083^{+0.022}_{-0.022}(0.086)$&$147.7^{+4.5}_{-4.8}(149.7)$&$[1250]$&$313^{+11}_{-10}(319)$\\
    \hline
\end{tabular}
\caption{Same as Table~\ref{Tab:dPIE-parameter} but for the mass model built for the mock cluster. Parameters of the mock are presented under the label \textit{mock}. The four other labels are associated with the reference model (\textit{Ref.}), the model with only the gas improvements (\textit{Gas only}), with the fundamental plane only (\textit{FP only}) and finally both with (\textit{All}).}
\label{Tab:dPIE-parameter-mock}
\end{table*}

Table~\ref{Tab:dPIE-parameter-mock} presents statistics from the sampling of the dPIE parameters associated with each modelling tested on the mock cluster as well as the mock parameters. The dPIE labels are the same as in Table~\ref{Tab:dPIE-parameter}. The relevant parameters for the dPIE modelled according to a galaxy relation are presented in Table~\ref{Tab:dPIE-parameter-mock-potfile}.

\begin{table*}
\centering
\begin{tabular}{cccccccc}
    \hline
    Model& $r^*_{\rm cut}$&$\sigma^*_{0\rm,lt}/\sigma^*_{e}$&$\nu$&$b$&$a$&$c$\\
    &$\rm{kpc}$&$\rm{km/s}$&&&&&\\
    \hline   Mock&$9.57$&$190$&$2.14$&$0.296$&$-0.565$&$1.473$\\
    &&&&&&&\\    
    Ref.&$7.1^{+2.38}_{-2.38}(8.19)$&$143^{+20}_{-10}(137)$&-&-&-&$[2]$\\
    &&&&&&&\\    
    Gas only&$5.9^{+2.22}_{-1.91}(6.09)$&$147^{+19}_{-14}(137)$&-&-&-&$[2]$\\
    &&&&&&&\\    
    FP only&$4.96^{+3.0}_{-1.88}(2.92)$&$193^{+6}_{-7}(197)$&$2.44^{+0.38}_{-0.36}(2.06)$&$0.345^{+0.024}_{-0.024}(0.362)$&$-0.537^{+0.028}_{-0.025}(-0.524)$&$1.357^{+0.11}_{-0.113}(1.571)$\\
    &&&&&&&\\    
    All&$3.88^{+1.91}_{-1.5}(1.74)$&$194^{+6}_{-6}(202)$&$1.86^{+0.35}_{-0.33}(1.62)$&$0.329^{+0.024}_{-0.024}(0.335)$&$-0.548^{+0.029}_{-0.028}(-0.539)$&$1.334^{+0.118}_{-0.115}(1.55)$\\
    \hline
\end{tabular}
\caption{\textit{From left to right:} Model's labels, cut radius ($r_{\rm cut}$) and velocity dispersion ($\sigma^*_{e\rm,lt}/\sigma^*_{0\rm,lt}$; Projected or central) of the reference galaxy. Parameters of the fundamental plane relations with the ratio between $R_e$ and $R_{\rm M_{\rm Tot}/2}$, and the fundamental plane parameters. Labels for each model are the same as in Table~\ref{Tab:dPIE-parameter-mock}, and we use the same presentation for the parameter values as in Table~\ref{Tab:dPIE-parameter}. The only exception is the use of "-" when the parameter is not relevant to the considered relations.}
\label{Tab:dPIE-parameter-mock-potfile}
\end{table*}

\bsp	
\label{lastpage}
\end{document}